\documentclass[twocolumn]{aastex63}

\usepackage{amsmath}
\usepackage{natbib}

\usepackage{anyfontsize}
\usepackage{indentfirst}
\usepackage{comment}
\usepackage{multirow} 
\usepackage{booktabs}
\usepackage{bigints}
\usepackage{mathrsfs,amsmath} 
\usepackage{makecell}
\usepackage{amsmath}
\usepackage{xcolor}

\usepackage[normalem]{ulem}
\useunder{\uline}{\ul}{}
\usepackage{hyperref}
\usepackage{lineno}

\usepackage{gensymb}

\hypersetup{linkcolor=cyan,citecolor=magenta,filecolor=yellow,urlcolor=blue}

\hypersetup{linkcolor=magenta, citecolor=cyan, filecolor=yellow, urlcolor=blue}

\shorttitle{Time-averaging Polarimetric and Spectral Properties of Gamma-Ray Bursts}
\shortauthors{Li \& Shakeri}

\begin{document}

\title{Time-averaging Polarimetric and Spectral Properties of Gamma-Ray Bursts}

\author{Liang Li}
\affiliation{Institute of Fundamental Physics and Quantum Technology, Ningbo University, Ningbo, Zhejiang 315211, People's Republic of China}
\affiliation{School of Physical Science and Technology, Ningbo University, Ningbo, Zhejiang 315211, People's Republic of China}
\affiliation{ICRANet, Piazza della Repubblica 10, I-65122 Pescara, Italy}
\affiliation{ICRA and Dipartimento di Fisica, Universit\`a  di Roma ``La Sapienza'', Piazzale Aldo Moro 5, I-00185 Roma, Italy}
\affiliation{INAF -- Osservatorio Astronomico d'Abruzzo, Via M. Maggini snc, I-64100, Teramo, Italy}

\author{Soroush~Shakeri}
\affiliation{Department of Physics, Isfahan University of Technology, Isfahan 84156-83111}
\affiliation{Iranian National Observatory, Institute for Research in Fundamental Sciences (IPM), P. O. Box 19395-5531 Tehran, Iran}
\affiliation{ICRANet-Isfahan, Isfahan University of Technology, Isfahan 84156-83111, Iran}

\correspondingauthor{Liang~Li, Soroush~Shakeri}
\email{liang.li@icranet.org; s.shakeri@iut.ac.ir}

\begin{abstract}

The composition and radiation mechanism of gamma-ray bursts (GRBs) within jets continue to be hotly debated. Investigating the joint polarimetric and spectral properties is crucial for understanding GRB composition and radiation mechanism. Various jet properties, such as ``kinetic-energy-dominated" (KED), ``Poynting-flux-dominated" (PFD), and ``hybrid-dominated" (HD) relativistic outflows, have been inferred from observed GRB spectra, with expectations of differing polarization levels among them. In this study, we analyzed a sample of 27 GRBs detected by the Gamma-ray Burst Monitor on board the NASA Fermi Gamma-ray Space Telescope, focusing on 26 bursts with significant polarization measurements. Our analysis revealed that 16 bursts (62\%) were predominantly associated with the ``PFD'' jet type, while 10 bursts (38\%) were classified as HD, implying that photosphere emission may also be a possible mechanism powering the high levels of polarization. Notably, no bursts were identified as KED-type. We found distinct polarization properties, with HD-type bursts exhibiting consistently higher polarization levels than PFD-type bursts. We proposed models incorporating ordered and random magnetic field configurations specific to hybrid jets.

\end{abstract}


\section{Introduction} \label{sec:intro}

Gamma-ray bursts (GRBs) are among the most energetic and luminous transient phenomena in the Universe, occurring at cosmological distances. Despite decades of investigation, the fundamental nature of jet composition (whether characterized by a hot baryonic-dominated fireball or a cold Poynting-flux-dominated, PFD, outflow), as well as the underlying radiation and energy dissipation mechanisms (such as synchrotron radiation or Comptonization of quasi-thermal emission originating from the photosphere) in GRB physics, remain unclear \citep[e.g.,][]{Rees1994, Meszaros2000, Rees2005, Peer2006a, Dai2006, Peer2015a, PeEr2017, Zhang2018, Begue2022}.

Two crucial clues can, in principle, help diagnose the jet composition, radiation mechanism, and energy dissipation mechanism of GRBs. A conventional approach involves examining the spectral properties of prompt emission. Theoretically, a thermal component originating from photosphere emission or a nonthermal component originating from synchrotron radiation, possibly also from inverse Compton scattering, is often expected to be present in GRB spectral analysis. Phenomenologically, GRB spectra in the keV-MeV energy range are typically well delineated by an empirical function, known as the Band function \citep{Band1993}, which is considered a nonthermal spectrum. This function features a smoothly broken power law, with peak energy $E_{\rm p} \simeq 200-300$ keV (the energy at which most of the energy is released) in $\nu F \nu$ space and asymptotic power-law photon indices below ($\alpha \simeq -0.8$) and above ($\beta \simeq -2.3$) the break energy \citep[e.g.,][]{Kaneko2006,Goldstein2012,Li2021b}. The low-energy spectra observed during the GRB prompt emission phase are closely related to the electron energy distribution \citep[e.g.,][]{Preece1998, Lloyd2000a, Geng2018}, providing insights into the radiation mechanisms and jet properties of GRBs. For instance, synchrotron emission predicts two distinct $\alpha$ values: $\alpha$=-3/2 and $\alpha$=-2/3 (the so-called line of death of synchrotron emission; \citealt{Preece1998}) correspond to the fast-cooling and slow-cooling synchrotron emission, respectively. It has been shown that synchrotron emission in the presence of a decaying magnetic field can reproduce the Band-like spectrum of the GRB prompt phase \citep{Lan:2021tja}. Despite years of observations revealing diverse spectral properties in GRBs, a single spectral model like the Band function alone may struggle to accurately capture all spectral shapes. For example, a recent study \citep{Acuner2020} suggests that the spectra that favor the photospheric model all have low-energy power-law indices $\alpha$ $\sim>$-0.5, as long as the data have a high significance. 

Decades of observations have revealed, however, that GRBs have diverse spectral properties, making it difficult for a single spectral model (such as Band) to accurately characterize all the spectral shapes. Time-resolved and time-integrated spectral analysis inferred from broadband Fermi observations have revealed the remarkable diversity in GRB prompt emission spectral properties \citep[e.g.,][]{Abdo2009a,Ryde2010,Axelsson2012,Ravasio2018,Acuner2019,Li2019a,Li2019b,Li2021b,Deng2022,Li2023a}. A kinetic-energy-dominated (KED) jet characterized by a quasi-thermal Planck-like spectrum has been detected in some bursts, such as GRBs 090902B and 220426A (e.g., \citealt{Abdo2009,Ryde2010,Deng2022,WangYun2022,SongXY2022}). On the other hand, a cold PFD outflow, characterized by a Band (or cutoff power law,\footnote{Recent studies \citep{Li2022,Li2023c} supported by several pieces of additional evidence (e.g., inconsistent spectral parameter distributions and distinct Amati and Yonetoku correlations) have shown that Band-like spectra and CPL-like spectra may originate from distinct radiation processes.} CPL) -like function, has also been inferred in other bursts, including GRB 080916C, GRB 130606B, and many others (e.g., \citealt{Abdo2009a,Zhang2009,Li2023c}). Moreover, a hybrid-dominated (HD) relativistic outflow that incorporates both a hot fireball component and a cold Poynting flux component has also been observed. This HD outflow is characterized either by a composite spectral scenario, combining a nonthermal component and a thermal component, as seen in GRBs 100724B, 110721A, 150314A, 190114C, and several others (e.g., \citealt{Axelsson2012,Guiriec2011,Wang2019,Li2023a,Li2023c}), or by a transition from a fireball to a PFD outflow within a single burst, as evidenced in GRBs 140206A, 160625B, and several others (e.g., \citealt{Li2019a}).
As such, we define the following
\begin{itemize}
\item The PFD-type outflow. Characterized by a single nonthermal (Band-like) spectral component, as exemplified by GRB 080916C \citep{Abdo2009a} and GRB 131231A \citep{Li2019b}.
\item The KED-type outflow. Dominated by a thermal (blackbody-like, BB-like) spectral component, observed in GRB 090902B \citep{Ryde2010} and GRB 220426A \citep{Deng2022}.
\item The HD-type outflow.  Featuring a hybrid spectrum (Band+BB-like) with both thermal and nonthermal components, as seen in GRB 110721A \citep{Axelsson2012} and GRB 140206A \citep{Li2019a}.
\end{itemize}

In practice, both time-integrated \cite[e.g.,][]{Li2023c} and time-resolved \cite[e.g.,][]{Li2021b} spectral analysis is frequently used to diagnose jet properties. The former represents average spectral properties, and thus the entire emission period is treated as a single-time event; the latter treats the entire emission period as divided into multiple-time events, and spectral analyses are therefore performed on each event individually. The time-integrated method relies on the statistical results from a large sample to ensure that a more trustworthy result is available, while a time-resolved technique ensures consistency (e.g., angular structure and magnetic field configuration) within the same burst for more reliable results. In this task, the former (time-integrated manner) is our primary interest, and the latter (time-resolved manner) will be used elsewhere. Our approach invokes identifying the optimal model by comparing various frequently used spectral models. This comparison is performed using algorithms such as reduced $\chi^{2}$ (referred to as the goodness of fit), or statistical information criteria such as the Akaike information criterion (AIC; \citealt{Akaike1974}, defined as ${\rm AIC}=2k-2{\rm ln}\mathcal{L}$, where $k$ is the number of estimated parameters in the model and $\mathcal{L}$ is the maximized value of the likelihood function for the model) and the Bayesian information criterion (BIC; \citealt{Schwarz1978}, defined as ${\rm BIC}=k{\rm ln} N-2{\rm ln}\mathcal{L}$, where $\mathcal{L}$ is the maximum likelihood, $k$ is the number of parameters of the model, and $N$ is the number of data points) to evaluate the selection of the best model. Once the best spectral model is identified, the jet properties inferred from the spectral properties can be evaluated.

An alternative method is to investigate GRB polarization properties. Theoretically, photon polarizations play a key role in understanding the composition, angular structure, geometric configuration, magnetic composition, and magnetic field configuration of GRB jets and the radiation mechanism of GRB jets \citep{Granot:2003dy,Lyutikov:2003bz,2003ApJ83G,2009ApJ...698.1042T, 2013MNRAS.428.2430L, 2014IJMPD..2330002Z, Zhang2019, Gill2021, Gill2024}. While magnetic field configurations with relatively large coherence lengths of more than gyroradius of charged particles can generate the same energy spectrum via synchrotron mechanism, the level of polarization may be significantly different for various magnetic field structures. Therefore, a joint spectral and polarization analysis is essential to determine the magnetic field structure in outflow materials of GRBs \citep{Kole2020}. For instance, the central engine is anticipated to generate strong magnetic fields (a highly magnetized jet) and launch them concurrently with the relativistic jets. It is unclear, nevertheless, whether the GRB emission is caused by shock dissipation or magnetic reconnection and whether the outflow is dominated by the photosphere or synchrotron emission \citep{2009ApJ...698.1042T}. 

Indeed, the generation of the polarization signal can be intrinsic to the emission process or due to the propagation effects \citep{Shakeri:2018qal,Teboul:2020yis}. Two major emission models, induced synchrotron emission \citep{Rybicki1979} and photosphere emission \citep{Lundman2014}, have been proposed to explain the intrinsic polarization properties of relativistic jets during the prompt emission phase. In the synchrotron emission model, previous studies \citep[e.g.,][]{Toma2009,Lan2020} have shown that higher values of linear polarized signal (with polarization degree $\pi$ ranging from $20\%$ to $70\%$) are expected when there is an ordered magnetic field in the synchrotron emission from a relativistic jet. In contrast, jets with random magnetic fields produce lower levels of polarization, primarily because the polarization tends to cancel out, resulting in a net polarization degree close to zero for an on-axis observer. Consequently, a polarization detection of less than $15\%$ is believed to originate from a random magnetic field configuration within the jet \citep{Mao:2013gha}. For example, if the emission is dominated by the internal shock model, $\pi$ is anticipated to range from $10\%$ (in the case of a mixed magnetic field configuration) to $70\%$ (with a large-scale ordered magnetic field configuration). On the other hand, the dissipative photosphere model predicts a relatively low degree of polarization in the $\gamma$-ray band. Furthermore, a structured jet photosphere model might generate polarized photons through Compton scattering, with the degree of polarization being energy-dependent, similar to the synchrotron model in ordered magnetic fields. For instance, it has been demonstrated that if the jet has a significant structure, the model may produce polarizations of up to 40\% within $\delta \Theta \sim \Gamma^{-1}$. However, in the absence of dissipation and below the photosphere, the polarization is typically limited to values below $15\%$-$20\%$ \citep{Gill:2018qrz}. To restrict these models, a high-sensitivity $\gamma$-ray polarimeter with a broad bandpass capable of detecting energy-dependent polarization signals is required \citep{2014IJMPD..2330002Z,Ito2014,Lundman2014,Lundman2018}.

It is highly speculated that the prompt emission is likely expected to be strongly polarized owing to its nonthermal origin, as indicated by the Band-like spectrum observed in most GRB prompt emission spectra. Observationally, higher levels of linear polarization measured from prompt $\gamma$-ray emission have been reported by several authors \citep[e.g.,][]{Coburn2003,Willis2005,McGlynn2007,2012ApJ...758L...1Y}. For instance, a higher polarization degree $\pi=80\pm20\%$ in GRB 021006 was claimed by \cite{Coburn2003} using the RHESSI data. Later, several other cases were also reported, such as GRB 930131 ($\pi>35\%$, \citealt{Willis2005}), GRB 960924 ($\pi>50\%$, \citealt{Willis2005}), GRB 041219A ($\pi=96\pm40\%$, \citealt{McGlynn2007}), GRB 100826A ($\pi=27\pm11\%$, \citealt{2011ApJ...743L..30Y}), GRB 110301A ($\pi=70\pm22\%$, \citealt{2012ApJ...758L...1Y}), and GRB 110721A ($\pi=84^{+16}_{-28}\%$, \citealt{2012ApJ...758L...1Y}). Subsequent observations in the optical band during the afterglow emission phase showed relatively low linear polarization levels compared to prompt $\gamma$-ray emission, for example, GRB 060418 ($\pi<8\%$, \citealt{Mundell2007Sci}), GRB 090102 ($\pi=10.1\pm1.3\%$, \citealt{2009Natur.462..767S}), GRB 091208B  ($\pi=10.4\pm2.5\%$, \citealt{2012ApJ...752L...6U}), and 120308A ($\pi=28\pm4\%$, \citealt{2013Natur.504..119M}). However, higher degrees of polarization are still expected to be measured from early reverse shocks, up to approximately 60\%.

Following these lines of argument, emissions dominated by different types of jets (KED, PFD, and HD) may exhibit varying levels of prompt GRB polarization measurements\footnote{Note that polarization measurement encompasses a range where $\pi$ varies from $0\%$ to $100\%$ ($0\% \leq \pi \leq 100\%$), including the case of nondetection (0\%). Therefore, obtaining a measurement that is consistent with zero is still considered a valid measurement. Polarization detection, however, reveals a different interpretation: $0\% < \pi \leq 100\%$, where $0\%$ is not included.}. Consequently, a different level of polarization degrees is expected due to different types of GRB jets, assuming other conditions are basically the same. This may provide a method to explore the correlations between polarization properties, spectral properties, and jet properties. A possible connection between the spectral and polarization properties has not yet been firmly established, though recent works provide some statistical results \citep{Chattopadhyay2019,Kole2020}. Therefore, this study aims to investigate possible connections that exist between polarization and jet properties. Practically, several important factors need to be taken into account in our analysis. Firstly, we focus on the bursts detected by the Gamma-ray Burst Monitor (GBM, 8 KeV-40 MeV, \citealt{Meegan2009}) on board the NASA Fermi Gamma-ray Space Telescope to well evaluate the frequently used GRB spectral models and thus to diagnose the jet properties. Secondly, we select bursts with available polarization observations and spectral data during the same period to enable a direct comparison of spectral and polarization properties. Lastly, the presence of systematic error in polarization measurements, which vary among different instruments, necessitates a high-significance signal for reliable results. In this paper, we collect a sample of the Fermi-GBM-detected bursts along with the polarized measurements reported in the literature using the time-integrated spectral and polarization analysis approach based on their statistical results, aiming to establish a connection between the polarization and jet properties of GRBs.

The paper is organized as follows. The sample and Methodology are presented in Section 2 and Section 3, respectively. Our results and their physical implications are summarized in Section 4 and Section 5, respectively. The conclusion is presented in Section 6. Throughout the paper, the standard $\Lambda$-CDM cosmology with the parameters $H_{0}= 67.4$ ${\rm km s^{-1}}$ ${\rm Mpc^{-1}}$, $\Omega_{M}=0.315$, and $\Omega_{\Lambda}=0.685$ are adopted \citep{PlanckCollaboration2018}.

\section{Methodology} \label{sec:Methodology}

\subsection{The sample} \label{sec:Sample}

A comprehensive database of GRB polarimetric observations has been created in a recent work \citep{Li2022a} by extensively searching for GRBs in the literature with reported polarization measurements. A total of 73 bursts with polarization detections were included in the database, covering a broad wavelength range from radio to optical, X-ray, and $\gamma$-ray emission (see Table 1 in \citealt {Li2022a}). The prompt emission data of these bursts were observed by various satellites (Fermi, \emph{Swift}, BeppoSAX, and BATSE). Among these satellites, Fermi covers the broadest energy range in the observation, making it crucial for evaluating current spectral models. The GBM (8 KeV-40 MeV, \citealt{Meegan2009}) and the Large Area Telescope (LAT, 20 MeV- 300 GeV, \citealt{Atwood2009}), on board the NASA Fermi Gamma-ray Space Telescope, together provide unprecedented spectral coverage for 7 orders of magnitude in energy (from $\sim$8 keV to $\sim$300 GeV). Our statistical analysis in the current work includes the prompt $\gamma$-ray emission spectral analysis and explores the connection between the spectrum and polarization based on the Fermi-detected bursts. We specifically focus on GRBs with polarization measurements taken during the prompt emission in the $\gamma$-ray band to compare polarization and spectral properties simultaneously. Our sample has been refined to include 27 bursts (refer to Table \ref{tab:Sample} for details).

\subsection{Spectral Analysis Techniques}

In order to diagnose the jet properties for a given burst, a refined time-integrated spectral analysis is required. Following the standard practice \citep{Yu2019, Li2019a, Burgess2019, Dereli-Begue2020, Li2021b} provided by the Fermi Science Term, the spectral analysis is performed using a pure Python package called the {\tt the Multi-Mission Maximum Likelihood Framework} ({\tt 3ML},\citealt{Vianello2015}), and a Bayesian approach and Markov Chain Monte Carlo (MCMC) iterations are used to explore the best parameter space. The main steps of our spectral analysis include selecting detectors, sources, and background intervals; choosing the same observed epoch during the prompt emission phase for polarization and spectral data; calculating the significance ($S$, \citealt{Vianello2018a}) for each burst; fitting all the spectral data by using various GRB spectral models and their hybrid version, like power law, BB, CPL, Band function, power law+BB, CPL+BB, and Band+BB; using a fully Bayesian approach to obtain the best model parameters; and comparing models by using the AIC (\citealt{Akaike1974}, defined as ${\rm AIC}=2k-2{\rm ln}\mathcal{L}$, where $k$ is the number of estimated parameters in the model and $\mathcal{L}$ is the maximized value of the likelihood function for the model) and BIC (\citealt{Schwarz1978}, defined as ${\rm BIC}=k{\rm ln} N-2{\rm ln}\mathcal{L}$, where $\mathcal{L}$ is the maximum likelihood, $k$ is the number of parameters of the model, and $N$ is the number of data points). We use both AIC and BIC because BIC is recommended for nested models (e.g., Band versus Band+BB), while AIC is favored for models that are not nested (e.g., Band versus CPL). For a given set of models, the preferred model is the one that provides the lowest AIC and BIC scores. We consequently accept the Band+BB model as the preferred one if the difference between the BIC of Band+BB and the BIC of Band is less than -10, i.e., $\Delta {\rm BIC = BIC_{Band+BB}-BIC_{Band}<-10}$ \citep{Li2023c}. We also double-check their corner–corner plots of the posteriors to determine well-constrained $\beta$ (Band-like) and unconstrained $\beta$ (CPL-like) to select the preferred model between Band and CPL models \citep{Li2022}.

\section{Results}\label{sec:Results}

\subsection{Spectral Properties and Their Inferred Jet Properties}\label{Spectral}

\begin{figure*}
\includegraphics[width=0.5\hsize,clip]{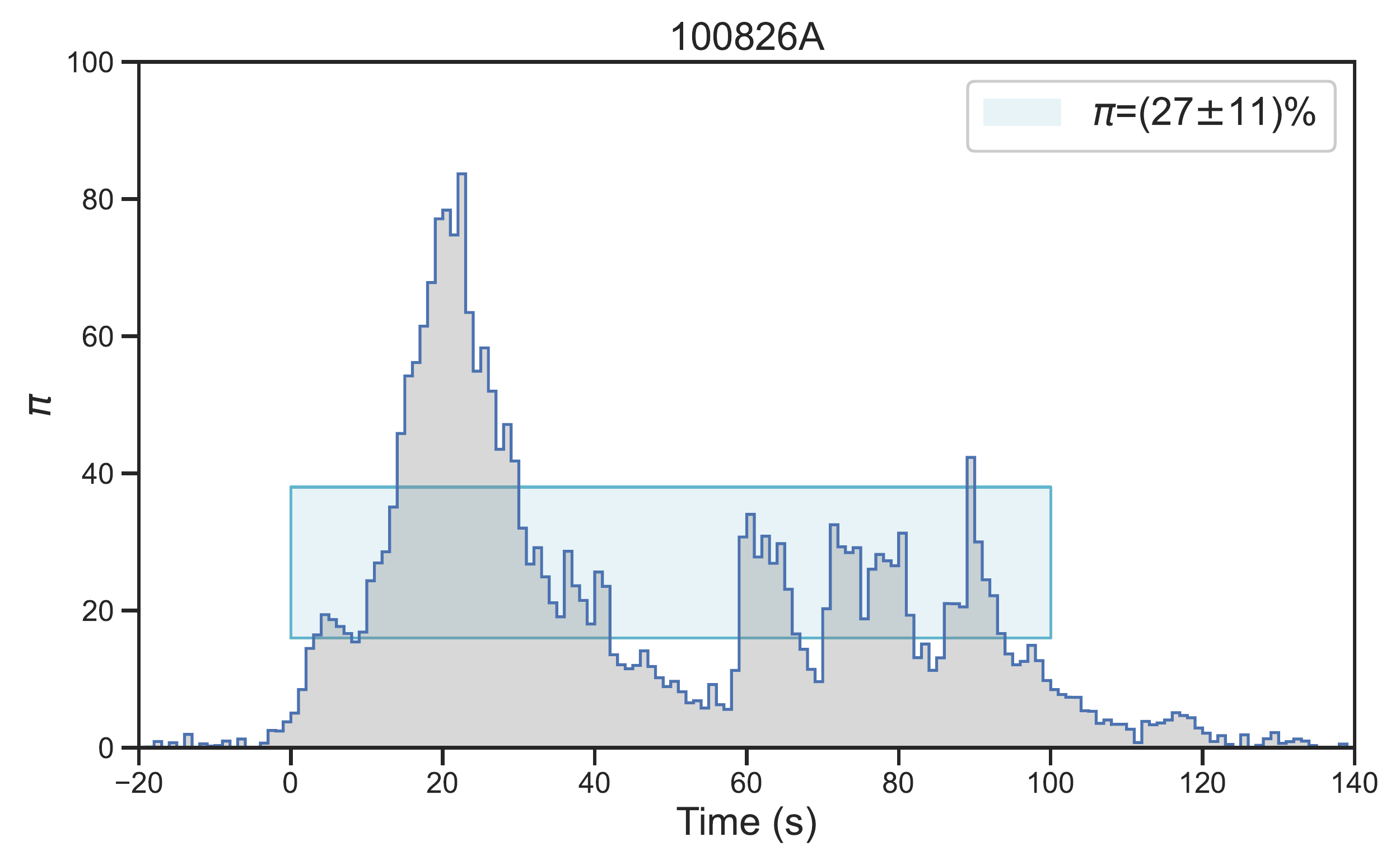}
\includegraphics[width=0.5\hsize,clip]{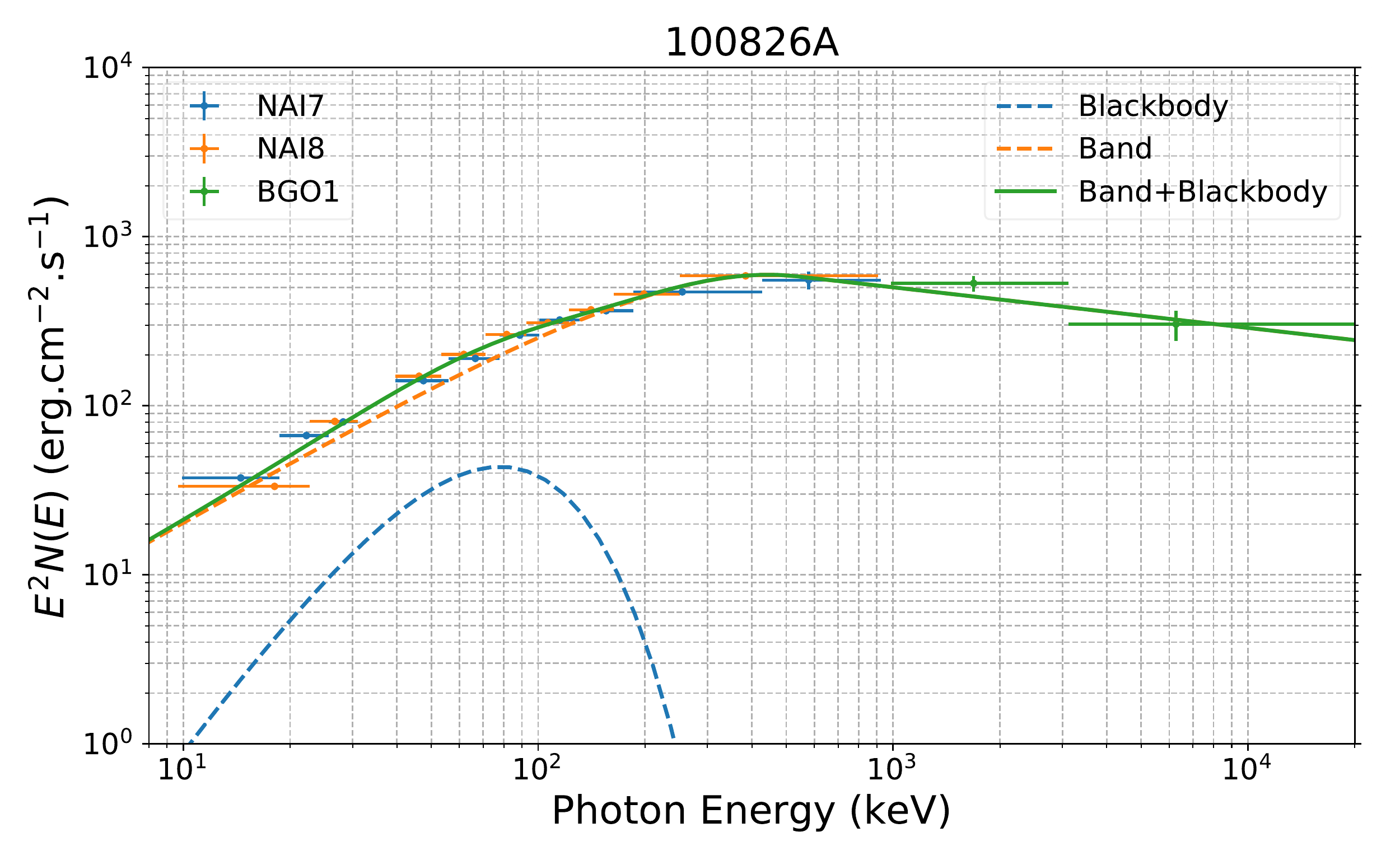}
\includegraphics[width=0.5\hsize,clip]{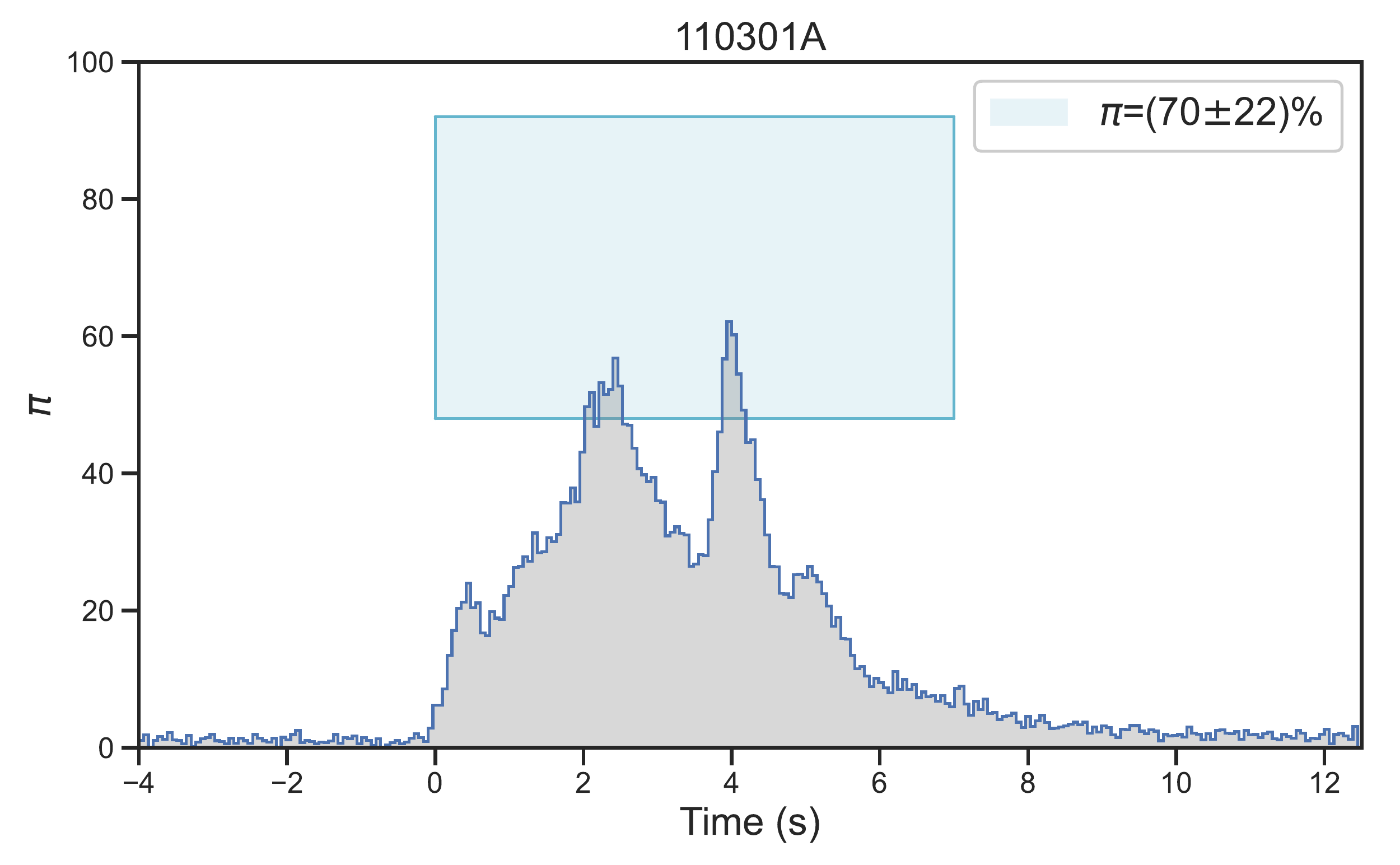}
\includegraphics[width=0.5\hsize,clip]{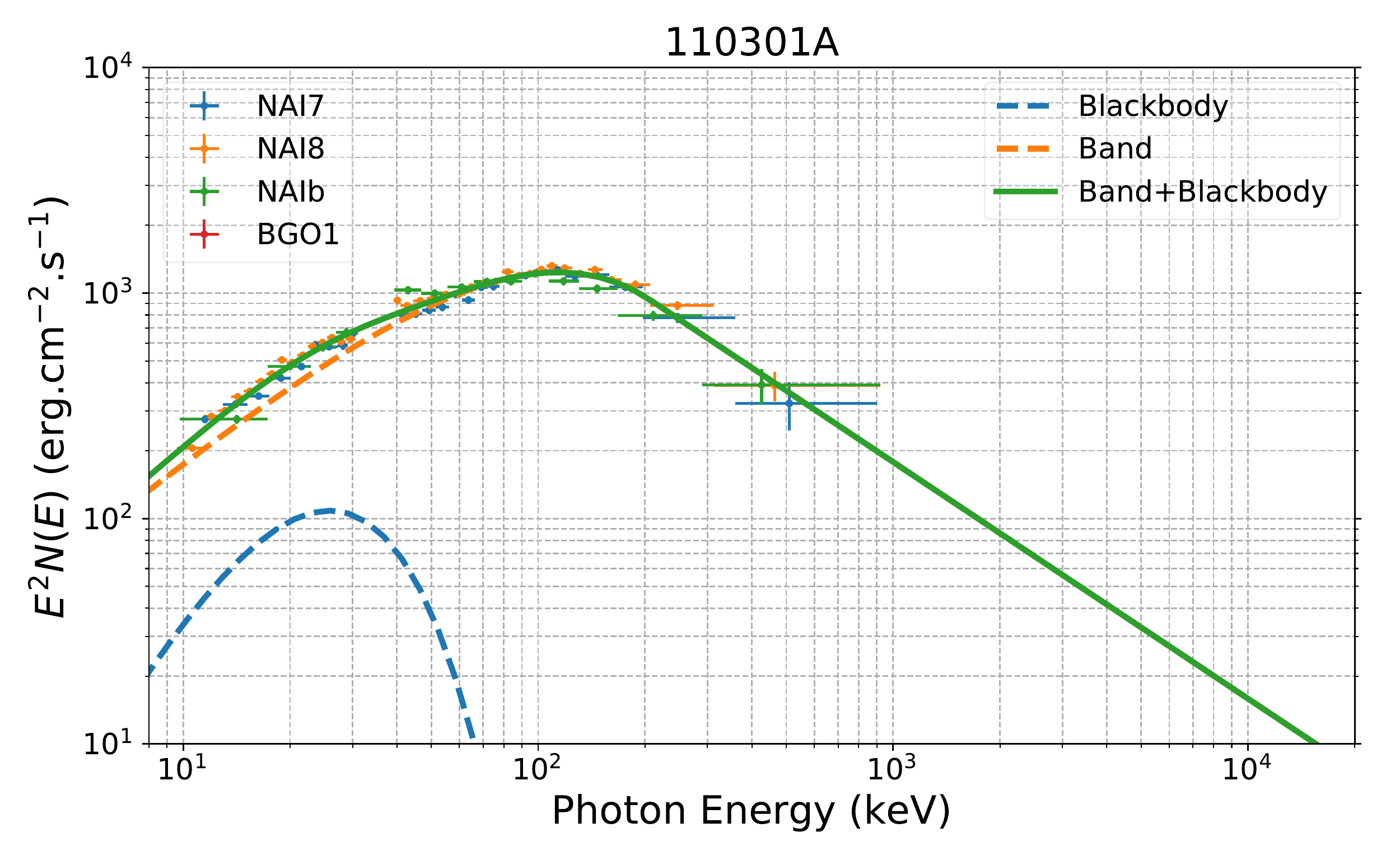}
\includegraphics[width=0.5\hsize,clip]{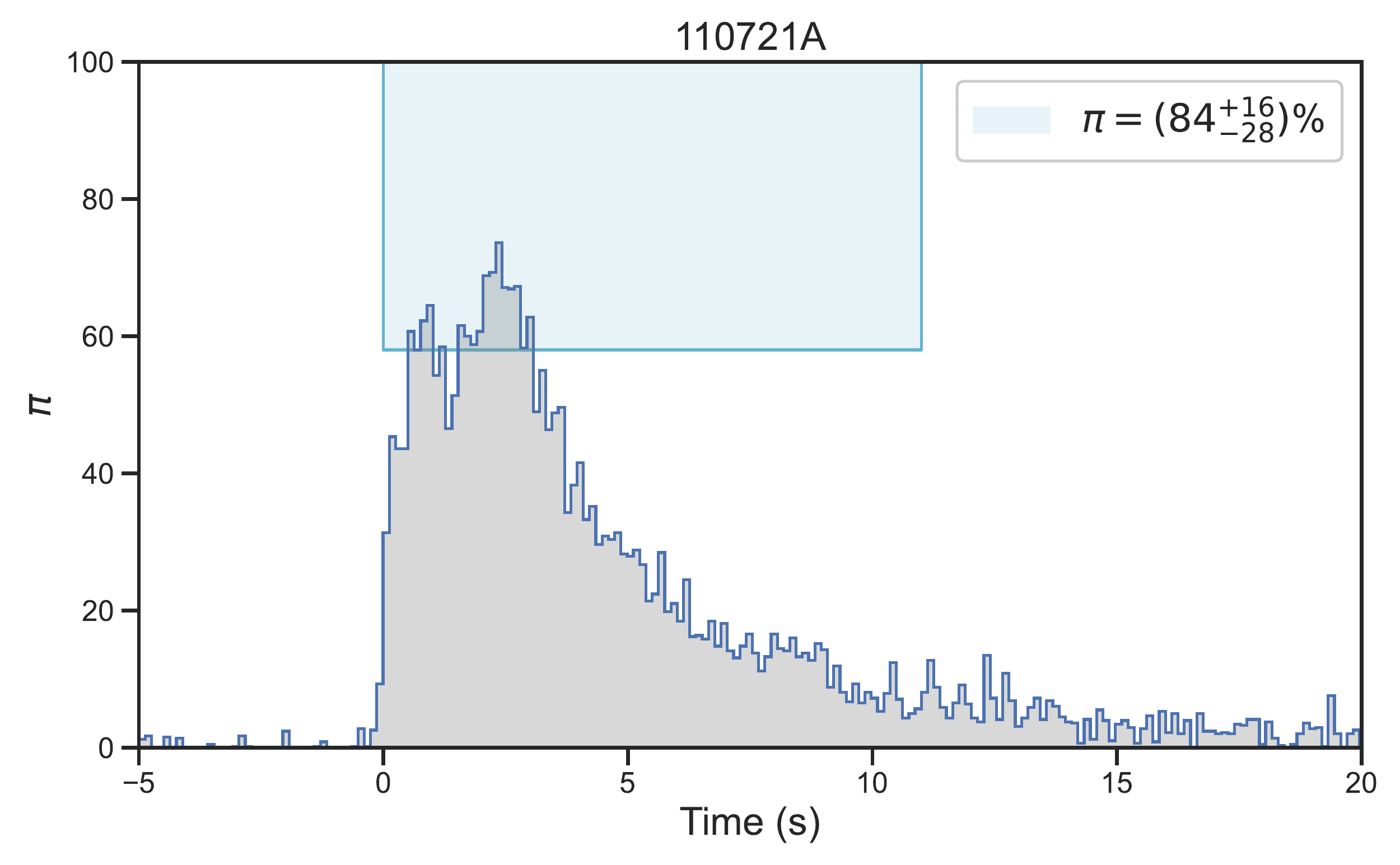}
\includegraphics[width=0.5\hsize,clip]{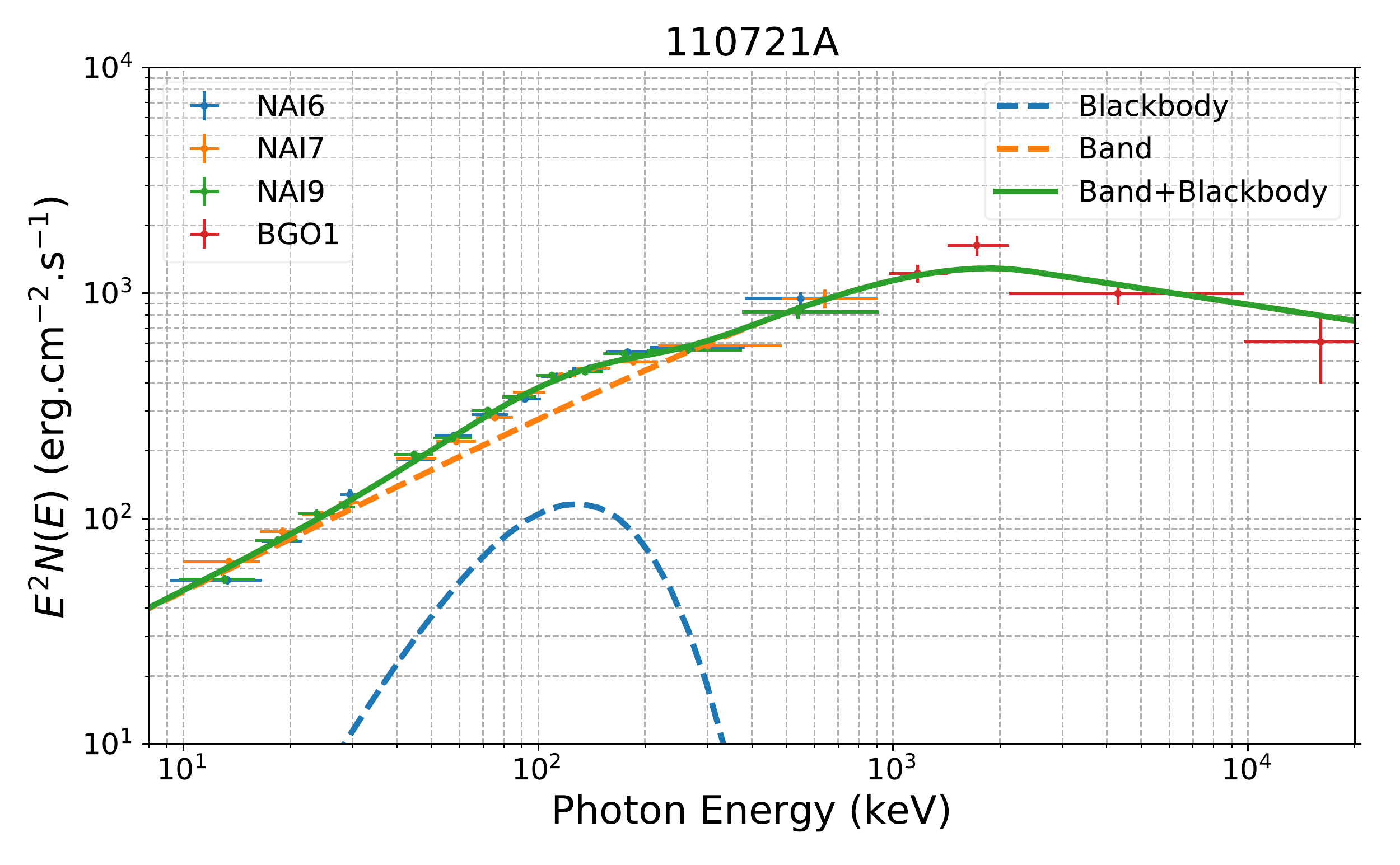}
\caption{The lightcurve, polarization, and spectral properties for the HD-type bursts. Left panels: the prompt emission lightcurve (overlaid in gray) and polarization observations in $\gamma$-ray/hard X-ray energy bands (cyan shaded area). Right panels: the spectral data and their best-fit model (Band+BB) during the time epoch (see Table \ref{tab:Sample}) of the matching polarization observations.}\label{fig:Spectrum_LC_HD}
\end{figure*}

Twenty-seven bursts were claimed to have a high-significance polarization measurement and GBM data taken at their prompt emission (Table \ref{tab:Sample}, and Figure \ref{fig:Spectrum_LC_HD}, Figure \ref{fig:Spectrum_LC_PFD_Band}, and Figure \ref{fig:Spectrum_LC_PFD_CPL}), providing an ``ideal''  sample to study the possible connection between jet properties and GRB polarization straightforwardly. 

Apart from one burst (GRB 160623A) found to have low statistical significance, which cannot ensure that the spectral fits are well determined and the type of outflow cannot be determined, the remaining 26 bursts are used in our analysis. For 10 bursts, the Band+BB model has an AIC/BIC-statistic improvement of at least 10 with respect to the Band alone and other models, which suggests Band+BB as the preferred model that would fit the data and a thermal component existing in the spectrum (Table \ref{tab:HD}). Our refined time-integrated spectral analysis suggests that the Band+BB model can best characterize the spectral shape of 10 out of 26 (38\%) bursts; therefore, the HD-type outflow can be identified (see Table \ref{tab:HD}). This implies that one cannot rule out that the photosphere emission may also be the possible mechanism powering the high levels of polarization. For the remaining 16 out of 26 (62\%) bursts, either a Band-like or CPL-like component has been observed and therefore can be attributed to the presence of a PFD-type outflow, accounting for the largest percentage of the sample (see Table \ref{tab:PFD_Band}). Notably, no bursts are identified as the “KED-type” outflow. 

Our time-integrated spectral analysis indicates that ten bursts (38\%) in our target sample belong to the HD-type outflow. This finding is quite interesting since only a subset of GRBs (a fairly low percentage) have an observed thermal component in their spectral analysis, as suggested by several statistical studies \citep[e.g.,][]{Li2023c}. Recently, \citep{Li2023c} has made a great effort to collect a complete GRB sample in which all bursts were detected by Fermi/GBM with known redshift and created a spectral parameter catalog based on their model-wise properties. He discovered that $\sim 5\%$ (7/153) of the analyzed bursts were found to require a subdominant thermal component in their time-integrated spectral analysis, including GRB 110721A ($\pi=84^{+16}_{-28}\%$). Our results imply that high-degree polarization measurements may still be dominated by the nonthermal component of the HD jets, originating well above the photosphere.

\subsection{A Comparison of the Polarization Properties of HD- and PFD-type Outflows}\label{comparison}

Using the same observed epoch during the prompt emission phase, our target sample allows for a thorough comparison of the polarization properties between the HD- and PFD-type bursts. Figure \ref{fig:Dis} shows the distribution of polarization degree $\pi$, comparing the PFD-type bursts (gray color) with the HD-type bursts (red color). It is evident that the HD- and PFD-type bursts exhibit inconsistent peaks, with the HD-type bursts typically exhibiting higher values compared to the PFD-type bursts. Notably, the PFD-type bursts display a bimodal distribution, with a lower peak at $\pi$=11.9\% and a higher peak at $\pi$=50.0\%. Observing such a high degree of polarization originating from the HD-type jets, particularly when it surpasses the polarization degree measurements from the PFD-type jets, is quite unexpected. According to current theoretical models, the expectation is the opposite: the PFD-type jets should exhibit higher-degree polarization measurements than the HD-type jets. Here we offer several remarks regarding these results. (1) The high-degree polarization measurements in the HD-type jets should still be dominated by their nonthermal components. (2) The HD-type jets contain both a KED thermal component and a PFD non-thermal component, and the thermal flux ratio between these thermal and nonthermal components varies from burst to burst, leading to a more intricate jet composition and magnetic field structure in the HD-type jets compared to pure PFD-type jets. If the finding that the HD-type bursts exhibit a higher degree of polarization than the PFD-type bursts truly holds, it is both intriguing and surprising. One possible explanation is that in the outflow of the HD-type bursts, in radiating regions dominated by nonthermal components, a large-scale ordered magnetic field (or a stronger ordered magnetic field) is more easily formed due to some mechanism, resulting in higher observed polarization measurements. (3) To answer this question better, a well-sampled collection of accumulating thermal-dominated bursts (similar to 090902B) may carry a key clue. By directly comparing the statistical results of polarization properties between the thermally dominated bursts and nonthermally dominated bursts, a firm conclusion can be drawn.

Figure \ref{fig:pipair} displays scatter plots between $\pi$ and various GRB observed quantities, for instance, $\pi$ correlated with (i) the peak energy ($E_{\rm p}$) of the $\nu F_\nu$ prompt emission spectrum for all the bursts (Fig.\ref{fig:pipair}a), (ii) the BB temperature $kT$ (Fig.\ref{fig:pipair}b) for the HD-type bursts, (iii) the corresponding energy fluence $S_{\gamma}$ for all bursts (Fig.\ref{fig:pipair}c), (iv) the magnetization parameter $\sigma_{0}$ for the HD-type bursts (Fig.\ref{fig:pipair}d), and (v) the redshift $z$ (Fig.\ref{fig:pipair}e). Despite the presence of a thermal component, the thermal flux ratio ($F_{\rm BB}$/$F_{\rm tot}$) for these bursts is found to be less than 50\% (see Table \ref{tab:HD}), indicating that the thermal components are subdominant. Interestingly, a recent study \citep{Chattopadhyay2019} of 11 bright bursts detected by CZTI during its first year of operation revealed that four bursts (GRB 160106A, GRB 160509A, GRB 160802A, and GRB 160910A) required an additional thermal BB component to accurately fit their spectra, deviating from the Band model. For the remaining 16 bursts, either the Band or CPL model provided a good fit to the spectral data. Following the spectral analysis, we obtain the peak energy ($E_{\rm p}$) and BB temperature ($kT$), allowing us to calculate the magnetization parameter $\sigma_{0}$ using hybrid-spectrum observed properties and the method described in \cite{Gao2015} and \cite{Li2020}. A robust correlation was not identified, as shown in Figure \ref{fig:pipair}. The most intriguing result that captures our attention is the distinct regions occupied by HD-type bursts and PFD-type bursts in the $\pi-E_{\rm p}$ and $\pi-S_{\gamma}$ planes. The former may be due to different $E_{\rm p}$ values expected by different polarization models \citep{Toma2009}. The latter, on the other hand, may be attributed to the more complicated spectral shape of the HD-type bursts compared to the PFD-type bursts. In practice, more complicated models, which have more free parameters, require more source photons to establish a reliable fit \citep{Li2022}.
\\
\\
\section{Physical Implication and modeling}

In this section, we aim to consider the physical origin of the polarization pattern observed in different types of bursts and try to give an extensive discussion about the connection between jet composition and other effective parameters in GRB polarization. There are several parameters to impact the degree of polarization in GRBs, including the geometry of the jet, its angular structure, the bulk Lorentz factor of outflow material, the magnetic field configuration, and the observer's point of view. Here, we consider an ultrarelativistic axisymmetric jet launched by a central engine, whether a black hole or a rapidly rotating magnetar \citep[e.g.,][]{Usov1992,Thompson1994,Dai1998,Wheeler2000,Zhang2001,Liu2007,Metzger2008,Lei2009,Metzger2011,Bucciantini2012,Lue2014,Li2018b}. 

\subsection{Linear polarization of GRB prompt emission}

The state of polarization of a radiation field can be expressed in terms of the Stokes parameters $I$ (total intensity), $Q$ and $U$ (linear polarizations), and $V$ (circular polarization). Stokes parameters $Q$ and $U$ are differences in the flux for two orthogonal directions on the sky that are coordinate-dependent quantities \citep{Rybicki:2008vo,1959ApJ...130..241W}, we define the local degree of linear polarization $\pi=\sqrt{Q^{2}+U^{2}}/I$ where

\begin{align}
\frac{U}{I}=\pi \sin 2\theta_p\ ,  \ \frac{Q}{I}=\pi \cos 2\theta_p\ ,  \  \theta_p=\frac{1}{2}\arctan\left(\frac{U}{Q}\right)\ ,  
\end{align}
and $\theta_p$ is the local polarization position angle (PA). The global Stokes parameters are evaluated by the integration over flux $dF_\nu$ of each fluid element that contributes to the radiation at any given observer time as 
\begin{equation}
\left\{
\frac{U_{g}}{I_{g}} \atop \frac{Q_{g}}{I_{g}}
\right\}
=\left( \int dF_\nu \right)^{-1}
\int dF_\nu
\left\{
\pi \sin 2\theta_p \atop \pi \cos 2\theta_p
\right\},
\end{equation}
leading to the global linear polarization $\Pi=\sqrt{Q_{g}^{2}+U_{g}^{2}}/I_{g}$  which is finally measured by an observer from the image of the GRB jet on the sky plane \citep{Granot2003a}.

During the prompt emission, we have an ultrarelativistic jet with bulk Lorentz factor $\Gamma \gg 1$, which leads to a strong beaming effect of the emitted radiation. In the  ultrarelativistic regime, the Doppler factor can be approximated as  
\begin{eqnarray}
\delta_{D}\approx \frac{2\Gamma}{1+y}, \ \ \ \ \ \text{where} \ \ \ \ y\equiv(\Gamma \tilde{\theta })^{2},
\end{eqnarray}
where the observed emission is mainly received from a region that is limited to a cone with angular size $\tilde{\theta }\lesssim 1/\Gamma$ around line of sight ({\it{LOS}}). In principle, different GRBs can be viewed from various observing angles $\theta_{\rm obs}$ with respect to the jet's central axis. Those observers whose LOS intersects the surface of the jet can detect the GRB's prompt phase. At the early-time prompt emission, when the LOS intersects the jet surface, if $\theta_{obs} /\theta_j \lesssim 1-(\Gamma \theta_j)^{-1} $ and $\Gamma \theta_j \gtrsim \mathcal{O}(10)$ with $\theta_j$ as the half-opening angle of the ejecta, the jet's edge remains invisible to the observer \cite{Gill:2018qrz}.

The emission region in the prompt emission can be approximated as an expanding thin spherical shell of width $\Delta\ll R/\Gamma^{2}$ (in the lab frame) in which particles cool relatively fast compared to the dynamical time scale of the system.  As the GRB jet has slowed down significantly during the afterglow, when the opening angle $\theta_j\simeq \Gamma^{-1}$ then the jet break happens, and the edge effects become important. The flux density measured by a distant observer from each fluid element in an infinite thin-shell approximation for the prompt emission is given by \citep{2005ApJ...631.1022G}
\begin{eqnarray}
F_{\nu}(t)=\frac{(1+z)}{16\pi^{2}d_{L}^{2}(z)}\int \delta_{D}^{3} L'_{\nu'}d\tilde{\Omega},
\end{eqnarray}
where $d_{L}(z)$ is the luminosity distance of the source and $d\tilde{\Omega}=d\tilde{\phi}d(\cos\tilde{\theta})$ is the solid angle with $\tilde{\theta}$ and $\tilde{\phi}$ as the polar angle and the azimuthal angle measured from the LOS, respectively.

The anisotropic synchrotron luminosity $L'_{\nu'}$ depends on the pitch angle  $\chi$ and the  frequency $\nu'$\citep{2003ApJ...597..998L,Rybicki:2008vo}, as 
 \begin{eqnarray}
 L'_{\nu'}\propto  \nu'^{-\alpha} \sin\chi^\epsilon,
\end{eqnarray}
where  $\chi$ is the angle between the electron's velocity vector (perpendicular to the local direction of the magnetic field $\hat B'$) and the observer's  LOS  $\hat n'$ in the comoving frame of the GRB jet  ($\cos\chi=\hat{n}'
\cdot \hat{B}'$). As long as the electron energy distribution is independent of the pitch angle, it was shown that $\epsilon=1+\alpha$, where $\alpha=-d\rm{log}(F_\nu)/d\rm{log}(\nu)$ is the spectral index \cite{Granot:2003dy}. Then, we have
\begin{eqnarray}\label{e6}
 L'_{\nu'} \propto \left(\frac{\nu'}{\nu_{p}'} \right)^{-\alpha} \left[ 1-(\hat{n}'
\cdot \hat{B}')^{2}\right]^{\frac{1+\alpha}{2}},
\end{eqnarray} 
This means that most of the power is emitted at the peak frequency $\nu_{p}'$   \cite{Granot:2003dy,Gill:2018qrz}. The above spectrum is associated with a specific LOS (fixed polar angle) for a thin shell radiating at a particular radius. In the case of considering the temporal evolution of prompt GRB polarization, a Band-like spectrum may apply to model the radiation over a broad range of radii \cite{Gill:2021jzc}.

\subsection{The Magnetic field structure and the spectral properties}
The polarization measurements can help to probe the magnetic field configuration inside the GRB shock wave. Moreover, the degree of polarization depends on the GRB jet's angular structure and the observer's viewing angle from the jet symmetry axis \citep{Lazzati2004}. Specifically, the observation of the early-time polarization from the prompt emission phase may indicate the magnetic field configuration close to the GRB progenitor. The direction of the local polarization vector in  the synchrotron emission is orthogonal to the LOS of the observer $\hat n$ and the local direction of the magnetic field $\hat B$ in the jet,
\begin{eqnarray}
\hat \pi =\frac{(\hat n\times \hat B)}{|\hat n\times \hat B)|},
\end{eqnarray}

The degree of linear polarization generated in the synchrotron emission from an isotropic electron distribution  with power-law energy spectrum ($n_{e}\propto \gamma^{-p} $), and for a given direction of the magnetic field is given by \citep{Rybicki:2008vo,1959ApJ...130..241W}:

\begin{eqnarray}\label{ord}
\Pi_{max}=\frac{\alpha+1}{\alpha+5/3}=\frac{p_{\rm eff}+1}{p_{\rm eff}+7/3},
\end{eqnarray}
where  $p_{\rm eff}=2\alpha+1$ is the effective power-law index of the electron distribution \citep{1998ApJ...497L..17S}.

\begin{eqnarray}
\label{n} p_{\rm eff}= \left\{ \begin{array}{ll}
2, \quad \nu_{c}<\nu<\nu_{m}, \quad \rm{fast\ cooling}\\
p, \quad \nu_{m}<\nu<\nu_{c}, \quad \rm{slow\ cooling}\\
p+1, \quad \nu>\rm{max}(\nu_{c},\nu_{m}),  \quad \rm{either\ cooling}
\end{array} \right.
\end{eqnarray}
and, therefore,
\begin{eqnarray}
\label{n2} \Pi^{lin}_{max}= \left\{ \begin{array}{ll}
9/13, \quad \nu_{c}<\nu<\nu_{m}, \quad \rm{fast\ cooling}\\
\frac{(p+1)}{(p+7/3)}, \quad \nu_{m}<\nu<\nu_{c}, \quad \rm{slow\ cooling}\\
\frac{(p+2)}{(p+10/3)}, \quad \nu>\rm{max}(\nu_{c},\nu_{m}), \ \rm{either\ cooling}
\end{array} \right.
\end{eqnarray}

This value of polarization may be associated with a very small region (pointlike emitter) in which the magnetic field has a specific orientation. Only in the case of an ordered magnetic field with the coherence length comparable to or larger than the visible surface of the emitting region can the highest value of the polarization can be generated. The photon index in the synchrotron radiation is limited to $-1/3\leqslant\alpha\lesssim 3/2$, which, regarding Eq. (\ref{ord}) leads to the maximum degree of polarization, $50\%\lesssim\Pi\lesssim 75\%$.

The local degree of linear polarization for a tangled or random field configuration for a thin ultrarelativistic shell modeling of the prompt emission by assuming $\alpha=1$ is obtained by averaging over all local magnetic field directions as \cite{1999ApJ...524L..43S,1999ApJ...525L..29G}
\begin{eqnarray}\label{rnd}
\Pi^{\text{lin}}_{\text{rnd}}=\Pi^{lin}_{max}
\frac{(b-1)\sin^{2}\theta_{B}}{2+(b-1)\sin^2\theta_{B}}
\end{eqnarray}
where $b\equiv 2 \langle B^{2}_{\|}\rangle /\langle B_{\perp}^2\rangle$ denotes the anisotropy of the magnetic field distribution as the ratio of the parallel $B_{\|}$ to the perpendicular $B_{\perp}$ components with respect to the shock direction and $\theta_{B}$ is the angle between the LOS from the observer and the direction of the shock. In the case of a globally ordered magnetic field configuration aligned with the jet direction ($B\rightarrow B_{\|}$, $b\rightarrow \infty$),  Eq. (\ref{rnd}) returns back to Eq. (\ref{ord}) and gives the maximum value of the linear polarization.

The magnetic field structure in KED and PFD flows has a different origin and can be classified into three categories \citep{Lyutikov:2003bz,Gill:2018qrz,Gill2021}: 
\begin{itemize}
    \item  A locally ordered magnetic field ($B_{\text{ord}}$)  with angular coherent length $\theta_j>\theta_{B} \gtrsim 1/\Gamma$.
    \item A toroidal magnetic field ($B_{\text{trod}}$) which has an ordered axisymmetric configuration in the transverse direction with respect to the jet.
    \item  A tangent magnetic field that could, in principle, be parallel ($B_{\|}$) or perpendicular ($B_{\perp}$) to the local fluid velocity.
\end{itemize}
 In the PFD, the magnetic field is dynamically dominated and usually has a large coherence length such as $B_{\text{trod}}$ which can be produced by a rotating central engine or in a high magnetized flow; other locally and globally ordered field configurations are also possible in this case. On the other hand, in KED we may have a tangled magnetic field structure with $B_{\perp}$ and/or $B_{\|}$ components; however, generating such an anisotropic field configuration in shock waves seems to be challenging \citep{Gill:2019tpt}. A globally ordered magnetic field may naturally be advected from near the central source, while random magnetic fields are generated in the shock dissipation region \citep{2015PhR...561....1K,Geng:2018dsd,Gill:2018qrz,Fan:2008zw}. The magnetic field structures that are generated at relativistic collisionless shocks, due to the two-stream instabilities are expected to be tangled within the shock plane \citep{Medvedev1999}.

\subsection{Time-integrated linear polarization  and the GRB sample}

In general, the measured polarization is obtained by integrating the local Stokes parameters over the flux of the GRB jet  as 
\begin{eqnarray}
\Pi(t_f)={\frac{\Bar{Q(t_f)}}{I(t_f)}} =\frac{\int dF_{\nu} \cos 2\theta_p}{\int dF_{\nu} }.
\end{eqnarray}
Assuming it to have an axisymmetric flow and taking into account symmetry consideration, we see that $\Bar{U}=0$, and consequently, the instantaneous total degree of the linear polarization is $\Bar{\Pi}={|\Bar{Q}|/I}$. We perform an integration over the equal time surface (EATS) for a single pulse as
\begin{eqnarray}
\Pi = \frac{\int\Bar{Q(t)} dt}{\int I(t)dt}=\frac{\int dF_{\nu} \cos 2\theta_p dt}{\int dF_{\nu} dt},
\end{eqnarray}
in order to obtain pulse-integrated polarization of the prompt emission, which leads to
 \begin{align}
\frac{\Pi_{\text{ord}}}{\Pi_{max}}=\frac{\int_{0}^{y_{max}} dy(1+y)^{-2-\alpha}\int d\phi \Lambda(y,\phi)  \cos 2\theta_p}{\int_{0}^{y_{max}}  dy(1+y)^{-2-\alpha} \int d\phi \Lambda(y,\phi) },
\end{align}
The above formula is valid for the prompt emission from an ultrarelativistic thin shell for an on-axis observer ($\theta_{\rm obs}=0$) where $y_{\rm max}=(\Gamma \theta_{\rm max})^{2}$ and $\theta_{\rm max}$ is defined as the maximum angle from the LOS \cite{Granot2003a}. The factor $\Lambda(y,\phi)$ is an average over the magnetic field orientations in the plane of the ejecta as 
\begin{eqnarray}\label{e14}
\Lambda(y,\phi)\equiv  \langle ( 
 1-(\hat{n}'\cdot \hat{B}')^{2})^{\frac{1+\alpha}{2}}
\rangle \ . 
\end{eqnarray}
The polarization angle $\theta_{p}$ and $\Lambda(y,\phi)$ take different forms regarding the configuration of the magnetic field in the plane of the GRB jet; it is instructive to compare Equations (\ref{e6}) and Eq. (\ref{e14}). In the case of an ordered magnetic field $B_{\text{ord}}$ we have :
\begin{eqnarray}
\Lambda_{\text{ord}}(y,\phi)\approx \left[ (\frac{1-y}{1+y})^{2}\cos^{2}\phi+\sin^{2}\phi
\right]^{\frac{1+\alpha}{2}},
\end{eqnarray}
\begin{eqnarray}
\theta_{p}= \phi+\arctan \left[(\frac{1-y}{1+y})\cot{\phi}\right] \ .
\label{ord222}
\end{eqnarray}
The time-integrated linear polarization in the presence of an ordered magnetic field in the plane normal to the jet velocity is plotted as a function of the spectral index in the right panel of Fig. (\ref{fig:orderd}). As it is seen, the polarization degree increases toward higher values of $\alpha$ and lower values of $y_{max}$, which can cover the observed polarization of GRB 110721A, GRB 110301A, GRB 160802A, GRB 170101B, GRB 180120A and GRB 180427A. For a configuration with the globally ordered magnetic field, high values of the linear polarization even larger than $50\%$ are obtainable.

In the left panel of Fig. (\ref{fig:orderd}), we see polarization values inferred from the theoretical model presented by Eq. (\ref{ord}) together with observed polarization data of our selected sample; here $\alpha$ is obtained from the spectral analysis given in \S \ref{sec:Results}. We indicate the polarization values of 16 GRBs, including 6 HD-type bursts and 10  PFD-type bursts, where 11 GRBs with low polarization significance are not displayed (see Table \ref{tab:Sample}). Interestingly, it is found that HD-type bursts are better distributed along with the line predicted by Eq. (\ref{ord}) and additionally show higher values of polarization compared to PFD-type bursts.

The degree of polarization for a magnetic field with a locally tangled or random configuration is obtained by averaging over all directions of the local magnetic field within the plane of the shock \citep{2003ApJ83G,1999ApJ...524L..43S,1999ApJ...525L..29G,2016MNRAS.455.1594N}. The presence of a random magnetic field leads to negligible values of net linear polarization measured by an on-axis observer. In the case of a random field behind the shock wave, only if the observer is off-axis and the circular symmetry is broken is nonzero net polarization measurable. The total linear polarization arising from the whole jet that is subjected to a random field with a direction perpendicular to the jet velocity is given by \cite{Granot2003a}
 \begin{align}
\frac{\Pi_{\perp}}{\Pi_{max}}=\frac{\int_{y_1}^{y_2}dy (1+y)^{-2-\alpha}\sin[2\Psi_{1}(y)]\mathcal{G}(y,\alpha)}
{\Theta(1-\zeta)\int_{0}^{y_1} \frac{dy \ \mathcal{H}(y,\alpha)}{(1+y)^{\alpha+2}} + \int_{y_1}^{y_2} dy\frac{dy \ \mathcal{H}(y,\alpha)}{(1+y)^{\alpha+2}}\left(\frac{\pi-\Psi(y)}{\pi}\right)},
\label{orto}   
\end{align}
where $\Theta (1-\zeta)$ is the Heaviside step function with $\zeta\equiv\theta_{\rm obs}/\theta_{j}$ as a parameter to define the observer's point of view, and 
 \begin{align}
\mathcal{G}(y,\alpha) = \frac{1}{2\pi}\int_{0}^{\pi} d\phi\left[\frac{(1-y)^{2}}{(1+y)^{2}}\cos^{2}\phi-\sin^{2}\phi\right]\\ \nonumber \times \left[1-\frac{4y\cos^{2}\phi}{(1+y)^{2}}\right]^{\frac{\alpha-1}{2}},
\end{align}
\begin{eqnarray}
 \mathcal{H}(y,\alpha) = \int_{0}^{\pi} d\phi\left[1-\frac{4y\cos^{2}\phi}{(1+y)^{2}}\right]^{\frac{1+\alpha}{2}},
 \end{eqnarray}
 \begin{eqnarray}
 \cos \Psi(y) = \frac{(1-\zeta^{2})y_{j}-y}{2\zeta \sqrt{y y_{j}}} .
 \end{eqnarray}
In the above expressions, $y_{1,2}=(1\mp\zeta)^{2} y_{j}$ and $y_{j}=(\Gamma \theta_{j})^{2}$. The variation of the linear polarization in the presence of a random field configuration measured by an off-axis observer is displayed in Fig. (\ref{fig9}). In the left panel, the spectral indices are selected to be consistent with the average values reported in Table (\ref{tab:HD}) for our target sample and for $y_{j}=10$. In the right panel,  $y_{j}$ is changed while $\alpha=1$, it is found that the appeared peak has a width in the order of $1/\sqrt{y_{j}}$.   
From Fig. (\ref{fig9}), we see that the polarization degree is limited to small values for $\zeta<1$ while it is sharply increased for $\zeta\approx 1$ and finally reaches an asymptotic limit at $\zeta>  \mathcal{O}(1)$. It is seen that the synchrotron radiation with $B_{\perp}$ can potentially generate a wide range of polarization values from low levels to moderate values that cover the observed values associated with our sample. In principle, various viewing angles $\theta_{\rm obs}$ and different angular structures of the jet affect the measured fluence of GRBs.  Note that the fluence significantly decreases from outside the jet's sharp edge, so high levels of polarization in off-axis jets may only be obtainable in very close bursts. The detectibility of GRB polarization needs high-fluence sources, and usually, the fluence rapidly drops below the detector threshold for a large off-axis observer. For example, the observed fluences of GRB 140206A are relatively higher than other sources (see Table \ref{tab:Sample}) and due to its higher redshift, $\zeta$ cannot get large values.

As another possibility, the polarized emission may also originate from independent magnetic patches with various field orientations \cite{Li2022a} where magnetic patches are locally coherent but distributed randomly in observed emission regions. In this case, the measured polarization from different patches is estimated as $\Pi=\Pi_{max}/\sqrt{N}$, where $N$ is the number of magnetic patches or, equivalently, multiple pulses where the coherence length of the magnetic field is as large as the emission region in a single pulse and the observed polarization is an average over multiple pulses \citep{Gruzinov:1998xt,2003ApJ...594L..83G}. The magnetic field that is generated within internal shock for KED jets usually has a coherence length much smaller than the angular size of the emission region, which causes negligible net polarization.

The time-resolved spectral analysis in \S \ref{Spectral} showed thermal-to-nonthermal (KED-to-PFD) transitions in our sample, where a subdominant component of the thermal emission during HD-type bursts is observed. The thermal component may also originate from photosphere emission due to repeated Compton scattering of photons with thermal electrons in plasma before escaping, and observing hard values of the spectral indices during the bursts can serve as hints that LOS is not highly off-axis, since high-latitude emission leads to a softer spectrum  \cite{2013MNRAS.428.2430L}. The local degree of polarization due to the Compton scattering process is given by \cite{Rybicki1979}: 
\begin{eqnarray}
\Pi_{sc}=\frac{1-\cos^{2}\theta_{sc}}{1+\cos^{2}\theta_{sc}}
\end{eqnarray}
where $\theta_{\rm sc}$ is the scattering angle between the incoming and scattered photons. For a plasma including charged particles with a velocity distribution, $\Pi_{\rm sc}$ is obtained by a weighted integrating over various directions of the incoming photons and velocities of the electrons. Compton scattering generates polarized emission especially close to the photosphere, where the anisotropy of the comoving scatter photons is large.

The observed high values of the polarization for HD bursts while they show the peak-KED pattern cannot be explained simply by the subphotospheric dissipation model based on Comptonization. The multiple scatterings at large optical depth regions lead to washing out the directionality of the polarization vectors \cite{Parsotan:2020ilh}. While the photospheric emission of a relativistically expanding fireball has been traditionally considered as a mechanism to produce small values of polarization  ($\Pi \lesssim 20\%$), however, introducing structures in the jet profile can increase the polarization up to $40\%$ \cite{Lundman2014}. A structured jet photosphere model is also considered in \citep{Chang:2013qya,Chang:2013yma,Chang:2013xp} as a source of  polarized photons via Compton scattering. If synchrotron emission is the origin of soft photons in the photospheric model, the observed polarization could even potentially be enhanced by approximately $50\%$ \citep{Lundman:2016fdl,Ito:2014ica}. To explain the strong polarized signals, models invoking dissipation of ordered magnetic fields are favored \citep{2003ApJ...597..998L, 2011ApJ...726...90Z, 2012MNRAS.419..573M}. 

The jitter radiation emitted by ultrarelativistic electrons accelerating in a small-scale random magnetic field \citep{Medvedev:2000gu}, can also generate a hard energy spectrum with a photon index as high as $\alpha=+0.5$. Due to the random distribution of the magnetic field, jitter radiation is highly symmetric in the electron radiative plane, leading to the vanishing polarization degree for an on-axis observer \citep{Mao:2013gha,Mao:2017dlb,Mao:2018rsr}. The maximum level of polarization is obtainable when the emitting plane is viewed from the edge-on; it can even reach up to $90\%$ \citep{Prosekin:2016caz}. However, for smaller off-axis viewing angles, which can yield measurable fluences, jitter radiation causes almost negligible polarization degree. Meanwhile, regardless of the viewing angle, the jitter radiation cannot produce the observed high degree of polarization close to the spectral peak energy of the jet. 

To summarize, polarization features can be explained either by the synchrotron radiation in the ordered/random magnetic field \citep{2003ApJ...596L..17G, 2003ApJ...594L..83G, 2003JCAP...10..005N}, the jet structure \citep{2009ApJ...700L.141L}, or the observer's viewing angle with respect to the jet \citep{Lazzati2004}, even in the case of thermal radiation from the jet photosphere \citep{2014MNRAS.440.3292L}. For a hybrid spectrum that includes thermal and nonthermal components, we expect to see relatively high values of the polarization in the prompt emission, which can be produced by the synchrotron emission mechanism in the ordered magnetic field of the jet and for random field configurations only for off-axis observers \citep{Gill2021}. However, the spectral properties of our target sample demonstrated that off-axis observations, especially for the large viewing angle, are not the case, and the observed values of the polarization most probably are a hint of the ordered magnetic field originating from the central engine. Since polarization washout effects are gradually increased from PFD jets towards HD and KED jets due to thermal photons, we would expect that the inequality $\pi_{\rm KED}\lesssim\pi_{\rm HD}\lesssim\pi_{\rm PFD}$ is satisfied if other conditions are fixed for a given jet. However, as it was noted in \S \ref{sec:Results}, the polarization pattern of GRBs in our sample displays a bimodal distribution for PFD-type bursts with two peaks at $\pi=11.9\%$ and $\pi=50.0\%$, and also a higher peak at $\pi=65.9\%$ for HD-type bursts. Although this observation is not consistent with our expectations, we would like to stress that this behavior can still be explained by the geometrical effects and the jet structure for different configurations of the magnetic fields. Due to the different degrees of polarization predicted by different emission models in various energy bands, it is essential to have a high-sensitivity gamma-ray polarimeter with a wide bandpass to detect energy-dependent polarization signals and constrain different models \citep{2014IJMPD..2330002Z}. However, because of several free parameters in polarization models, upcoming more precise observations and theoretical investigations are needed to discriminate between competing models in order to explain the observed joint polarization and spectral properties.

It should be noted that the time-integrated assumption could potentially smear out important details about the spectral evolution in GRBs, making it challenging to identify the nature of the outflow. Therefore, the time-resolved analysis, as discussed in \citep{2019A&A...627A.105B}, is crucial for capturing the dynamic changes in the spectrum and accurately identifying the emission mechanisms and outflow properties. By considering the temporal evolution of the spectral parameters, one can better distinguish between different physical processes and improve the classification of the outflow type in GRBs; this will be the subject of our future paper.

\section{Conclusion}

Early polarization observations during the prompt emission phase play a crucial role in understanding the radiation mechanism and jet composition of GRBs. Observations over the past few decades suggest that the jet composition of GRBs may have diverse properties. If the outflow is matter-dominated (i.e., a fireball), the GRB prompt emission spectra would include a bright thermal component originating from the fireball photosphere. Alternatively, if the outflow is PFD, the GRB prompt emission spectra would include a dominant nonthermal component originating from the synchrotron radiation. In cases where the outflow is HD, the GRB prompt emission spectra would contain both a thermal component from the fireball photosphere and a nonthermal component from the synchrotron radiation. It is highly speculated that the prompt emission is expected to be strongly polarized owing to its nonthermal origin. Consequently, varying levels of polarization degrees ($\pi_{\rm KED}\lesssim\pi_{\rm HD}\lesssim\pi_{\rm PFD}$) during the prompt emission phase are naturally expected due to the different outflow types. In this paper, we have collected a GRB sample comprising all bursts detected by Fermi/GBM with reported polarization detection in the emission region in the literature, containing 27 interesting bursts. By analyzing time-averaging polarization observations and selecting the same epoch for the GBM data collected during the prompt emission phase, we then attempted to investigate the correlations between the jet properties and polarization properties of GRBs.

We conducted a detailed time-averaged spectral analysis for each burst in our target sample using several typical GRB spectral models. The best model was selected based on information criteria and therefore to infer the type of the outflow. Apart from one burst (GRB 160623A) with a low statistical significance, which cannot ensure that the spectral fits are well determined and the jet type cannot be determined, the remaining 26 bursts are used in the analysis. By using a refined spectral analysis using the time epoch of polarization measurement, we identified that 16 out of 26 (62\%) bursts were dominated by the ``PFD'' jet type, making up the largest percentage of the sample. We also discover that 10 out of 26 (38\%) bursts are classified as the ``HD" jet type, implying that one cannot rule out that the photosphere emission may also be a possible mechanism powering the high levels of polarization. None of the bursts were identified as the ``KED" jet type. Interestingly, HD- and PFD- jet-type bursts exhibit distinct polarization properties, with HD-type bursts consistently showing higher levels of polarization compared to PFD-type bursts.

Using the same observed epoch during the prompt emission phase, our target sample allows for a reasonable comparison between the different outflow types. We initially compared the distribution of polarization degree $\pi$ for different outflow types. It was observed that the HD- and PFD-type bursts exhibited inconsistent peaks, with the HD-type bursts generally exhibiting higher values compared to the PFD-type bursts. More interestingly, the PFD-type bursts show a bimodal distribution with a lower peak around $\pi$=11.9\% and a higher peak around $\pi$=50.0\%. Furthermore, we conducted an attempt to explore the correlations between GRB polarization and several typical GRB observed quantities. The correlations we attempted to study included the polarization degree $\pi$ correlated with (i) the peak energy ($E_{\rm p}$) of the $\nu F_\nu$ prompt emission spectrum for all the bursts, (ii) the BB temperature $kT$ for the HD-type bursts, (iii) the corresponding energy fluence $S_{\gamma}$ for all bursts, (iv) the magnetization parameter $\sigma_{0}$ for the HD-type bursts, and (v) the redshift $z$. A robust correlation was not identified, and the HD-type bursts and PFD-type bursts are not situated in the same region in the $\pi-E_{\rm p}$ and $\pi-S_{\gamma}$ planes. As a result, the most intriguing result that captures our attention is that the polarization properties seem to be different between the HD-type and PFD-type bursts.

Lastly, we discovered that 10 bursts in our target sample have a relatively high degree of $\pi$ that seems to correlate with the ``HD'' jet type. If it is an intrinsic characteristic of GRBs, this could provide a clue to studying the radiation mechanism and composition of GRB jets. We have also discussed some physical interpretations of this interesting phenomenon. Since the configuration of the magnetic field inside the jet is one of the crucial parameters to determine the polarization degree, we discussed two main configurations (i.e. ordered and random fields), and their connection to the jet composition is clarified. We considered polarization patterns as a function of different dynamical parameters associated with the outflow materials, the spectral indices, and the observer's LOS concerning the jet. Combining the spectral analysis and the polarization measurements allowed us to find out that the detection of polarization values $\Pi>50\%$ during prompt emission of GRB 160802A, GRB 110721A, and GRB 110301A is a piece of strong evidence for the synchrotron emission mechanism in the presence of an ordered magnetic field that can be advected from the GRB central engine. Regarding the different properties of our target sample, we conclude that geometrical effects and large off-axis observations are unlikely to be responsible for the measured polarizations assuming random magnetic fields within the jets.

Finally, some caveats that are worth mentioning when applying our analysis.
(i) Spectrum. We have resolved the jet properties based on the low-energy spectrum. However, it may be difficult to classify jets as either KED or PFD jets based on the spectral index alone. Indeed, a photospheric quasi-thermal component would have a harder low-energy spectral index as compared to an optically thin synchrotron, but that does not guarantee that the jet is KED \citep[an example, see][]{Gill2020}. 
(ii) Polarization. The degree of polarization ultimately probes the (local) structure of the $B$-field in the emission region. 
An ordered field would necessarily yield high polarization whereas a tangled field would yield a very small polarization. It is unclear, however, whether these field configurations are exclusive to a given jet configuration (or a particular level of magnetization). In addition, the angular structure of the jet also plays an important role in governing the observed polarization. Thus, due to the large range of model parameters, it is difficult to attribute a given level of polarization to a given jet composition. More discussion is provided in a recent review article \citep{Gill2021}.
(iii) Different instrument analysis. Currently, it is not clear why different instruments, namely, POLAR, IKAROS-GAP, and ASTROSAT/CZTI, are finding different levels of polarization for a small sample of GRBs \citep{Chattopadhyay2019}. POLAR is finding a rather low-level polarization, which is consistent with zero within 3$\sigma$ of their quoted central values, whereas both IKAROS and AstroSAT are finding higher levels of polarization. However, in a recent study \citep{Chattopadhyay:2022yzx}, AstroSAT/CZTI measurements show less polarization for time-integrated emissions of a sample of 20 GRBs which looks more consistent with  POLAR measurements. A larger number of bursts with relatively high polarization degrees detected by CZTI might be a result of the higher energy window of observations in this instrument compared to the POLAR. Hard X-ray to soft gamma-ray polarization measurements are very tricky and the analysis has to be carried out very carefully. As such, some of these measurements are probably not representative of GRBs and need to be further verified by future more precise instruments.
(iv) Time-resolved polarization analysis. In the current analysis, none of the cases have shown time-resolved polarization measurements. Even though the GRBs in our target sample have time-resolved spectral indices, not having corresponding polarization measurements makes it difficult to ascertain the properties of the B-field and outflows. 

\vspace{5mm}
{\bf Acknowledgements.} We thank the anonymous referee for the valuable comments and suggestions. We also thank Ramandeep~Gill, Jonathan~Granot, Mi-Xiang~Lan, Asaf~Pe{\textquoteright}er, Jin-Jun~Geng, Christoffer~Lundman, Tanmoy Chattopadhyay, and ICRANet members for many discussions on GRB physics and phenomena. L.L. acknowledges support from the Natural Science Foundation of China (grants No. 11874033) and the KC Wong Magna Foundation at Ningbo University and made use of the High Energy Astrophysics Science Archive Research Center (HEASARC) Online Service at the NASA/Goddard Space Flight Center (GSFC). The computations were supported by the high-performance computing center at Ningbo University.

\bibliography{MyReferences.bib}

\begin{thebibliography}{}
\expandafter\ifx\csname natexlab\endcsname\relax\def\natexlab#1{#1}\fi
\providecommand{\url}[1]{\href{#1}{#1}}
\providecommand{\dodoi}[1]{doi:~\href{http://doi.org/#1}{\nolinkurl{#1}}}
\providecommand{\doeprint}[1]{\href{http://ascl.net/#1}{\nolinkurl{http://ascl.net/#1}}}
\providecommand{\doarXiv}[1]{\href{https://arxiv.org/abs/#1}{\nolinkurl{https://arxiv.org/abs/#1}}}

\bibitem[{Abdo {et~al.}(2009)Abdo, Ackermann, Arimoto, Asano, Atwood, Axelsson,
  Baldini, Ballet, Band, Barbiellini, Baring, Bastieri, Battelino, Baughman,
  Bechtol, Bellardi, Bellazzini, Berenji, Bhat, Bissaldi, Blandford, Bloom,
  Bogaert, Bogart, Bonamente, Bonnell, Borgland, Bouvier, Bregeon, Brez,
  Briggs, Brigida, Bruel, Burnett, Burrows, Busetto, Caliandro, Cameron,
  Caraveo, Casandjian, Ceccanti, Cecchi, Celotti, Charles, Chekhtman, Cheung,
  Chiang, Ciprini, Claus, Cohen-Tanugi, Cominsky, Connaughton, Conrad,
  Costamante, Cutini, DeKlotz, Dermer, de~Angelis, de~Palma, Digel, Dingus,
  do~Couto~e Silva, Drell, Dubois, Dumora, Edmonds, Evans, Fabiani, Farnier,
  Favuzzi, Finke, Fishman, Focke, Frailis, Fukazawa, Funk, Fusco, Gargano,
  Gasparrini, Gehrels, Germani, Giebels, Giglietto, Giommi, Giordano, Glanzman,
  Godfrey, Goldstein, Granot, Greiner, Grenier, Grondin, Grove, Guillemot,
  Guiriec, Haller, Hanabata, Harding, Hayashida, Hays, Morata, Hoover, Hughes,
  Johannesson, Johnson, Johnson, Johnson, Johnson, Kamae, Katagiri, Kataoka,
  Kavelaars, Kawai, Kelly, Kennea, Kerr, Kippen, Knodlseder, Kocevski, Kocian,
  Komin, Kouveliotou, Kuehn, Kuss, Lande, Landriu, Larsson, Latronico,
  Lavalley, Lee, Lee, Lemoine-Goumard, Lichti, Longo, Loparco, Lott,
  Lovellette, Lubrano, Madejski, Makeev, Marangelli, Mazziotta, McBreen,
  McEnery, McGlynn, Meegan, M{\'e}sz{\'a}ros, Meurer, Michelson, Minuti,
  Mirizzi, Mitthumsiri, Mizuno, Moiseev, Monte, Monzani, Moretti, Morselli,
  Moskalenko, Murgia, Nakamori, Nelson, Nolan, Norris, Nuss, Ohno, Ohsugi,
  Okumura, Omodei, Orlando, Ormes, Ozaki, Paciesas, Paneque, Panetta, Parent,
  Pelassa, Pepe, Perri, Pesce-Rollins, Petrosian, Pinchera, Piron, Porter,
  Preece, Raino, Ramirez-Ruiz, Rando, Rapposelli, Razzano, Razzaque, Rea,
  Reimer, Reimer, Reposeur, Reyes, Ritz, Rochester, Rodriguez, Roth, Ryde,
  Sadrozinski, Sanchez, Sander, Parkinson, Scargle, Schalk, Segal, Sgro,
  Shimokawabe, Siskind, Smith, Smith, Spandre, Spinelli, Stamatikos, Starck,
  Stecker, Steinle, Stephens, Strickman, Suson, Tagliaferri, Tajima, Takahashi,
  Takahashi, Tanaka, Tenze, Thayer, Thayer, Thompson, Tibaldo, Torres, Tosti,
  Tramacere, Turri, Tuvi, Usher, Van~der Horst, Vigiani, Vilchez, Vitale, von
  Kienlin, Waite, Williams, Wilson-Hodge, Winer, Wood, Wu, Yamazaki, Ylinen,
  Ziegler, Collaboration, \& Collaboration}]{Abdo2009a}
Abdo, A.~A., Ackermann, M., Arimoto, M., {et~al.} 2009, Science, 323, 1688

\bibitem[{{Abdo} {et~al.}(2009){Abdo}, {Ackermann}, {Ajello}, {Asano},
  {Atwood}, {Axelsson}, {Baldini}, {Ballet}, {Barbiellini}, {Baring},
  {Bastieri}, {Bechtol}, {Bellazzini}, {Berenji}, {Bhat}, {Bissaldi},
  {Blandford}, {Bloom}, {Bonamente}, {Borgland}, {Bouvier}, {Bregeon}, {Brez},
  {Briggs}, {Brigida}, {Bruel}, {Burgess}, {Burrows}, {Buson}, {Caliandro},
  {Cameron}, {Caraveo}, {Casandjian}, {Cecchi}, {{\c C}elik}, {Chekhtman},
  {Cheung}, {Chiang}, {Ciprini}, {Claus}, {Cohen-Tanugi}, {Cominsky},
  {Connaughton}, {Conrad}, {Cutini}, {d'Elia}, {Dermer}, {de Angelis}, {de
  Palma}, {Digel}, {Dingus}, {Silva}, {Drell}, {Dubois}, {Dumora}, {Farnier},
  {Favuzzi}, {Fegan}, {Finke}, {Fishman}, {Focke}, {Fortin}, {Frailis},
  {Fukazawa}, {Funk}, {Fusco}, {Gargano}, {Gehrels}, {Germani}, {Giavitto},
  {Giebels}, {Giglietto}, {Giordano}, {Glanzman}, {Godfrey}, {Goldstein},
  {Granot}, {Greiner}, {Grenier}, {Grove}, {Guillemot}, {Guiriec}, {Hanabata},
  {Harding}, {Hayashida}, {Hays}, {Horan}, {Hughes}, {Jackson},
  {J{\'o}hannesson}, {Johnson}, {Johnson}, {Johnson}, {Kamae}, {Katagiri},
  {Kataoka}, {Kawai}, {Kerr}, {Kippen}, {Kn{\"o}dlseder}, {Kocevski}, {Komin},
  {Kouveliotou}, {Kuss}, {Lande}, {Latronico}, {Lemoine-Goumard}, {Longo},
  {Loparco}, {Lott}, {Lovellette}, {Lubrano}, {Madejski}, {Makeev},
  {Mazziotta}, {McBreen}, {McEnery}, {McGlynn}, {Meegan}, {M{\'e}sz{\'a}ros},
  {Meurer}, {Michelson}, {Mitthumsiri}, {Mizuno}, {Moiseev}, {Monte},
  {Monzani}, {Moretti}, {Morselli}, {Moskalenko}, {Murgia}, {Nakamori},
  {Nolan}, {Norris}, {Nuss}, {Ohno}, {Ohsugi}, {Omodei}, {Orlando}, {Ormes},
  {Paciesas}, {Paneque}, {Panetta}, {Pelassa}, {Pepe}, {Pesce-Rollins},
  {Petrosian}, {Piron}, {Porter}, {Preece}, {Rain{\`o}}, {Rando}, {Rau},
  {Razzano}, {Razzaque}, {Reimer}, {Reimer}, {Reposeur}, {Ritz}, {Rochester},
  {Rodriguez}, {Roming}, {Roth}, {Ryde}, {Sadrozinski}, {Sanchez}, {Sander},
  {Saz Parkinson}, {Scargle}, {Schalk}, {Sgr{\`o}}, {Siskind}, {Smith},
  {Spinelli}, {Stamatikos}, {Stecker}, {Stratta}, {Strickman}, {Suson},
  {Swenson}, {Tajima}, {Takahashi}, {Tanaka}, {Thayer}, {Thayer}, {Thompson},
  {Tibaldo}, {Torres}, {Tosti}, {Tramacere}, {Uchiyama}, {Uehara}, {Usher},
  {van der Horst}, {Vasileiou}, {Vilchez}, {Vitale}, {von Kienlin}, {Waite},
  {Wang}, {Wilson-Hodge}, {Winer}, {Wood}, {Yamazaki}, {Ylinen}, \&
  {Ziegler}}]{Abdo2009}
{Abdo}, A.~A., {Ackermann}, M., {Ajello}, M., {et~al.} 2009, \apjl, 706, L138,
  \dodoi{10.1088/0004-637X/706/1/L138}

\bibitem[{{Acuner} {et~al.}(2020){Acuner}, {Ryde}, {Pe'er}, {Mortlock}, \&
  {Ahlgren}}]{Acuner2020}
{Acuner}, Z., {Ryde}, F., {Pe'er}, A., {Mortlock}, D., \& {Ahlgren}, B. 2020,
  \apj, 893, 128, \dodoi{10.3847/1538-4357/ab80c7}

\bibitem[{{Acuner} {et~al.}(2019){Acuner}, {Ryde}, \& {Yu}}]{Acuner2019}
{Acuner}, Z., {Ryde}, F., \& {Yu}, H.-F. 2019, \mnras, 487, 5508,
  \dodoi{10.1093/mnras/stz1356}

\bibitem[{{Akaike}(1974)}]{Akaike1974}
{Akaike}, H. 1974, IEEE Transactions on Automatic Control, 19, 716

\bibitem[{{Atwood} {et~al.}(2009){Atwood}, {Abdo}, {Ackermann}, {Althouse},
  {Anderson}, {Axelsson}, {Baldini}, {Ballet}, {Band}, {Barbiellini}, \&
  et~al.}]{Atwood2009}
{Atwood}, W.~B., {Abdo}, A.~A., {Ackermann}, M., {et~al.} 2009, \apj, 697,
  1071, \dodoi{10.1088/0004-637X/697/2/1071}

\bibitem[{{Axelsson} {et~al.}(2012){Axelsson}, {Baldini}, {Barbiellini},
  {Baring}, {Bellazzini}, {Bregeon}, {Brigida}, {Bruel}, {Buehler},
  {Caliandro}, {Cameron}, {Caraveo}, {Cecchi}, {Chaves}, {Chekhtman}, {Chiang},
  {Claus}, {Conrad}, {Cutini}, {D'Ammando}, {de Palma}, {Dermer}, {Silva},
  {Drell}, {Favuzzi}, {Fegan}, {Ferrara}, {Focke}, {Fukazawa}, {Fusco},
  {Gargano}, {Gasparrini}, {Gehrels}, {Germani}, {Giglietto}, {Giroletti},
  {Godfrey}, {Guiriec}, {Hadasch}, {Hanabata}, {Hayashida}, {Hou}, {Iyyani},
  {Jackson}, {Kocevski}, {Kuss}, {Larsson}, {Larsson}, {Longo}, {Loparco},
  {Lundman}, {Mazziotta}, {McEnery}, {Mizuno}, {Monzani}, {Moretti},
  {Morselli}, {Murgia}, {Nuss}, {Nymark}, {Ohno}, {Omodei}, {Pesce-Rollins},
  {Piron}, {Pivato}, {Racusin}, {Rain{\`o}}, {Razzano}, {Razzaque}, {Reimer},
  {Roth}, {Ryde}, {Sanchez}, {Sgr{\`o}}, {Siskind}, {Spandre}, {Spinelli},
  {Stamatikos}, {Tibaldo}, {Tinivella}, {Usher}, {Vandenbroucke}, {Vasileiou},
  {Vianello}, {Vitale}, {Waite}, {Winer}, {Wood}, {Burgess}, {Bhat},
  {Bissaldi}, {Briggs}, {Connaughton}, {Fishman}, {Fitzpatrick}, {Foley},
  {Gruber}, {Kippen}, {Kouveliotou}, {Jenke}, {McBreen}, {McGlynn}, {Meegan},
  {Paciesas}, {Pelassa}, {Preece}, {Tierney}, {von Kienlin}, {Wilson-Hodge},
  {Xiong}, \& {Pe'er}}]{Axelsson2012}
{Axelsson}, M., {Baldini}, L., {Barbiellini}, G., {et~al.} 2012, \apjl, 757,
  L31, \dodoi{10.1088/2041-8205/757/2/L31}

\bibitem[{{Band} {et~al.}(1993){Band}, {Matteson}, {Ford}, {Schaefer},
  {Palmer}, {Teegarden}, {Cline}, {Briggs}, {Paciesas}, {Pendleton}, {Fishman},
  {Kouveliotou}, {Meegan}, {Wilson}, \& {Lestrade}}]{Band1993}
{Band}, D., {Matteson}, J., {Ford}, L., {et~al.} 1993, \apj, 413, 281,
  \dodoi{10.1086/172995}

\bibitem[{{B{\'e}gu{\'e}} {et~al.}(2022){B{\'e}gu{\'e}}, {Samuelsson}, \&
  {Pe'er}}]{Begue2022}
{B{\'e}gu{\'e}}, D., {Samuelsson}, F., \& {Pe'er}, A. 2022, \apj, 937, 101,
  \dodoi{10.3847/1538-4357/ac85b7}

\bibitem[{{Bucciantini} {et~al.}(2012){Bucciantini}, {Metzger}, {Thompson}, \&
  {Quataert}}]{Bucciantini2012}
{Bucciantini}, N., {Metzger}, B.~D., {Thompson}, T.~A., \& {Quataert}, E. 2012,
  \mnras, 419, 1537, \dodoi{10.1111/j.1365-2966.2011.19810.x}

\bibitem[{{Burgess} {et~al.}(2019{\natexlab{a}}){Burgess}, {Greiner},
  {B{\'e}gu{\'e}}, \& {Berlato}}]{Burgess2019}
{Burgess}, J.~M., {Greiner}, J., {B{\'e}gu{\'e}}, D., \& {Berlato}, F.
  2019{\natexlab{a}}, \mnras, 490, 927, \dodoi{10.1093/mnras/stz2589}

\bibitem[{{Burgess} {et~al.}(2019{\natexlab{b}}){Burgess}, {Kole}, {Berlato},
  {Greiner}, {Vianello}, {Produit}, {Li}, \& {Sun}}]{2019A&A...627A.105B}
{Burgess}, J.~M., {Kole}, M., {Berlato}, F., {et~al.} 2019{\natexlab{b}}, \aap,
  627, A105, \dodoi{10.1051/0004-6361/201935056}

\bibitem[{{Chand} {et~al.}(2018){Chand}, {Chattopadhyay}, {Iyyani}, {Basak},
  {Aarthy}, {Rao}, {Vadawale}, {Bhattacharya}, \&
  {Bhalerao}}]{2018ApJ...862..154C}
{Chand}, V., {Chattopadhyay}, T., {Iyyani}, S., {et~al.} 2018, \apj, 862, 154,
  \dodoi{10.3847/1538-4357/aacd12}

\bibitem[{Chang {et~al.}(2013)Chang, Jiang, \& Lin}]{Chang:2013xp}
Chang, Z., Jiang, Y., \& Lin, H.~N. 2013, Astrophys. J., 769, 70,
  \dodoi{10.1088/0004-637X/769/1/70}

\bibitem[{Chang {et~al.}(2014{\natexlab{a}})Chang, Jiang, \&
  Lin}]{Chang:2013yma}
Chang, Z., Jiang, Y., \& Lin, H.-N. 2014{\natexlab{a}}, Astrophys. J., 780, 68,
  \dodoi{10.1088/0004-637X/780/1/68}

\bibitem[{Chang {et~al.}(2014{\natexlab{b}})Chang, Lin, \&
  Jiang}]{Chang:2013qya}
Chang, Z., Lin, H.-N., \& Jiang, Y. 2014{\natexlab{b}}, Astrophys. J., 783, 30,
  \dodoi{10.1088/0004-637X/783/1/30}

\bibitem[{{Chattopadhyay} {et~al.}(2019){Chattopadhyay}, {Vadawale}, {Aarthy},
  {Mithun}, {Chand}, {Ratheesh}, {Basak}, {Rao}, {Bhalerao}, {Mate}, {Arvind},
  {Sharma}, \& {Bhattacharya}}]{Chattopadhyay2019}
{Chattopadhyay}, T., {Vadawale}, S.~V., {Aarthy}, E., {et~al.} 2019, \apj, 884,
  123, \dodoi{10.3847/1538-4357/ab40b7}

\bibitem[{Chattopadhyay {et~al.}(2022)}]{Chattopadhyay:2022yzx}
Chattopadhyay, T., {et~al.} 2022, Astrophys. J., 936, 12,
  \dodoi{10.3847/1538-4357/ac82ef}

\bibitem[{{Chattopadhyay} {et~al.}(2022){Chattopadhyay}, {Gupta}, {Iyyani},
  {Saraogi}, {Sharma}, {Tsvetkova}, {Ratheesh}, {Gupta}, {Mithun}, {Vaishnava},
  {Prasad}, {Aarthy}, {Kumar}, {Rao}, {Vadawale}, {Bhalerao}, {Bhattacharya},
  {Vibhute}, \& {Frederiks}}]{Chattopadhyay2022}
{Chattopadhyay}, T., {Gupta}, S., {Iyyani}, S., {et~al.} 2022, \apj, 936, 12,
  \dodoi{10.3847/1538-4357/ac82ef}

\bibitem[{{Coburn} \& {Boggs}(2003)}]{Coburn2003}
{Coburn}, W., \& {Boggs}, S.~E. 2003, \nat, 423, 415,
  \dodoi{10.1038/nature01612}

\bibitem[{{Dai} \& {Lu}(1998)}]{Dai1998}
{Dai}, Z.~G., \& {Lu}, T. 1998, \aap, 333, L87

\bibitem[{{Dai} {et~al.}(2006){Dai}, {Wang}, {Wu}, \& {Zhang}}]{Dai2006}
{Dai}, Z.~G., {Wang}, X.~Y., {Wu}, X.~F., \& {Zhang}, B. 2006, Science, 311,
  1127, \dodoi{10.1126/science.1123606}

\bibitem[{{Deng} {et~al.}(2022){Deng}, {Lin}, {Zhou}, {Wang}, {Yang}, {Hou},
  {Li}, {Wang}, {Lu}, \& {Liang}}]{Deng2022}
{Deng}, L.-T., {Lin}, D.-B., {Zhou}, L., {et~al.} 2022, \apjl, 934, L22,
  \dodoi{10.3847/2041-8213/ac8169}

\bibitem[{{Dereli-B{\'e}gu{\'e}} {et~al.}(2020){Dereli-B{\'e}gu{\'e}}, {Pe'er},
  \& {Ryde}}]{Dereli-Begue2020}
{Dereli-B{\'e}gu{\'e}}, H., {Pe'er}, A., \& {Ryde}, F. 2020, \apj, 897, 145,
  \dodoi{10.3847/1538-4357/ab9a2d}

\bibitem[{Fan {et~al.}(2008)Fan, Xu, \& Wei}]{Fan:2008zw}
Fan, Y.-Z., Xu, D., \& Wei, D.-M. 2008, Mon. Not. Roy. Astron. Soc., 387, 92,
  \dodoi{10.1111/j.1365-2966.2008.12886.x}

\bibitem[{{Gao} \& {Zhang}(2015)}]{Gao2015}
{Gao}, H., \& {Zhang}, B. 2015, \apj, 801, 103,
  \dodoi{10.1088/0004-637X/801/2/103}

\bibitem[{Geng {et~al.}(2018)Geng, Huang, Wu, Song, \& Zong}]{Geng:2018dsd}
Geng, J.-J., Huang, Y.-F., Wu, X.-F., Song, L.-M., \& Zong, H.-S. 2018,
  Astrophys. J., 862, 115, \dodoi{10.3847/1538-4357/aacd05}

\bibitem[{{Geng} {et~al.}(2018){Geng}, {Huang}, {Wu}, {Zhang}, \&
  {Zong}}]{Geng2018}
{Geng}, J.-J., {Huang}, Y.-F., {Wu}, X.-F., {Zhang}, B., \& {Zong}, H.-S. 2018,
  \apjs, 234, 3, \dodoi{10.3847/1538-4365/aa9e84}

\bibitem[{Gill \& Granot(2020)}]{Gill:2019tpt}
Gill, R., \& Granot, J. 2020, Mon. Not. Roy. Astron. Soc., 491, 5815,
  \dodoi{10.1093/mnras/stz3340}

\bibitem[{Gill \& Granot(2021)}]{Gill:2021jzc}
---. 2021, Mon. Not. Roy. Astron. Soc., 504, 1939,
  \dodoi{10.1093/mnras/stab1013}

\bibitem[{{Gill} \& {Granot}(2024)}]{Gill2024}
{Gill}, R., \& {Granot}, J. 2024, \mnras, 527, 12178,
  \dodoi{10.1093/mnras/stad3991}

\bibitem[{{Gill} {et~al.}(2020){Gill}, {Granot}, \& {Beniamini}}]{Gill2020}
{Gill}, R., {Granot}, J., \& {Beniamini}, P. 2020, \mnras, 499, 1356,
  \dodoi{10.1093/mnras/staa2870}

\bibitem[{Gill {et~al.}(2018)Gill, Granot, \& Kumar}]{Gill:2018qrz}
Gill, R., Granot, J., \& Kumar, P. 2018.
\newblock \doarXiv{1811.11555}

\bibitem[{{Gill} {et~al.}(2021){Gill}, {Kole}, \& {Granot}}]{Gill2021}
{Gill}, R., {Kole}, M., \& {Granot}, J. 2021, Galaxies, 9, 82,
  \dodoi{10.3390/galaxies9040082}

\bibitem[{{Goldstein} {et~al.}(2012){Goldstein}, {Burgess}, {Preece}, {Briggs},
  {Guiriec}, {van der Horst}, {Connaughton}, {Wilson-Hodge}, {Paciesas},
  {Meegan}, {von Kienlin}, {Bhat}, {Bissaldi}, {Chaplin}, {Diehl}, {Fishman},
  {Fitzpatrick}, {Foley}, {Gibby}, {Giles}, {Greiner}, {Gruber}, {Kippen},
  {Kouveliotou}, {McBreen}, {McGlynn}, {Rau}, \& {Tierney}}]{Goldstein2012}
{Goldstein}, A., {Burgess}, J.~M., {Preece}, R.~D., {et~al.} 2012, \apjs, 199,
  19, \dodoi{10.1088/0067-0049/199/1/19}

\bibitem[{{G{\"o}tz} {et~al.}(2014){G{\"o}tz}, {Laurent}, {Antier}, {Covino},
  {D'Avanzo}, {D'Elia}, \& {Melandri}}]{2014MNRAS.444.2776G}
{G{\"o}tz}, D., {Laurent}, P., {Antier}, S., {et~al.} 2014, \mnras, 444, 2776,
  \dodoi{10.1093/mnras/stu1634}

\bibitem[{Granot(2003)}]{Granot:2003dy}
Granot, J. 2003, Astrophys. J. Lett., 596, L17, \dodoi{10.1086/379110}

\bibitem[{{Granot}(2003{\natexlab{a}})}]{Granot2003a}
{Granot}, J. 2003{\natexlab{a}}, \apjl, 596, L17, \dodoi{10.1086/379110}

\bibitem[{{Granot}(2003{\natexlab{b}})}]{2003ApJ...596L..17G}
---. 2003{\natexlab{b}}, \apjl, 596, L17, \dodoi{10.1086/379110}

\bibitem[{{Granot}(2005)}]{2005ApJ...631.1022G}
---. 2005, \apj, 631, 1022, \dodoi{10.1086/432676}

\bibitem[{{Granot} \& {K{\"o}nigl}(2003{\natexlab{a}})}]{2003ApJ83G}
{Granot}, J., \& {K{\"o}nigl}, A. 2003{\natexlab{a}}, APJ, 594, L83,
  \dodoi{10.1086/378733}

\bibitem[{{Granot} \& {K{\"o}nigl}(2003{\natexlab{b}})}]{2003ApJ...594L..83G}
---. 2003{\natexlab{b}}, \apjl, 594, L83, \dodoi{10.1086/378733}

\bibitem[{{Gruzinov}(1999)}]{1999ApJ...525L..29G}
{Gruzinov}, A. 1999, \apjl, 525, L29, \dodoi{10.1086/312323}

\bibitem[{Gruzinov \& Waxman(1999)}]{Gruzinov:1998xt}
Gruzinov, A., \& Waxman, E. 1999, Astrophys. J., 511, 852,
  \dodoi{10.1086/306720}

\bibitem[{{Guiriec} {et~al.}(2011){Guiriec}, {Connaughton}, {Briggs},
  {Burgess}, {Ryde}, {Daigne}, {M{\'e}sz{\'a}ros}, {Goldstein}, {McEnery},
  {Omodei}, {Bhat}, {Bissaldi}, {Camero-Arranz}, {Chaplin}, {Diehl}, {Fishman},
  {Foley}, {Gibby}, {Giles}, {Greiner}, {Gruber}, {von Kienlin}, {Kippen},
  {Kouveliotou}, {McBreen}, {Meegan}, {Paciesas}, {Preece}, {Rau}, {Tierney},
  {van der Horst}, \& {Wilson-Hodge}}]{Guiriec2011}
{Guiriec}, S., {Connaughton}, V., {Briggs}, M.~S., {et~al.} 2011, \apjl, 727,
  L33, \dodoi{10.1088/2041-8205/727/2/L33}

\bibitem[{{Ito} {et~al.}(2014){Ito}, {Nagataki}, {Matsumoto}, {Lee}, {Tolstov},
  {Mao}, {Dainotti}, \& {Mizuta}}]{Ito2014}
{Ito}, H., {Nagataki}, S., {Matsumoto}, J., {et~al.} 2014, \apj, 789, 159,
  \dodoi{10.1088/0004-637X/789/2/159}

\bibitem[{Ito {et~al.}(2014)Ito, Nagataki, Matsumoto, Lee, Tolstov, Mao,
  Dainotti, \& Mizuta}]{Ito:2014ica}
Ito, H., Nagataki, S., Matsumoto, J., {et~al.} 2014, Astrophys. J., 789, 159,
  \dodoi{10.1088/0004-637X/789/2/159}

\bibitem[{{Kaneko} {et~al.}(2006){Kaneko}, {Preece}, {Briggs}, {Paciesas},
  {Meegan}, \& {Band}}]{Kaneko2006}
{Kaneko}, Y., {Preece}, R.~D., {Briggs}, M.~S., {et~al.} 2006, \apjs, 166, 298,
  \dodoi{10.1086/505911}

\bibitem[{{Kole} {et~al.}(2020){Kole}, {De Angelis}, {Berlato}, {Burgess},
  {Gauvin}, {Greiner}, {Hajdas}, {Li}, {Li}, {Pollo}, {Produit}, {Rybka},
  {Song}, {Sun}, {Szabelski}, {Tymieniecka}, {Wang}, {Wu}, {Wu}, {Xiong},
  {Zhang}, \& {Zhang}}]{Kole2020}
{Kole}, M., {De Angelis}, N., {Berlato}, F., {et~al.} 2020, \aap, 644, A124,
  \dodoi{10.1051/0004-6361/202037915}

\bibitem[{{Kumar} \& {Zhang}(2015)}]{2015PhR...561....1K}
{Kumar}, P., \& {Zhang}, B. 2015, \physrep, 561, 1,
  \dodoi{10.1016/j.physrep.2014.09.008}

\bibitem[{{Lan} \& {Dai}(2020)}]{Lan2020}
{Lan}, M.-X., \& {Dai}, Z.-G. 2020, \apj, 892, 141,
  \dodoi{10.3847/1538-4357/ab7b5d}

\bibitem[{Lan {et~al.}(2021)Lan, Wang, Xu, Liu, \& Wu}]{Lan:2021tja}
Lan, M.-X., Wang, H.-B., Xu, S., Liu, S., \& Wu, X.-F. 2021, Astrophys. J.,
  909, 184, \dodoi{10.3847/1538-4357/abe3fb}

\bibitem[{{Lazzati} \& {Begelman}(2009)}]{2009ApJ...700L.141L}
{Lazzati}, D., \& {Begelman}, M.~C. 2009, \apjl, 700, L141,
  \dodoi{10.1088/0004-637X/700/2/L141}

\bibitem[{{Lazzati} {et~al.}(2004){Lazzati}, {Rossi}, {Ghisellini}, \&
  {Rees}}]{Lazzati2004}
{Lazzati}, D., {Rossi}, E., {Ghisellini}, G., \& {Rees}, M.~J. 2004, \mnras,
  347, L1, \dodoi{10.1111/j.1365-2966.2004.07387.x}

\bibitem[{{Lei} {et~al.}(2009){Lei}, {Wang}, {Zhang}, {Gan}, {Zou}, \&
  {Xie}}]{Lei2009}
{Lei}, W.~H., {Wang}, D.~X., {Zhang}, L., {et~al.} 2009, \apj, 700, 1970,
  \dodoi{10.1088/0004-637X/700/2/1970}

\bibitem[{{Li}(2019)}]{Li2019a}
{Li}, L. 2019, \apjs, 242, 16, \dodoi{10.3847/1538-4365/ab1b78}

\bibitem[{{Li}(2020)}]{Li2020}
---. 2020, \apj, 894, 100, \dodoi{10.3847/1538-4357/ab8014}

\bibitem[{{Li}(2022)}]{Li2022}
---. 2022, \apj, 941, 27, \dodoi{10.3847/1538-4357/ac3d89}

\bibitem[{{Li}(2023)}]{Li2023c}
---. 2023, \apjs, 266, 31, \dodoi{10.3847/1538-4365/acc867}

\bibitem[{{Li} {et~al.}(2021){Li}, {Ryde}, {Pe'er}, {Yu}, \&
  {Acuner}}]{Li2021b}
{Li}, L., {Ryde}, F., {Pe'er}, A., {Yu}, H.-F., \& {Acuner}, Z. 2021, \apjs,
  254, 35, \dodoi{10.3847/1538-4365/abee2a}

\bibitem[{{Li} {et~al.}(2018){Li}, {Wu}, {Lei}, {Dai}, {Liang}, \&
  {Ryde}}]{Li2018b}
{Li}, L., {Wu}, X.-F., {Lei}, W.-H., {et~al.} 2018, \apjs, 236, 26,
  \dodoi{10.3847/1538-4365/aabaf3}

\bibitem[{{Li} {et~al.}(2022){Li}, {Xue}, \& {Dai}}]{Li2022a}
{Li}, L., {Xue}, S.-S., \& {Dai}, Z.-G. 2022, arXiv e-prints, arXiv:2208.03583.
\newblock \doarXiv{2208.03583}

\bibitem[{{Li} {et~al.}(2019){Li}, {Geng}, {Meng}, {Wu}, {Huang}, {Wang},
  {Moradi}, {Uhm}, \& {Zhang}}]{Li2019b}
{Li}, L., {Geng}, J.-J., {Meng}, Y.-Z., {et~al.} 2019, \apj, 884, 109,
  \dodoi{10.3847/1538-4357/ab40b9}

\bibitem[{{Li} {et~al.}(2023){Li}, {Wang}, {Ryde}, {Pe'er}, {Zhang}, {Guiriec},
  {Castro-Tirado}, {Kann}, {Axelsson}, {Page}, {Veres}, \& {Bhat}}]{Li2023a}
{Li}, L., {Wang}, Y., {Ryde}, F., {et~al.} 2023, \apjl, 944, L57,
  \dodoi{10.3847/2041-8213/acb99d}

\bibitem[{{Liu} {et~al.}(2007){Liu}, {Gu}, {Xue}, \& {Lu}}]{Liu2007}
{Liu}, T., {Gu}, W.-M., {Xue}, L., \& {Lu}, J.-F. 2007, \apj, 661, 1025,
  \dodoi{10.1086/513689}

\bibitem[{{Lloyd} \& {Petrosian}(2000)}]{Lloyd2000a}
{Lloyd}, N.~M., \& {Petrosian}, V. 2000, \apj, 543, 722, \dodoi{10.1086/317125}

\bibitem[{{L{\"u}} \& {Zhang}(2014)}]{Lue2014}
{L{\"u}}, H.-J., \& {Zhang}, B. 2014, \apj, 785, 74,
  \dodoi{10.1088/0004-637X/785/1/74}

\bibitem[{{Lundman} {et~al.}(2013){Lundman}, {Pe'er}, \&
  {Ryde}}]{2013MNRAS.428.2430L}
{Lundman}, C., {Pe'er}, A., \& {Ryde}, F. 2013, \mnras, 428, 2430,
  \dodoi{10.1093/mnras/sts219}

\bibitem[{{Lundman} {et~al.}(2014{\natexlab{a}}){Lundman}, {Pe'er}, \&
  {Ryde}}]{Lundman2014}
---. 2014{\natexlab{a}}, \mnras, 440, 3292, \dodoi{10.1093/mnras/stu457}

\bibitem[{{Lundman} {et~al.}(2014{\natexlab{b}}){Lundman}, {Pe'er}, \&
  {Ryde}}]{2014MNRAS.440.3292L}
---. 2014{\natexlab{b}}, \mnras, 440, 3292, \dodoi{10.1093/mnras/stu457}

\bibitem[{Lundman {et~al.}(2018{\natexlab{a}})Lundman, Vurm, \&
  Beloborodov}]{Lundman2018}
Lundman, C., Vurm, I., \& Beloborodov, A.~M. 2018{\natexlab{a}}, Astrophys. J.,
  856, 145, \dodoi{10.3847/1538-4357/aab3e8}

\bibitem[{Lundman {et~al.}(2018{\natexlab{b}})Lundman, Vurm, \&
  Beloborodov}]{Lundman:2016fdl}
---. 2018{\natexlab{b}}, Astrophys. J., 856, 145,
  \dodoi{10.3847/1538-4357/aab3e8}

\bibitem[{Lyutikov {et~al.}(2003)Lyutikov, Pariev, \&
  Blandford}]{Lyutikov:2003bz}
Lyutikov, M., Pariev, V.~I., \& Blandford, R.~D. 2003, Astrophys. J., 597, 998,
  \dodoi{10.1086/378497}

\bibitem[{{Lyutikov} {et~al.}(2003){Lyutikov}, {Pariev}, \&
  {Blandford}}]{2003ApJ...597..998L}
{Lyutikov}, M., {Pariev}, V.~I., \& {Blandford}, R.~D. 2003, \apj, 597, 998,
  \dodoi{10.1086/378497}

\bibitem[{Mao {et~al.}(2018)Mao, Covino, \& Wang}]{Mao:2018rsr}
Mao, J., Covino, S., \& Wang, J. 2018, Astrophys. J., 860, 153,
  \dodoi{10.3847/1538-4357/aac5d9}

\bibitem[{Mao \& Wang(2013)}]{Mao:2013gha}
Mao, J., \& Wang, J. 2013, Astrophys. J., 776, 17,
  \dodoi{10.1088/0004-637X/776/1/17}

\bibitem[{Mao \& Wang(2017)}]{Mao:2017dlb}
---. 2017, Astrophys. J., 838, 78, \dodoi{10.3847/1538-4357/aa6628}

\bibitem[{{McGlynn} {et~al.}(2007){McGlynn}, {Clark}, {Dean}, {Hanlon},
  {McBreen}, {Willis}, {McBreen}, {Bird}, \& {Foley}}]{McGlynn2007}
{McGlynn}, S., {Clark}, D.~J., {Dean}, A.~J., {et~al.} 2007, \aap, 466, 895,
  \dodoi{10.1051/0004-6361:20066179}

\bibitem[{{McKinney} \& {Uzdensky}(2012)}]{2012MNRAS.419..573M}
{McKinney}, J.~C., \& {Uzdensky}, D.~A. 2012, \mnras, 419, 573,
  \dodoi{10.1111/j.1365-2966.2011.19721.x}

\bibitem[{Medvedev(2000)}]{Medvedev:2000gu}
Medvedev, M.~V. 2000, Astrophys. J., 540, 704, \dodoi{10.1086/309374}

\bibitem[{{Medvedev} \& {Loeb}(1999)}]{Medvedev1999}
{Medvedev}, M.~V., \& {Loeb}, A. 1999, \apj, 526, 697, \dodoi{10.1086/308038}

\bibitem[{{Meegan} {et~al.}(2009){Meegan}, {Lichti}, {Bhat}, {Bissaldi},
  {Briggs}, {Connaughton}, {Diehl}, {Fishman}, {Greiner}, {Hoover}, {van der
  Horst}, {von Kienlin}, {Kippen}, {Kouveliotou}, {McBreen}, {Paciesas},
  {Preece}, {Steinle}, {Wallace}, {Wilson}, \& {Wilson-Hodge}}]{Meegan2009}
{Meegan}, C., {Lichti}, G., {Bhat}, P.~N., {et~al.} 2009, \apj, 702, 791,
  \dodoi{10.1088/0004-637X/702/1/791}

\bibitem[{{M{\'e}sz{\'a}ros} \& {Rees}(2000)}]{Meszaros2000}
{M{\'e}sz{\'a}ros}, P., \& {Rees}, M.~J. 2000, \apj, 530, 292,
  \dodoi{10.1086/308371}

\bibitem[{{Metzger} {et~al.}(2011){Metzger}, {Giannios}, {Thompson},
  {Bucciantini}, \& {Quataert}}]{Metzger2011}
{Metzger}, B.~D., {Giannios}, D., {Thompson}, T.~A., {Bucciantini}, N., \&
  {Quataert}, E. 2011, \mnras, 413, 2031,
  \dodoi{10.1111/j.1365-2966.2011.18280.x}

\bibitem[{{Metzger} {et~al.}(2008){Metzger}, {Quataert}, \&
  {Thompson}}]{Metzger2008}
{Metzger}, B.~D., {Quataert}, E., \& {Thompson}, T.~A. 2008, \mnras, 385, 1455,
  \dodoi{10.1111/j.1365-2966.2008.12923.x}

\bibitem[{{Mundell} {et~al.}(2007){Mundell}, {Steele}, {Smith}, {Kobayashi},
  {Melandri}, {Guidorzi}, {Gomboc}, {Mottram}, {Clarke}, {Monfardini},
  {Carter}, \& {Bersier}}]{Mundell2007Sci}
{Mundell}, C.~G., {Steele}, I.~A., {Smith}, R.~J., {et~al.} 2007, Science, 315,
  1822, \dodoi{10.1126/science.1138484}

\bibitem[{{Mundell} {et~al.}(2013){Mundell}, {Kopa{\v c}}, {Arnold}, {Steele},
  {Gomboc}, {Kobayashi}, {Harrison}, {Smith}, {Guidorzi}, {Virgili},
  {Melandri}, \& {Japelj}}]{2013Natur.504..119M}
{Mundell}, C.~G., {Kopa{\v c}}, D., {Arnold}, D.~M., {et~al.} 2013, \nat, 504,
  119, \dodoi{10.1038/nature12814}

\bibitem[{{Nakar} {et~al.}(2003){Nakar}, {Piran}, \&
  {Waxman}}]{2003JCAP...10..005N}
{Nakar}, E., {Piran}, T., \& {Waxman}, E. 2003, \jcap, 10, 005,
  \dodoi{10.1088/1475-7516/2003/10/005}

\bibitem[{Nava {et~al.}(2016)Nava, Nakar, \& Piran}]{2016MNRAS.455.1594N}
Nava, L., Nakar, E., \& Piran, T. 2016, Monthly Notices of the Royal
  Astronomical Society, 455, 1594

\bibitem[{Parsotan {et~al.}(2020)Parsotan, Lopez-Camara, \&
  Lazzati}]{Parsotan:2020ilh}
Parsotan, T., Lopez-Camara, D., \& Lazzati, D. 2020, Astrophys. J., 896, 139,
  \dodoi{10.3847/1538-4357/ab910f}

\bibitem[{{Pe'er}(2015)}]{Peer2015a}
{Pe'er}, A. 2015, Advances in Astronomy, 2015, 907321,
  \dodoi{10.1155/2015/907321}

\bibitem[{{Pe'er} {et~al.}(2006){Pe'er}, {M{\'e}sz{\'a}ros}, \&
  {Rees}}]{Peer2006a}
{Pe'er}, A., {M{\'e}sz{\'a}ros}, P., \& {Rees}, M.~J. 2006, \apj, 642, 995,
  \dodoi{10.1086/501424}

\bibitem[{{Pe'Er} \& {Ryde}(2017)}]{PeEr2017}
{Pe'Er}, A., \& {Ryde}, F. 2017, International Journal of Modern Physics D, 26,
  1730018, \dodoi{10.1142/S021827181730018X}

\bibitem[{{Planck Collaboration} {et~al.}(2018){Planck Collaboration},
  {Aghanim}, {Akrami}, {Ashdown}, {Aumont}, {Baccigalupi}, {Ballardini},
  {Banday}, {Barreiro}, {Bartolo}, {Basak}, {Battye}, {Benabed}, {Bernard},
  {Bersanelli}, {Bielewicz}, {Bock}, {Bond}, {Borrill}, {Bouchet}, {Boulanger},
  {Bucher}, {Burigana}, {Butler}, {Calabrese}, {Cardoso}, {Carron},
  {Challinor}, {Chiang}, {Chluba}, {Colombo}, {Combet}, {Contreras}, {Crill},
  {Cuttaia}, {de Bernardis}, {de Zotti}, {Delabrouille}, {Delouis}, {Di
  Valentino}, {Diego}, {Dor{\'e}}, {Douspis}, {Ducout}, {Dupac}, {Dusini},
  {Efstathiou}, {Elsner}, {En{\ss}lin}, {Eriksen}, {Fantaye}, {Farhang},
  {Fergusson}, {Fernandez-Cobos}, {Finelli}, {Forastieri}, {Frailis},
  {Franceschi}, {Frolov}, {Galeotta}, {Galli}, {Ganga}, {G{\'e}nova-Santos},
  {Gerbino}, {Ghosh}, {Gonz{\'a}lez-Nuevo}, {G{\'o}rski}, {Gratton},
  {Gruppuso}, {Gudmundsson}, {Hamann}, {Hand ley}, {Herranz}, {Hivon}, {Huang},
  {Jaffe}, {Jones}, {Karakci}, {Keih{\"a}nen}, {Keskitalo}, {Kiiveri}, {Kim},
  {Kisner}, {Knox}, {Krachmalnicoff}, {Kunz}, {Kurki-Suonio}, {Lagache},
  {Lamarre}, {Lasenby}, {Lattanzi}, {Lawrence}, {Le Jeune}, {Lemos},
  {Lesgourgues}, {Levrier}, {Lewis}, {Liguori}, {Lilje}, {Lilley}, {Lindholm},
  {L{\'o}pez-Caniego}, {Lubin}, {Ma}, {Mac{\'\i}as-P{\'e}rez}, {Maggio},
  {Maino}, {Mandolesi}, {Mangilli}, {Marcos-Caballero}, {Maris}, {Martin},
  {Martinelli}, {Mart{\'\i}nez-Gonz{\'a}lez}, {Matarrese}, {Mauri}, {McEwen},
  {Meinhold}, {Melchiorri}, {Mennella}, {Migliaccio}, {Millea}, {Mitra},
  {Miville-Desch{\^e}nes}, {Molinari}, {Montier}, {Morgante}, {Moss}, {Natoli},
  {N{\o}rgaard-Nielsen}, {Pagano}, {Paoletti}, {Partridge}, {Patanchon},
  {Peiris}, {Perrotta}, {Pettorino}, {Piacentini}, {Polastri}, {Polenta},
  {Puget}, {Rachen}, {Reinecke}, {Remazeilles}, {Renzi}, {Rocha}, {Rosset},
  {Roudier}, {Rubi{\~n}o-Mart{\'\i}n}, {Ruiz-Granados}, {Salvati}, {Sandri},
  {Savelainen}, {Scott}, {Shellard}, {Sirignano}, {Sirri}, {Spencer},
  {Sunyaev}, {Suur-Uski}, {Tauber}, {Tavagnacco}, {Tenti}, {Toffolatti},
  {Tomasi}, {Trombetti}, {Valenziano}, {Valiviita}, {Van Tent}, {Vibert},
  {Vielva}, {Villa}, {Vittorio}, {Wand elt}, {Wehus}, {White}, {White},
  {Zacchei}, \& {Zonca}}]{PlanckCollaboration2018}
{Planck Collaboration}, {Aghanim}, N., {Akrami}, Y., {et~al.} 2018, arXiv
  e-prints, arXiv:1807.06209.
\newblock \doarXiv{1807.06209}

\bibitem[{{Preece} {et~al.}(1998){Preece}, {Briggs}, {Mallozzi}, {Pendleton},
  {Paciesas}, \& {Band}}]{Preece1998}
{Preece}, R.~D., {Briggs}, M.~S., {Mallozzi}, R.~S., {et~al.} 1998, \apjl, 506,
  L23, \dodoi{10.1086/311644}

\bibitem[{Prosekin {et~al.}(2016)Prosekin, Kelner, \&
  Aharonian}]{Prosekin:2016caz}
Prosekin, A.~{\relax Yu}., Kelner, S.~R., \& Aharonian, F.~A. 2016, Phys. Rev.,
  D94, 063010, \dodoi{10.1103/PhysRevD.94.063010}

\bibitem[{{Ravasio} {et~al.}(2018){Ravasio}, {Oganesyan}, {Ghirlanda}, {Nava},
  {Ghisellini}, {Pescalli}, \& {Celotti}}]{Ravasio2018}
{Ravasio}, M.~E., {Oganesyan}, G., {Ghirlanda}, G., {et~al.} 2018, \aap, 613,
  A16, \dodoi{10.1051/0004-6361/201732245}

\bibitem[{{Rees} \& {Meszaros}(1994)}]{Rees1994}
{Rees}, M.~J., \& {Meszaros}, P. 1994, \apjl, 430, L93, \dodoi{10.1086/187446}

\bibitem[{{Rees} \& {M{\'e}sz{\'a}ros}(2005)}]{Rees2005}
{Rees}, M.~J., \& {M{\'e}sz{\'a}ros}, P. 2005, \apj, 628, 847,
  \dodoi{10.1086/430818}

\bibitem[{{Rybicki} \& {Lightman}(1979)}]{Rybicki1979}
{Rybicki}, G.~B., \& {Lightman}, A.~P. 1979, {Radiative processes in
  astrophysics}

\bibitem[{Rybicki \& Lightman(2008)}]{Rybicki:2008vo}
Rybicki, G.~B., \& Lightman, A.~P. 2008, {Radiative Processes in Astrophysics}
  (John Wiley {\&} Sons)

\bibitem[{{Ryde} {et~al.}(2010){Ryde}, {Axelsson}, {Zhang}, {McGlynn}, {Pe'er},
  {Lundman}, {Larsson}, {Battelino}, {Zhang}, {Bissaldi}, {Bregeon}, {Briggs},
  {Chiang}, {de Palma}, {Guiriec}, {Larsson}, {Longo}, {McBreen}, {Omodei},
  {Petrosian}, {Preece}, \& {van der Horst}}]{Ryde2010}
{Ryde}, F., {Axelsson}, M., {Zhang}, B.~B., {et~al.} 2010, \apjl, 709, L172,
  \dodoi{10.1088/2041-8205/709/2/L172}

\bibitem[{{Sari}(1999)}]{1999ApJ...524L..43S}
{Sari}, R. 1999, APJ, 524, L43, \dodoi{10.1086/312294}

\bibitem[{{Sari} {et~al.}(1998){Sari}, {Piran}, \&
  {Narayan}}]{1998ApJ...497L..17S}
{Sari}, R., {Piran}, T., \& {Narayan}, R. 1998, \apjl, 497, L17,
  \dodoi{10.1086/311269}

\bibitem[{{Schwarz}(1978)}]{Schwarz1978}
{Schwarz}, G. 1978, Annals of Statistics, 6, 461

\bibitem[{Shakeri \& Allahyari(2018)}]{Shakeri:2018qal}
Shakeri, S., \& Allahyari, A. 2018, JCAP, 11, 042,
  \dodoi{10.1088/1475-7516/2018/11/042}

\bibitem[{{Song} {et~al.}(2022){Song}, {Zhang}, {Ge}, \& {Zhang}}]{SongXY2022}
{Song}, X.-Y., {Zhang}, S.-N., {Ge}, M.-Y., \& {Zhang}, S. 2022, \mnras, 517,
  2088, \dodoi{10.1093/mnras/stac2764}

\bibitem[{{Steele} {et~al.}(2009){Steele}, {Mundell}, {Smith}, {Kobayashi}, \&
  {Guidorzi}}]{2009Natur.462..767S}
{Steele}, I.~A., {Mundell}, C.~G., {Smith}, R.~J., {Kobayashi}, S., \&
  {Guidorzi}, C. 2009, \nat, 462, 767, \dodoi{10.1038/nature08590}

\bibitem[{Teboul \& Shaviv(2021)}]{Teboul:2020yis}
Teboul, O., \& Shaviv, N.~J. 2021, Mon. Not. Roy. Astron. Soc., 507, 5340,
  \dodoi{10.1093/mnras/stab2491}

\bibitem[{{Thompson}(1994)}]{Thompson1994}
{Thompson}, C. 1994, \mnras, 270, 480, \dodoi{10.1093/mnras/270.3.480}

\bibitem[{{Toma} {et~al.}(2009{\natexlab{a}}){Toma}, {Sakamoto}, {Zhang},
  {Hill}, {McConnell}, {Bloser}, {Yamazaki}, {Ioka}, \&
  {Nakamura}}]{2009ApJ...698.1042T}
{Toma}, K., {Sakamoto}, T., {Zhang}, B., {et~al.} 2009{\natexlab{a}}, \apj,
  698, 1042, \dodoi{10.1088/0004-637X/698/2/1042}

\bibitem[{{Toma} {et~al.}(2009{\natexlab{b}}){Toma}, {Sakamoto}, {Zhang},
  {Hill}, {McConnell}, {Bloser}, {Yamazaki}, {Ioka}, \& {Nakamura}}]{Toma2009}
---. 2009{\natexlab{b}}, \apj, 698, 1042, \dodoi{10.1088/0004-637X/698/2/1042}

\bibitem[{{Uehara} {et~al.}(2012){Uehara}, {Toma}, {Kawabata}, {Chiyonobu},
  {Fukazawa}, {Ikejiri}, {Inoue}, {Itoh}, {Komatsu}, {Miyamoto}, {Mizuno},
  {Nagae}, {Nakaya}, {Ohsugi}, {Sakimoto}, {Sasada}, {Tanaka}, {Uemura},
  {Yamanaka}, {Yamashita}, {Yamazaki}, \& {Yoshida}}]{2012ApJ...752L...6U}
{Uehara}, T., {Toma}, K., {Kawabata}, K.~S., {et~al.} 2012, \apjl, 752, L6,
  \dodoi{10.1088/2041-8205/752/1/L6}

\bibitem[{{Usov}(1992)}]{Usov1992}
{Usov}, V.~V. 1992, \nat, 357, 472, \dodoi{10.1038/357472a0}

\bibitem[{{Vianello}(2018)}]{Vianello2018a}
{Vianello}, G. 2018, \apjs, 236, 17, \dodoi{10.3847/1538-4365/aab780}

\bibitem[{{Vianello} {et~al.}(2015){Vianello}, {Lauer}, {Younk}, {Tibaldo},
  {Burgess}, {Ayala}, {Harding}, {Hui}, {Omodei}, \& {Zhou}}]{Vianello2015}
{Vianello}, G., {Lauer}, R.~J., {Younk}, P., {et~al.} 2015, arXiv e-prints.
\newblock \doarXiv{1507.08343}

\bibitem[{{Wang} {et~al.}(2019){Wang}, {Li}, {Moradi}, \& {Ruffini}}]{Wang2019}
{Wang}, Y., {Li}, L., {Moradi}, R., \& {Ruffini}, R. 2019, arXiv e-prints.
\newblock \doarXiv{1901.07505}

\bibitem[{{Wang} {et~al.}(2022){Wang}, {Zheng}, \& {Jin}}]{WangYun2022}
{Wang}, Y., {Zheng}, T.-C., \& {Jin}, Z.-P. 2022, \apj, 940, 142,
  \dodoi{10.3847/1538-4357/aca017}

\bibitem[{{Westfold}(1959)}]{1959ApJ...130..241W}
{Westfold}, K.~C. 1959, \apj, 130, 241, \dodoi{10.1086/146713}

\bibitem[{{Wheeler} {et~al.}(2000){Wheeler}, {Yi}, {H{\"o}flich}, \&
  {Wang}}]{Wheeler2000}
{Wheeler}, J.~C., {Yi}, I., {H{\"o}flich}, P., \& {Wang}, L. 2000, \apj, 537,
  810, \dodoi{10.1086/309055}

\bibitem[{{Willis} {et~al.}(2005){Willis}, {Barlow}, {Bird}, {Clark}, {Dean},
  {McConnell}, {Moran}, {Shaw}, \& {Sguera}}]{Willis2005}
{Willis}, D.~R., {Barlow}, E.~J., {Bird}, A.~J., {et~al.} 2005, \aap, 439, 245,
  \dodoi{10.1051/0004-6361:20052693}

\bibitem[{{Yonetoku} {et~al.}(2011){Yonetoku}, {Murakami}, {Gunji}, {Mihara},
  {Toma}, {Sakashita}, {Morihara}, {Takahashi}, {Toukairin}, {Fujimoto},
  {Kodama}, {Kubo}, \& {IKAROS Demonstration Team}}]{2011ApJ...743L..30Y}
{Yonetoku}, D., {Murakami}, T., {Gunji}, S., {et~al.} 2011, \apjl, 743, L30,
  \dodoi{10.1088/2041-8205/743/2/L30}

\bibitem[{{Yonetoku} {et~al.}(2012){Yonetoku}, {Murakami}, {Gunji}, {Mihara},
  {Toma}, {Morihara}, {Takahashi}, {Wakashima}, {Yonemochi}, {Sakashita},
  {Toukairin}, {Fujimoto}, \& {Kodama}}]{2012ApJ...758L...1Y}
---. 2012, \apjl, 758, L1, \dodoi{10.1088/2041-8205/758/1/L1}

\bibitem[{{Yu} {et~al.}(2019){Yu}, {Dereli-B{\'e}gu{\'e}}, \& {Ryde}}]{Yu2019}
{Yu}, H.-F., {Dereli-B{\'e}gu{\'e}}, H., \& {Ryde}, F. 2019, \apj, 886, 20,
  \dodoi{10.3847/1538-4357/ab488a}

\bibitem[{{Zhang}(2014)}]{2014IJMPD..2330002Z}
{Zhang}, B. 2014, International Journal of Modern Physics D, 23, 1430002,
  \dodoi{10.1142/S021827181430002X}

\bibitem[{{Zhang}(2018)}]{Zhang2018}
---. 2018, {The Physics of Gamma-Ray Bursts}, \dodoi{10.1017/9781139226530}

\bibitem[{{Zhang} \& {M{\'e}sz{\'a}ros}(2001)}]{Zhang2001}
{Zhang}, B., \& {M{\'e}sz{\'a}ros}, P. 2001, \apjl, 552, L35,
  \dodoi{10.1086/320255}

\bibitem[{{Zhang} \& {Pe'er}(2009)}]{Zhang2009}
{Zhang}, B., \& {Pe'er}, A. 2009, \apjl, 700, L65,
  \dodoi{10.1088/0004-637X/700/2/L65}

\bibitem[{{Zhang} \& {Yan}(2011)}]{2011ApJ...726...90Z}
{Zhang}, B., \& {Yan}, H. 2011, \apj, 726, 90,
  \dodoi{10.1088/0004-637X/726/2/90}

\bibitem[{{Zhang} {et~al.}(2019){Zhang}, {Kole}, {Bao}, {Batsch}, {Bernasconi},
  {Cadoux}, {Chai}, {Dai}, {Dong}, {Gauvin}, {Hajdas}, {Lan}, {Li}, {Li}, {Li},
  {Liu}, {Liu}, {Marcinkowski}, {Produit}, {Orsi}, {Pohl}, {Rybka}, {Shi},
  {Song}, {Sun}, {Szabelski}, {Tymieniecka}, {Wang}, {Wang}, {Wen}, {Wu}, {Wu},
  {Wu}, {Xiao}, {Xiong}, {Zhang}, {Zhang}, {Zhang}, {Zhang}, \&
  {Zwolinska}}]{Zhang2019}
{Zhang}, S.-N., {Kole}, M., {Bao}, T.-W., {et~al.} 2019, Nature Astronomy,
  \dodoi{10.1038/s41550-018-0664-0}

\end{thebibliography}

\clearpage
\begin{deluxetable}{cccccccccccc}
\tablewidth{0pt}
\centering
\rotate
\tabletypesize{\scriptsize}
\tablecaption{A Sample of GRB Polarimetric Observations Used in Our Analysis}\label{tab:Sample}
\tablehead{
\colhead{GRB}
&\colhead{$t_{1}$$\sim$$t_{2}$}
&\colhead{PD}
&\colhead{PA}
&\colhead{Energy Band}
&\colhead{Significance}
&\colhead{Instrument}
&\colhead{References}
&\colhead{$T_{90}$}
&\colhead{Type}
&\colhead{$S_{\gamma}$}
&\colhead{$z$}\\
(Fermi ID)&(s)&($\pi$\%)&($^{\circ}$)&&&($\sigma$)&(for polarization)&(s)&&(erg cm$^{-2}$)
}
\colnumbers
\startdata
100826A(957)&0$\sim$100&$27 \pm 11$&75$\pm$20&20 keV-10 MeV&2.9$\sigma$&IKAROS-GAP&\cite{2011ApJ...743L..30Y}&85.0$\pm$0.7&Long&(1.6$\pm$0.0)$\times$10$^{-4}$&\nodata\\
110301A(214)&0$\sim$7&$70 \pm 22$&73$\pm$11&10 keV-1 MeV&3.7$\sigma$&IKAROS-GAP&\cite{2012ApJ...758L...1Y}&5.7$\pm$0.4&Long&(3.6$\pm$0.0)$\times$10$^{-5}$&\nodata\\
110721A(200)&0$\sim$11&$84^{+16}_{-28}$&160$\pm$11&10 keV-1 MeV&3.3$\sigma$&IKAROS-GAP&\cite{2012ApJ...758L...1Y}&21.8$\pm$0.6&Long&(3.7$\pm$0.0)$\times$10$^{-5}$&0.382\\
140206A(275)&4$\sim$26&$>28$&80$\pm$15&15 keV-1 MeV&90\% confidence&INTEGRAL-IBIS&\cite{2014MNRAS.444.2776G}&146.7$\pm$4.4&Long&(1.2$\pm$0.0)$\times$10$^{-4}$&2.73\\
160325A(291)&2.28$\sim$46.10&$<$45.02&NA&100-600 keV&$\sim$2.2$\sigma$&$AstroSat$-CZTI&\cite{Chattopadhyay2022}&42.9$\pm$0.6&Long&(1.9$\pm$0.0)$\times$10$^{-5}$&\nodata\\
160623A(209)&1.16$\sim$18.21&$<56.51$&NA&100-600 keV&NA&$AstroSat$-CZTI&\cite{Chattopadhyay2022}&107.8$\pm$8.7&Long&(4.0$\pm$0.1)$\times$10$^{-6}$&0.367\\
160802A(259)&0$\sim$20.34&85$\pm$29&$\sim$-32&100-600 keV&$\sim$3$\sigma$&$AstroSat$-CZTI&\cite{2018ApJ...862..154C}&16.4$\pm$0.4&Long&(6.8$\pm$0.0)$\times$10$^{-5}$&\nodata\\
160821A(857)&117.18$\sim$155.135&$<33.87$&NA&100-600 keV&NA&$AstroSat$-CZTI&\cite{Chattopadhyay2022}&43.0$\pm$0.7&Long&(5.2$\pm$0.0)$\times$10$^{-4}$&\nodata\\
161218B(356)&0$\sim$25.1(0.2)&$13^{+28}_{-13}$&$68^{+36}_{-54}$&50-500 keV&NA&POLAR&\cite{Kole2020}&25.9$\pm$0.4&Long&(8.6$\pm$0.0)$\times$10$^{-5}$&\nodata\\
161229A(878)&0$\sim$31.26(0.44)&$17^{+24}_{-13}$&$106^{+55}_{-22}$&50-500 keV&NA&POLAR&\cite{Kole2020}&33.5$\pm$0.4&Long&(3.8$\pm$0.0)$\times$10$^{-5}$&\nodata\\
170101B(116)&0$\sim$11.14(0.38)&$60^{+24}_{-36}$&$109^{+34}_{-21}$&50-500 keV&NA&POLAR&\cite{Kole2020}&12.8$\pm$0.4&Long&(1.3$\pm$0.0)$\times$10$^{-5}$&\nodata\\
170114A(917)&0$\sim$10.48(0.16)&$10.1^{+10.5}_{-7.4}$&$166^{+25}_{-22}$&50-500 keV&NA&POLAR&\cite{Kole2020}&12.0$\pm$1.3&Long&(1.8$\pm$0.0)$\times$10$^{-5}$&\nodata\\
170127C(067)&0$\sim$0.14(0.01)&$9.9^{+19.3}_{-8.4}$&$45^{+39}_{-36}$&50-500 keV&NA&POLAR&\cite{Kole2020}&0.13$\pm$0.05&Short&(5.6$\pm$0.1)$\times$10$^{-6}$&\nodata\\
170206A(453)&0$\sim$1.26(0.01)&$13.5^{+7.4}_{-8.6}$&NA&50-500 keV&NA&POLAR&\cite{Kole2020}&1.17$\pm$0.10&Short&(1.0$\pm$0.0)$\times$10$^{-5}$&\nodata\\
170207A(906)&0$\sim$38.76(0.26)&$5.9^{+9.6}_{-5.9}$&NA&50-500 keV&NA&POLAR&\cite{Kole2020}&38.9$\pm$0.6&Long&(5.4$\pm$0.0)$\times$10$^{-5}$&\nodata\\
170210A(116)&0$\sim$47.63(2.51)&$11.4^{+35.7}_{-9.7}$&$13^{+24}_{-21}$&50-500 keV&NA&POLAR&\cite{Kole2020}&76.5$\pm$1.4&Long&(9.6$\pm$0.0)$\times$10$^{-5}$&\nodata\\
170305A(256)&0$\sim$0.45(0.01)&$40^{+25}_{-25}$&$88^{+30}_{-20}$&50-500 keV&NA&POLAR&\cite{Kole2020}&0.45$\pm$0.07&Short&(1.1$\pm$0.0)$\times$10$^{-6}$&\nodata\\
170527A(480)&-0.76$\sim$37.18&$<36.46$&NA&100-600 keV&NA&$AstroSat$-CZTI&\cite{Chattopadhyay2022}&49.2$\pm$1.6&Long&(8.4$\pm$0.0)$\times$10$^{-5}$&\nodata\\
171010A(792)&7.14$\sim$106.20&$<30.02$&NA&100-600 keV&NA&$AstroSat$-CZTI&\cite{Chattopadhyay2022}&107.3$\pm$0.8&Long&(6.3$\pm$0.0)$\times$10$^{-4}$&0.3285\\
171227A(000)&0.26$\sim$30.27&$<55.62$&NA&100-600 keV&NA&$AstroSat$-CZTI&\cite{Chattopadhyay2022}&37.6$\pm$0.6&Long&(2.9$\pm$0.0)$\times$10$^{-4}$&\nodata\\
180120A(207)&0.09$\sim$24.10&62.37$\pm$29.79&-3.65$\pm$26.00&100-600 keV&NA&$AstroSat$-CZTI&\cite{Chattopadhyay2022}&28.9$\pm$0.7&Long&(6.5$\pm$0.0)$\times$10$^{-5}$&\nodata\\
180427A(442)&0.15$\sim$13.16&60.01$\pm$22.32&16.91$\pm$23.00&100-600 keV&NA&$AstroSat$-CZTI&\cite{Chattopadhyay2022}&25.9$\pm$0.7&Long&(5.1$\pm$0.0)$\times$10$^{-5}$&\nodata\\
180806A(944)&-0.01$\sim$10.32&$<95.80$&NA&100-600 keV&NA&$AstroSat$-CZTI&\cite{Chattopadhyay2022}&15.6$\pm$0.8&Long&(2.2$\pm$0.0)$\times$10$^{-5}$&\nodata\\
180914A(522)&5.34$\sim$133.35&$<33.55$&NA&100-600 keV&NA&$AstroSat$-CZTI&\cite{Chattopadhyay2022}&122.4$\pm$0.6&Long&(8.2$\pm$0.0)$\times$10$^{-5}$&\nodata\\
190530A(430)&7.24$\sim$34.10&46.85$\pm$18.53&43.58$\pm$5.00&100-600 keV&NA&$AstroSat$-CZTI&\cite{Chattopadhyay2022}&18.4$\pm$0.4&Long&(3.7$\pm$0.0)$\times$10$^{-4}$&0.9386\\
200311A(636)&0.14$\sim$39.20&$<33.10$&NA&100-600 keV&NA&$AstroSat$-CZTI&\cite{Chattopadhyay2022}&52.5$\pm$0.8&Long&(4.3$\pm$0.0)$\times$10$^{-5}$&\nodata\\
200412A(290)&-2.85$\sim$35.18&$<53.73$&NA&100-600 keV&NA&$AstroSat$-CZTI&\cite{Chattopadhyay2022}&12.6$\pm$0.4&Long&(2.9$\pm$0.0)$\times$10$^{-5}$&\nodata\\
\enddata 
\vspace{0mm}
\tablecomments{No Fermi data available for GRB160703A, GRB161203, GRB161217C, GRB161218A, GRB170101A, GRB170320A, GRB180103A, GRB180809B, GRB180914B, GRB190928A, and GRB200806A.}
\end{deluxetable}

\clearpage
\setlength{\tabcolsep}{0.05em}
\begin{deluxetable}{ccccccccccccccc}
\centering
\rotate
\tabletypesize{\tiny}
\tablecaption{Spectral Fit Results of the HD-type Bursts with Band+BB-like Spectra.}\label{tab:HD}
\tablehead{
\specialrule{0em}{5pt}{5pt} 
&
&
&
&
&\multirow{2}{*}{Spectral Models}
&\multicolumn{4}{c}{Base Component}
&\multicolumn{2}{c}{Additional Component}\\  
\cmidrule(lr){7-10} 
\cmidrule(lr){11-13}\\
\colhead{GRB}
&\colhead{Detectors}
&\colhead{$[\Delta T_{\rm (bkg,1)},\Delta T_{\rm (bkg,2)}]$}
&\colhead{$\Delta T_{\rm src}$}
&\colhead{$S$}
&\colhead{Preferred}
&\colhead{$K$}
&\colhead{$\alpha$}
&\colhead{$E_{\rm p}$}
&\colhead{$\beta$}
&\colhead{$K$}
&\colhead{$k$T}
&\colhead{$F_{\rm BB}/F_{\rm tot}$}
&\colhead{$F_{\gamma}$}
&\colhead{$\Delta$BIC}\\
\hline
&&[($t^{\rm bkg,1}_{1}$$\sim$$t^{\rm bkg,1}_{2}$), ($t^{\rm bkg,2}_{1}$$\sim$$t^{\rm bkg,2}_{2}$)]&($t_{1}$$\sim$$t_{2}$)&&&(Band)&(Band)&(Band)&(Band)&(BB)&(BB)&(Flux Ratio)&(Total Flux)&[(Band+BB)-Band)]\\
&&(s)&(s)&&&
(ph.s$^{-1}$.cm$^{-2}$.keV$^{-1}$)&&(keV)&&(ph.s$^{-1}$.cm$^{-2}$.keV$^{-1}$)&(keV)&&(erg.cm$^{-2}$.s$^{-1}$)
}
\colnumbers
\startdata
100826A&n7(n8)b1&[(-20$\sim$-10), (200$\sim$250)]&(0$\sim$82)&103&Band+BB&(3.0$^{+0.2}_{-0.2}$)$\times 10^{-2}$&-0.84$^{+0.04}_{-0.04}$&518$^{+46}_{-46}$&-2.28$^{+0.07}_{-0.07}$&(5.5$^{+2.0}_{-1.9}$)$\times 10^{-5}$&21$^{+2}_{-2}$&0.03$^{+0.03}_{-0.03}$&(3.2$^{+0.3}_{-0.3}$)$\times 10^{-6}$&-60\\
110301A&n7(n8)nbb1&[(-20 $\sim$ -10), (40 $\sim$ 60)]&(0$\sim$7)&277&Band+BB&(3.7$^{+0.3}_{-0.3}$)$\times 10^{-1}$&-0.72$^{+0.07}_{-0.07}$&114$^{+2}_{-3}$&-2.87$^{+0.08}_{-0.07}$&(1.1$^{+0.6}_{-0.5}$)$\times 10^{-2}$&7$^{+1}_{-1}$&0.04$^{+0.03}_{-0.03}$&(5.9$^{+0.7}_{-0.7}$)$\times 10^{-6}$&-23\\
110721A&(n6)n7n9b1&[(-20 $\sim$ -10), (40 $\sim$ 60)]&(0$\sim$11)&115&Band+BB&(3.0$^{+0.1}_{-0.1}$)$\times 10^{-2}$&-1.20$^{+0.02}_{-0.02}$&1620$^{+234}_{-229}$&-2.19$^{+0.10}_{-0.10}$&(1.7$^{+0.3}_{-0.3}$)$\times 10^{-5}$&33$^{+2}_{-2}$&0.03$^{+0.01}_{-0.01}$&(6.9$^{+0.7}_{-0.6}$)$\times 10^{-6}$&-29\\
140206A&n0(n1)n3b0&[(-40 $\sim$ -20), (70 $\sim$ 90)]&(4$\sim$26)&170&Band+BB&(3.9$^{+0.1}_{-0.1}$)$\times 10^{-2}$&-1.06$^{+0.02}_{-0.02}$&679$^{+43}_{-43}$&-2.32$^{+0.08}_{-0.08}$&(4.8$^{+0.5}_{-0.5}$)$\times 10^{-5}$&27$^{+1}_{-1}$&0.05$^{+0.01}_{-0.01}$&(4.6$^{+0.3}_{-0.2}$)$\times 10^{-6}$&-131\\
160802A&(n2)b0&[(-20 $\sim$ -10), (60 $\sim$ 80)]&(0$\sim$20.34)&152&Band+BB&(3.9$^{+0.2}_{-0.2}$)$\times 10^{-2}$&-1.00$^{+0.03}_{-0.03}$&515$^{+44}_{-44}$&-3.23$^{+0.73}_{-0.74}$&(9.1$^{+1.2}_{-1.2}$)$\times 10^{-5}$&25$^{+1}_{-1}$&0.09$^{+0.02}_{-0.02}$&(3.9$^{+0.6}_{-0.4}$)$\times 10^{-6}$&-51\\
160821A&(n6)n7n9b1&[(-50 $\sim$ -10), (200 $\sim$ 220)]&(117.18$\sim$155.135)&516&Band+BB&(1.2$^{+0.0}_{-0.0}$)$\times$10$^{-1}$&-1.03$^{+0.01}_{-0.01}$&1288$^{+33}_{-33}$&-2.45$^{+0.03}_{-0.03}$&(2.5$^{+0.2}_{-0.2}$)$\times$10$^{-5}$&37$^{+1}_{-1}$&0.02$^{+0.00}_{-0.00}$&(2.7$^{+0.1}_{-0.1}$)$\times$10$^{-5}$&-263\\
171010A&(n8)nbb1&[(-50 $\sim$ -10), (200 $\sim$ 250)]&(7.14$\sim$106.20)&638&Band+BB&(2.3$^{+0.0}_{-0.0}$)$\times$10$^{-1}$&-1.03$^{+0.01}_{-0.01}$&119$^{+2}_{-2}$&-2.36$^{+0.03}_{-0.03}$&(3.5$^{+0.5}_{-0.5}$)$\times$10$^{-6}$&71$^{+3}_{-2}$&0.12$^{+0.03}_{-0.03}$&(7.5$^{+0.3}_{-0.3}$)$\times$10$^{-6}$&-67\\
171227A&(n5)b0&[(-50 $\sim$ -10), (60 $\sim$ 100)]&(0.26$\sim$30.27)&231&Band+BB&(5.9$^{+0.1}_{-0.1}$)$\times$10$^{-2}$&-0.75$^{+0.02}_{-0.02}$&1192$^{+52}_{-54}$&-2.81$^{+0.08}_{-0.07}$&(2.4$^{+0.3}_{-0.3}$)$\times$10$^{-5}$&40$^{+2}_{-2}$&0.04$^{+0.01}_{-0.01}$&(1.7$^{+0.1}_{-0.1}$)$\times$10$^{-5}$&-169\\
180120A&n9na(nb)b1&[(-40 $\sim$ -10), (60 $\sim$ 100)]&(0.09$\sim$24.10)&217&Band+BB&(1.4$^{+0.1}_{-0.1}$)$\times$10$^{-1}$&-0.91$^{+0.03}_{-0.03}$&92$^{+8}_{-8}$&-2.66$^{+0.17}_{-0.15}$&(5.0$^{+2.6}_{-2.4}$)$\times$10$^{-6}$&56$^{+5}_{-6}$&0.17$^{+0.14}_{-0.14}$&(2.7$^{+0.5}_{-0.4}$)$\times$10$^{-6}$&-291\\
190530A&(n0)n5b0&[(-30 $\sim$ -10), (50 $\sim$ 80)]&(7.24$\sim$34.10)&374&Band+BB&(1.3$^{+0.0}_{-0.0}$)$\times$10$^{-1}$&-0.96$^{+0.01}_{-0.01}$&971$^{+17}_{-17}$&-3.90$^{+0.22}_{-0.23}$&(1.9$^{+0.3}_{-0.3}$)$\times$10$^{-5}$&39$^{+2}_{-2}$&0.02$^{+0.01}_{-0.01}$&(2.0$^{+0.0}_{-0.0}$)$\times$10$^{-5}$&-106\\
\enddata 
\end{deluxetable}

\clearpage
\setlength{\tabcolsep}{0.05em}
\begin{deluxetable}{ccccccccccc}
\centering
\tabletypesize{\scriptsize}
\tablecaption{Spectral Fit Results of the PFD-type Bursts with Band-like Spectra.}\label{tab:PFD_Band}
\tablehead{
\specialrule{0em}{5pt}{5pt}  
\colhead{GRB}
&\colhead{Detectors}
&\colhead{$[\Delta T_{\rm (bkg,1)},\Delta T_{\rm (bkg,2)}]$}
&\colhead{$\Delta T_{\rm src}$}
&\colhead{$S$}
&\colhead{Spectral model}
&\colhead{$K$}
&\colhead{$\alpha$}
&\colhead{$E_{\rm p}$}
&\colhead{$\beta$}
&\colhead{$F_{\gamma}$}\\
\hline
&&[($t^{\rm bkg,1}_{1}$$\sim$$t^{\rm bkg,1}_{2}$), ($t^{\rm bkg,2}_{1}$$\sim$$t^{\rm bkg,2}_{2}$)]&($t_{1}$$\sim$$t_{2}$)&&(Preferred)&&&&&(Total flux)\\
&&(s)&(s)&&&
(ph.s$^{-1}$.cm$^{-2}$.keV$^{-1}$)&&(keV)&&(erg.cm$^{-2}$.s$^{-1}$)
}
\colnumbers
\startdata
160325A&(n6)n7n9b1&[(-50 $\sim$ -10), (200 $\sim$ 250)]&(2.28$\sim$46.10)&44&Band&(1.1$^{+0.1}_{-0.1}$)$\times$10$^{-2}$&-0.84$^{+0.05}_{-0.06}$&264$^{+28}_{-28}$&-2.12$^{+0.16}_{-0.15}$&(7.7$^{+2.0}_{-1.4}$)$\times$10$^{-7}$\\
170101B&(n9)nanbb1&[(-50 $\sim$ -10), (40 $\sim$ 80)]&(0$\sim$11.14)&50&Band&(3.5$^{+0.3}_{-0.3}$)$\times$10$^{-2}$&-0.46$^{+0.06}_{-0.07}$&206$^{+13}_{-13}$&-2.48$^{+0.21}_{-0.21}$&(1.2$^{+0.3}_{-0.2}$)$\times$10$^{-6}$\\
170114A&n1(n2)nab0&[(-50 $\sim$ -10), (50 $\sim$ 100)]&(0$\sim$10.48)&84&Band&(4.6$^{+0.3}_{-0.4}$)$\times$10$^{-2}$&-0.74$^{+0.05}_{-0.05}$&225$^{+18}_{-18}$&-2.04$^{+0.07}_{-0.07}$&(2.9$^{+0.5}_{-0.4}$)$\times$10$^{-6}$\\
170206A&n9na(nb)b1&[(-20 $\sim$ -5), (10 $\sim$ 30)]&(0$\sim$1.26)&80&Band&(1.8$^{+0.1}_{-0.1}$)$\times$10$^{-1}$&-0.20$^{+0.05}_{-0.05}$&313$^{+15}_{-15}$&-2.64$^{+0.15}_{-0.14}$&(1.0$^{+0.1}_{-0.1}$)$\times$10$^{-5}$\\
170207A&n1(n2)n5b0&[(-40 $\sim$ -10), (60 $\sim$ 100)]&(0$\sim$38.76)&92&Band&(2.1$^{+0.1}_{-0.1}$)$\times$10$^{-2}$&-0.91$^{+0.03}_{-0.03}$&468$^{+32}_{-31}$&-2.63$^{+0.29}_{-0.28}$&(1.9$^{+0.3}_{-0.2}$)$\times$10$^{-6}$\\
170210A&(n2)nab0&[(-50 $\sim$ -10), (150 $\sim$ 250)]&(0$\sim$47.63)&78&Band&(1.9$^{+0.1}_{-0.1}$)$\times$10$^{-2}$&-0.94$^{+0.03}_{-0.04}$&416$^{+44}_{-42}$&-2.13$^{+0.11}_{-0.12}$&(2.0$^{+0.3}_{-0.3}$)$\times$10$^{-6}$\\
180806A&n7(n8)b1&[(-50 $\sim$ -10, 50 $\sim$ 100)]&(-0.01$\sim$10.32)&67&Band&(3.2$^{+0.2}_{-0.1}$)$\times$10$^{-2}$&-0.89$^{+0.04}_{-0.03}$&421$^{+39}_{-42}$&-2.42$^{+0.18}_{-0.17}$&(2.9$^{+0.4}_{-0.4}$)$\times$10$^{-6}$\\
180914A&n6(n8)b1&[(-100 $\sim$ -10), (180 $\sim$ 280)]&(5.34$\sim$133.35)&102&Band&(1.9$^{+0.1}_{-0.1}$)$\times$10$^{-2}$&-0.63$^{+0.03}_{-0.03}$&261$^{+13}_{-13}$&-2.35$^{+0.13}_{-0.13}$&(1.0$^{+0.1}_{-0.1}$)$\times$10$^{-6}$\\
200412A&n9(na)b1&[(-50 $\sim$ -10), (50 $\sim$ 100)]&(-2.85$\sim$35.18)&74&Band&(2.6$^{+0.2}_{-0.2}$)$\times$10$^{-2}$&-0.68$^{+0.04}_{-0.04}$&233$^{+13}_{-14}$&-2.55$^{+0.23}_{-0.22}$&(1.1$^{+0.2}_{-0.1}$)$\times$10$^{-6}$\\
\enddata 
\end{deluxetable}

\clearpage
\setlength{\tabcolsep}{0.05em}
\begin{deluxetable}{cccccccccc}
\centering
\tabletypesize{\scriptsize}
\tablecaption{Spectral Fit Results of the PFD-type Bursts with CPL-like Spectra.}\label{tab:PFD_CPL}
\tablehead{
\specialrule{0em}{5pt}{5pt}  
\colhead{GRB}
&\colhead{Detectors}
&\colhead{$[\Delta T_{\rm (bkg,1)},\Delta T_{\rm (bkg,2)}]$}
&\colhead{$\Delta T_{\rm src}$}
&\colhead{$S$}
&\colhead{Spectral model}
&\colhead{$K$}
&\colhead{$\alpha$}
&\colhead{$E_{\rm p}$}
&\colhead{$F_{\gamma}$}\\
\hline
&&[($t^{\rm bkg,1}_{1}$$\sim$$t^{\rm bkg,1}_{2}$), ($t^{\rm bkg,2}_{1}$$\sim$$t^{\rm bkg,2}_{2}$)]&($t_{1}$$\sim$$t_{2}$)&&(Preferred)&&&&(Total Flux)\\
&&(s)&(s)&&&
(ph.s$^{-1}$.cm$^{-2}$.keV$^{-1}$)&&(keV)&(erg.cm$^{-2}$.s$^{-1}$)
}
\colnumbers
\startdata
161218B&n3(n4)n8b0&[(-100 $\sim$ -50), (200 $\sim$ 250)]&(0$\sim$25.1)&223&CPL&1.5$^{+0.1}_{-0.1}$&-0.56$^{+0.02}_{-0.02}$&223$^{+6}_{-6}$&(3.1$^{+0.3}_{-0.3}$)$\times$10$^{-6}$\\
161229A&n9na(nb)b1&[(-50 $\sim$ -10), (60 $\sim$ 100)]&(0$\sim$31.26)&65&CPL&(3.6$^{+0.6}_{-0.5}$)$\times$10$^{-1}$&-0.62$^{+0.05}_{-0.05}$&335$^{+33}_{-30}$&(1.0$^{+0.4}_{-0.3}$)$\times$10$^{-6}$\\
170127C&(n4)b0&[(-40 $\sim$ -10), (20 $\sim$ 50)]&(0$\sim$0.14)&38&CPL&(1.2$^{+1.3}_{-0.6}$)$\times$10$^{-1}$&0.11$^{+0.15}_{-0.15}$&889$^{+124}_{-117}$&(6.7$^{+17.0}_{-4.7}$)$\times$10$^{-5}$\\
170305A&n0n1(n2)b0&[(-50 $\sim$ -10), (10 $\sim$ 50)]&(0$\sim$0.45)&27&CPL&(6.6$^{+1.6}_{-2.9}$)$\times$10$^{-1}$&-0.42$^{+0.17}_{-0.06}$&234$^{+34}_{-51}$&(2.4$^{+2.7}_{-1.1}$)$\times$10$^{-6}$\\
170527A&n0n3(n4)b0&[(-50 $\sim$ -10), (100 $\sim$ 150)]&(0$\sim$37.18)&139&CPL&2.5$^{+0.1}_{-0.1}$&-1.01$^{+0.01}_{-0.01}$&1034$^{+88}_{-84}$&(3.8$^{+0.6}_{-0.5}$)$\times$10$^{-6}$\\
180427A&(n4)b0&[(-30 $\sim$ -10), (60 $\sim$ 80)]&(0.15$\sim$13.16)&139&CPL&(3.2$^{+0.6}_{-0.5}$)$\times$10$^{-1}$&0.06$^{+0.05}_{-0.05}$&133$^{+6}_{-6}$&(2.8$^{+1.0}_{-0.7}$)$\times$10$^{-6}$\\
200311A&n4(n8)b1&[(-50 $\sim$ -10), (50 $\sim$ 100)]&(0.14$\sim$39.20)&57&CPL&(7.7$^{+1.3}_{-1.1}$)$\times$10$^{-1}$&-0.90$^{+0.04}_{-0.04}$&1228$^{+176}_{-155}$&(2.6$^{+1.0}_{-0.8}$)$\times$10$^{-6}$\\
\enddata 
\vspace{3mm}
\tablecomments{The peak energy $E_{\rm p}$ here is obtained by the equation $E_{\rm p}$=$E_{\rm c}$(2+$\alpha$).}
\end{deluxetable}

\clearpage
\begin{figure*}
\includegraphics[width=0.5\hsize,clip]{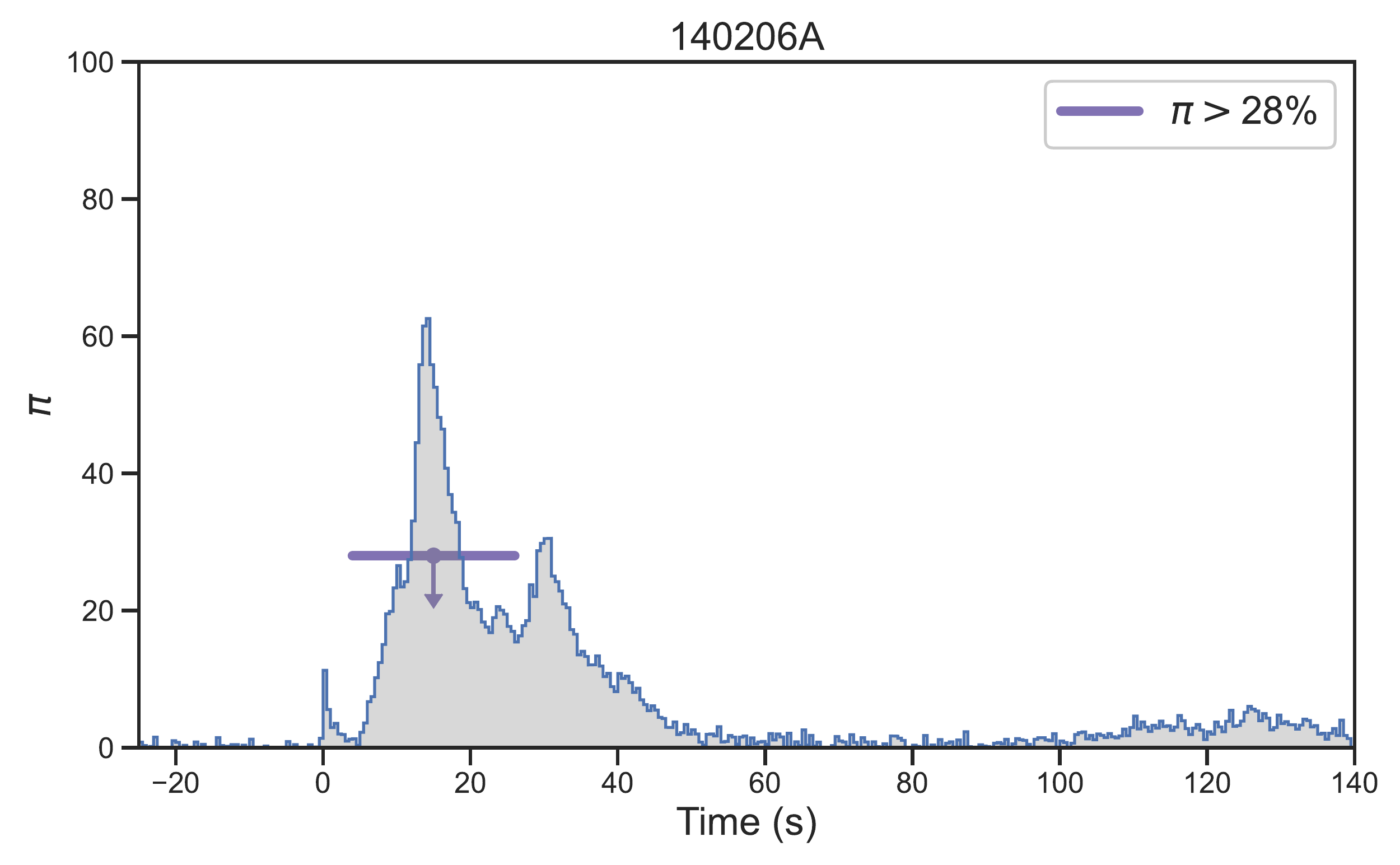}
\includegraphics[width=0.5\hsize,clip]{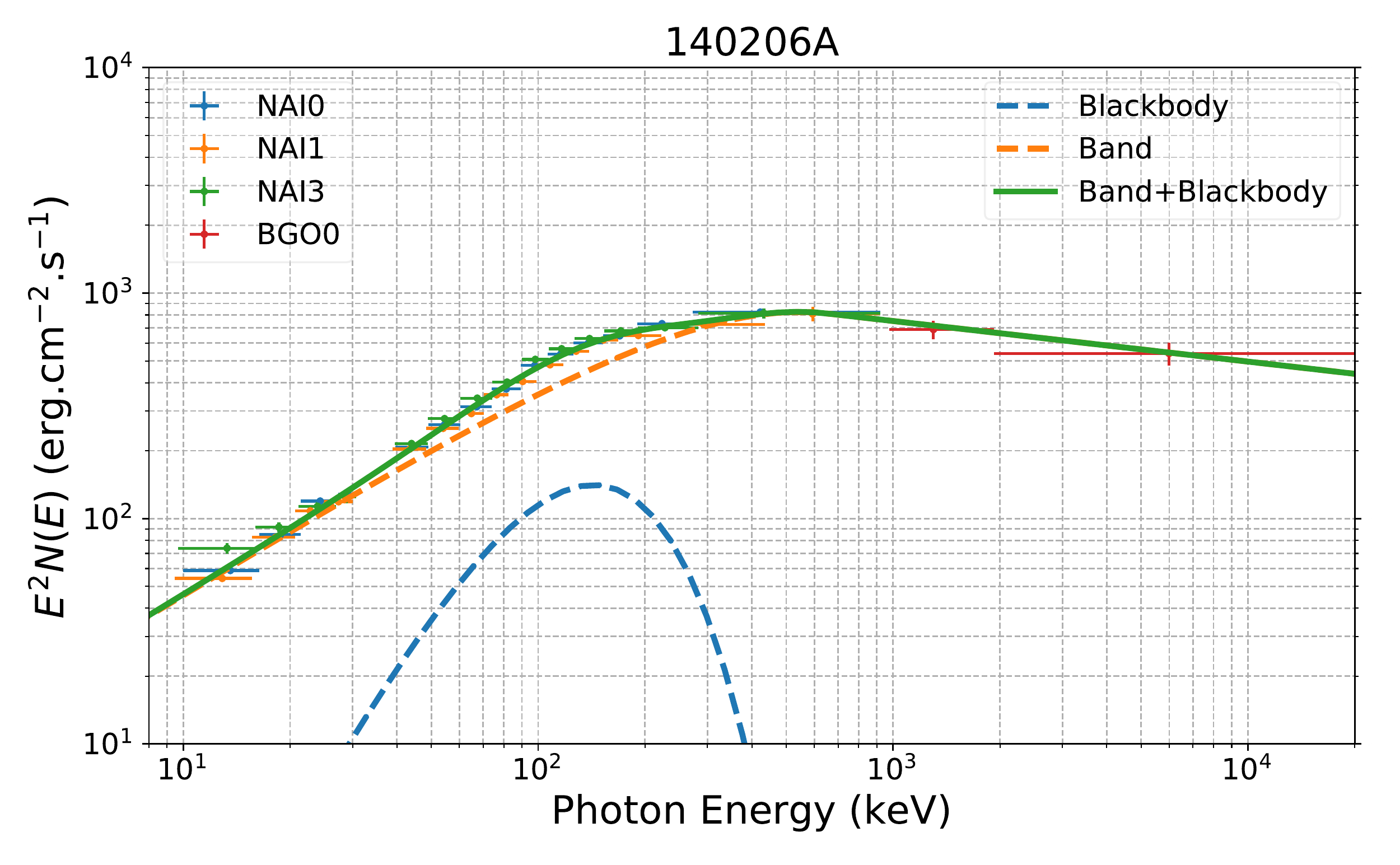}
\includegraphics[width=0.5\hsize,clip]{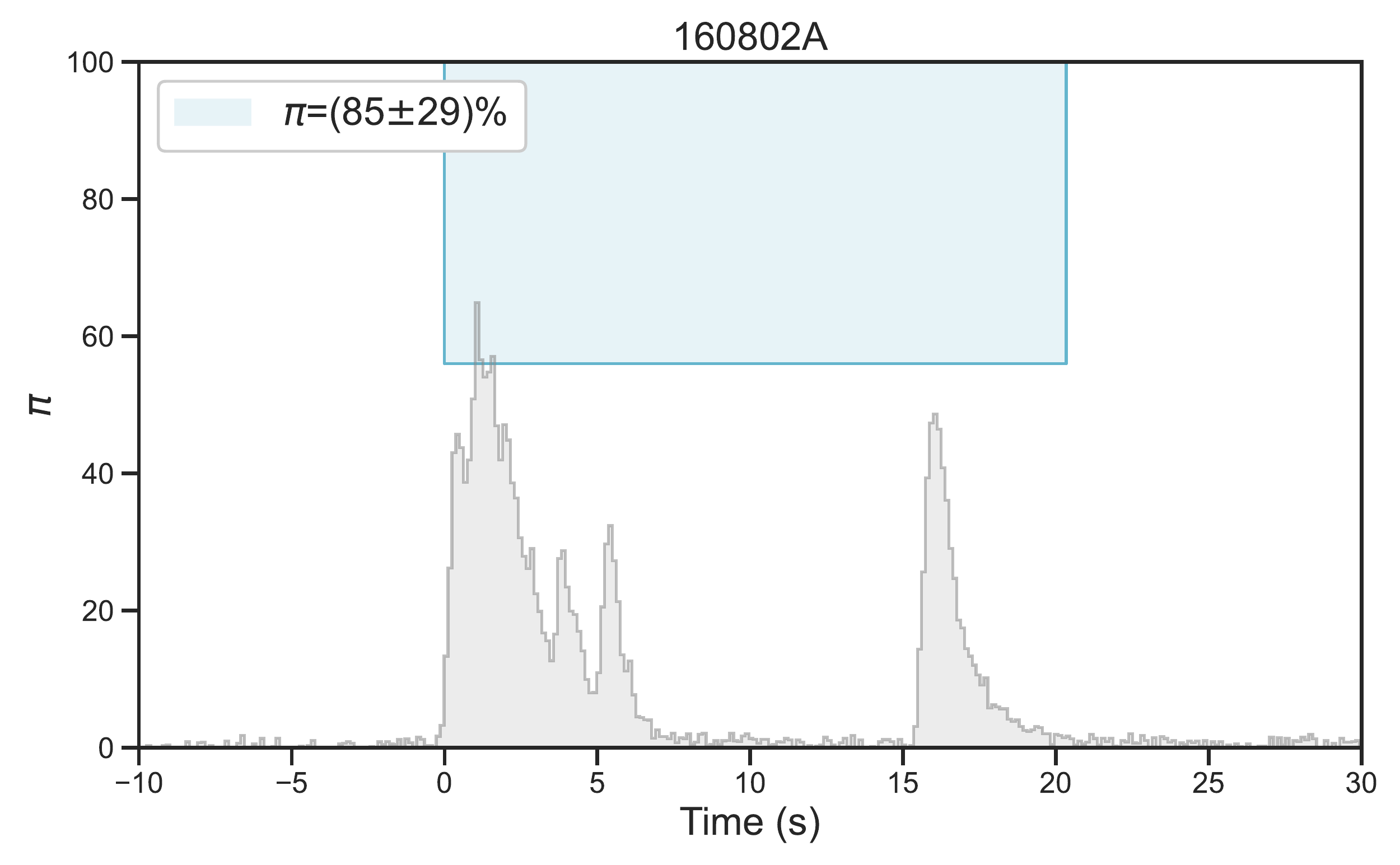}
\includegraphics[width=0.5\hsize,clip]{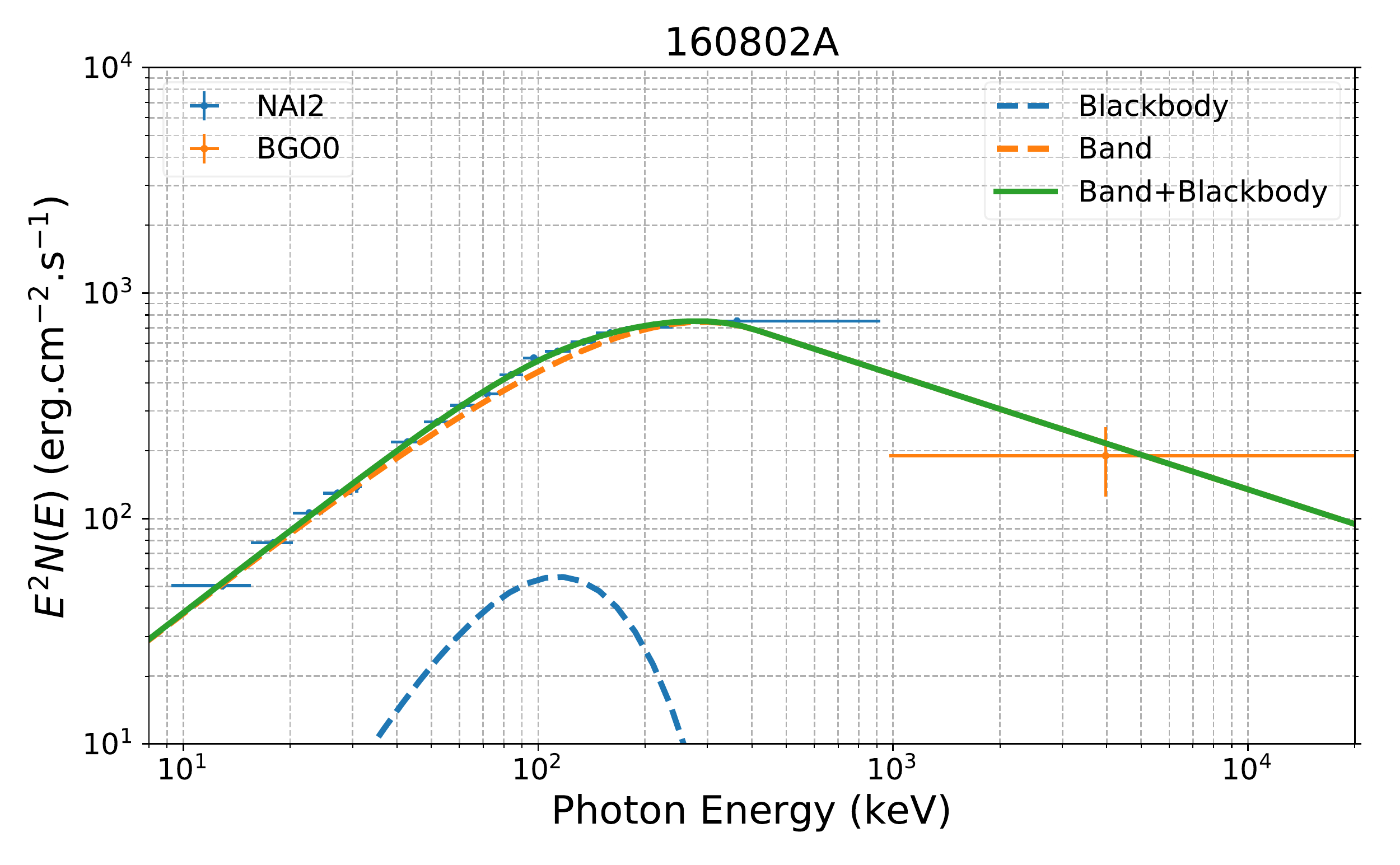}
\includegraphics[width=0.5\hsize,clip]{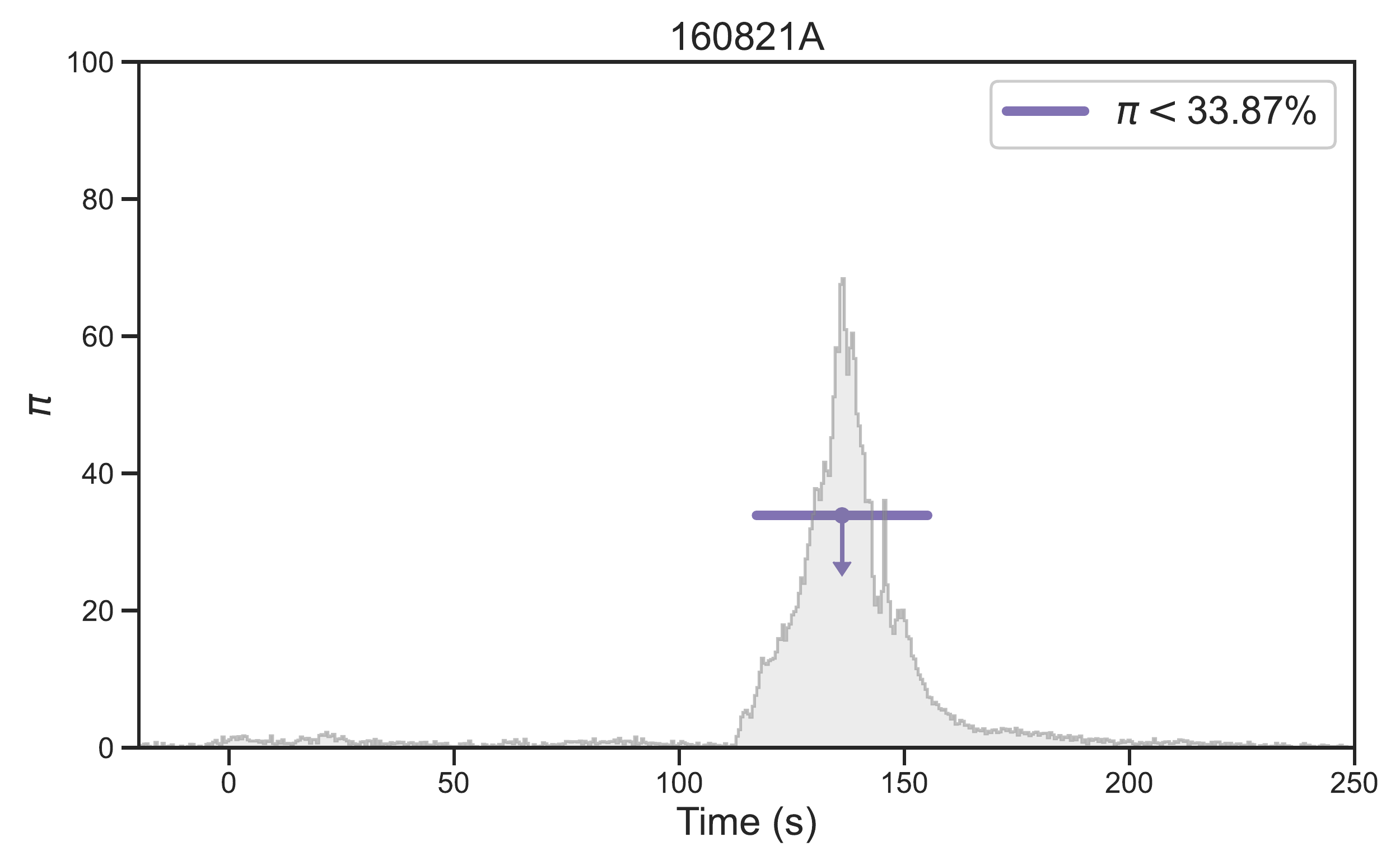}
\includegraphics[width=0.5\hsize,clip]{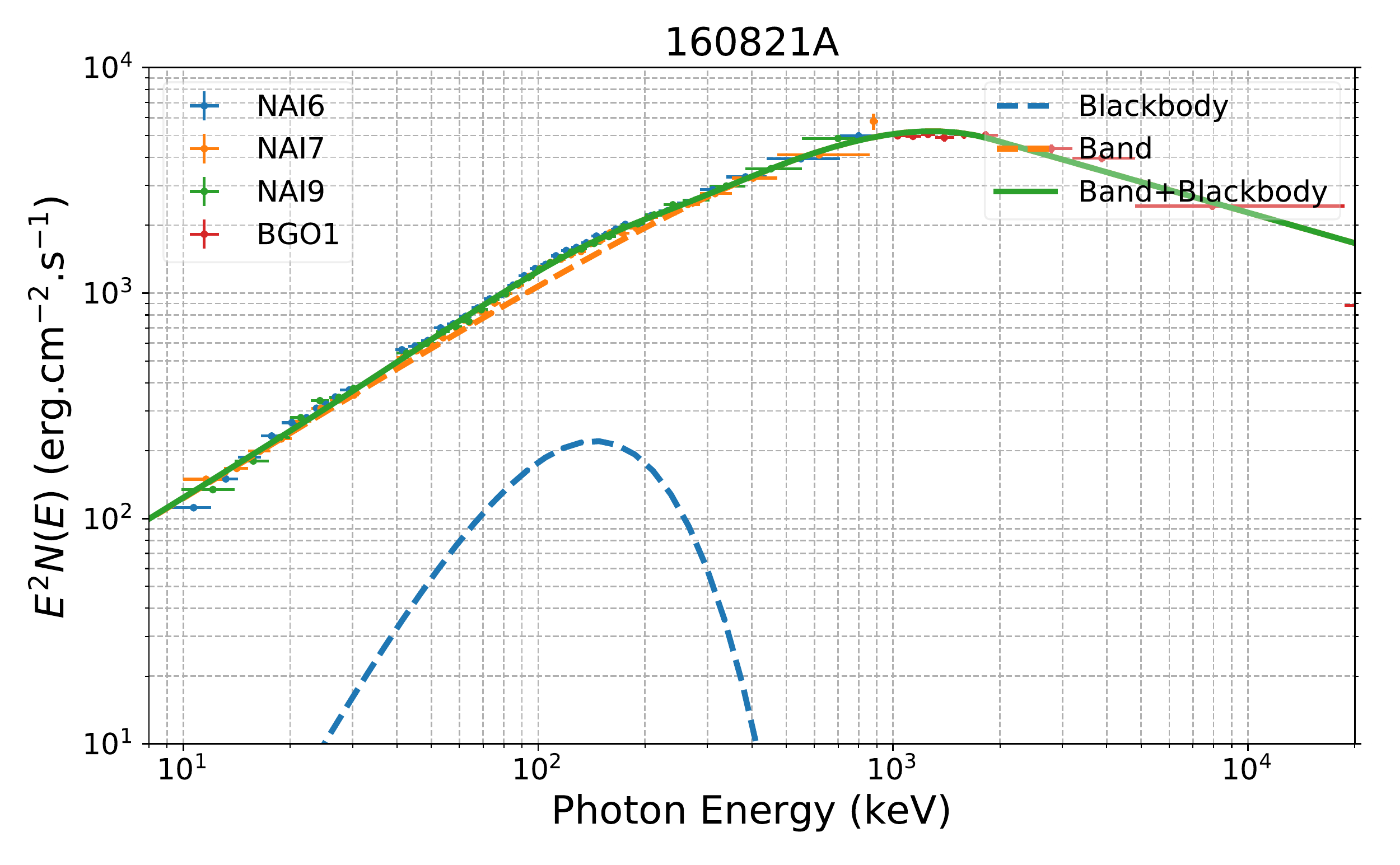}
\includegraphics[width=0.5\hsize,clip]{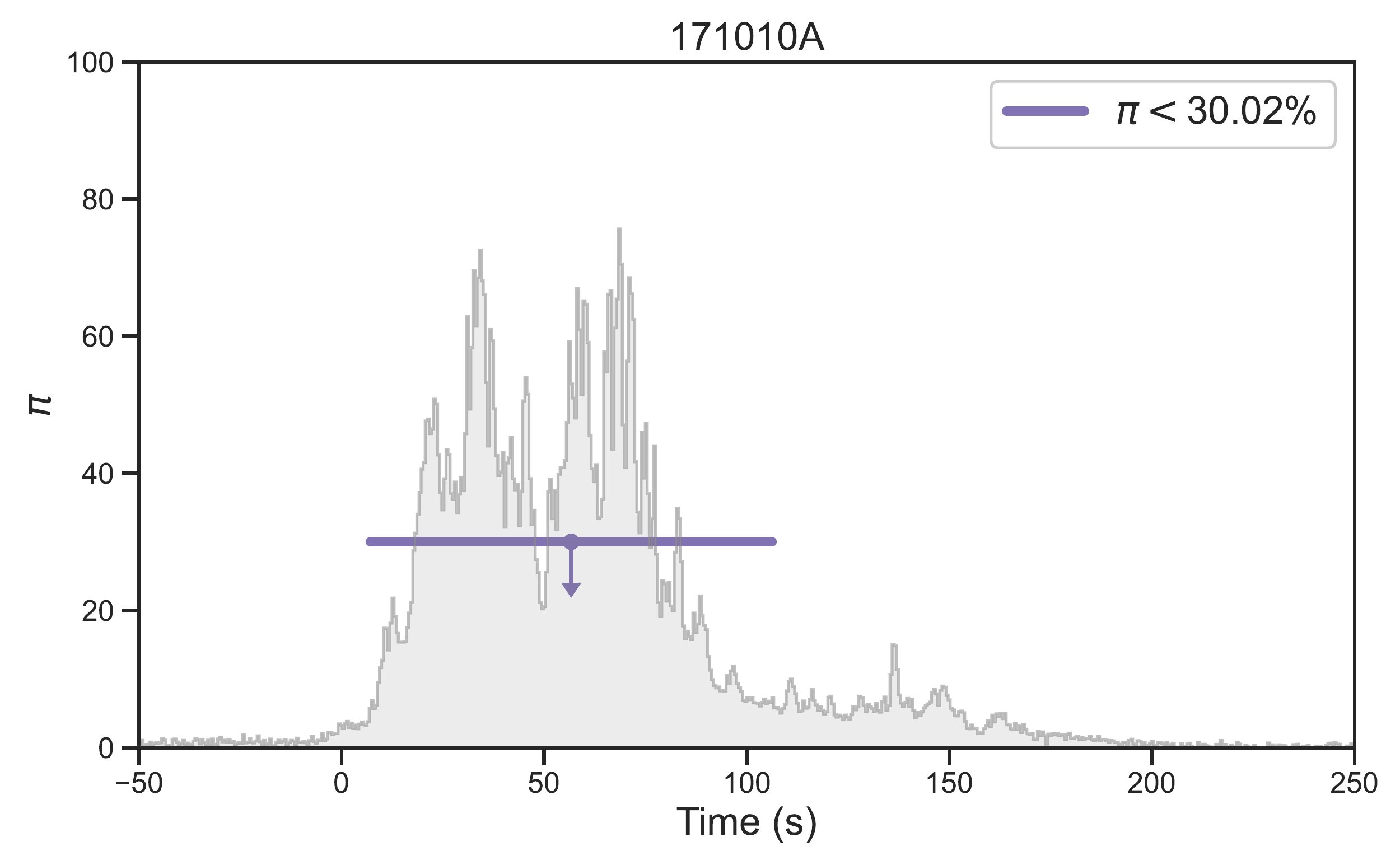}
\includegraphics[width=0.5\hsize,clip]{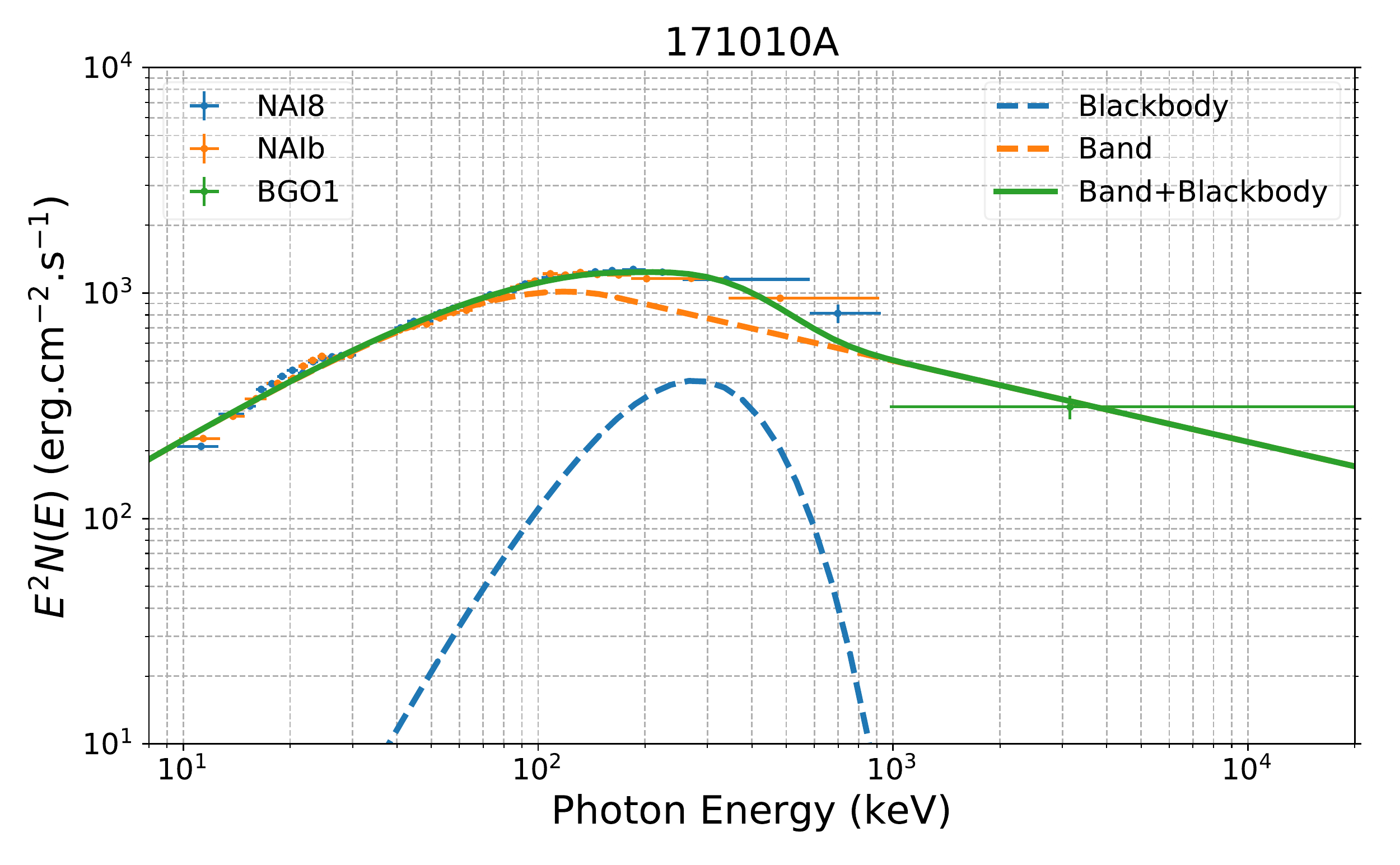}
\center{Fig. \ref{fig:Spectrum_LC_HD}--- Continued}
\end{figure*}
\begin{figure*}
\includegraphics[width=0.5\hsize,clip]{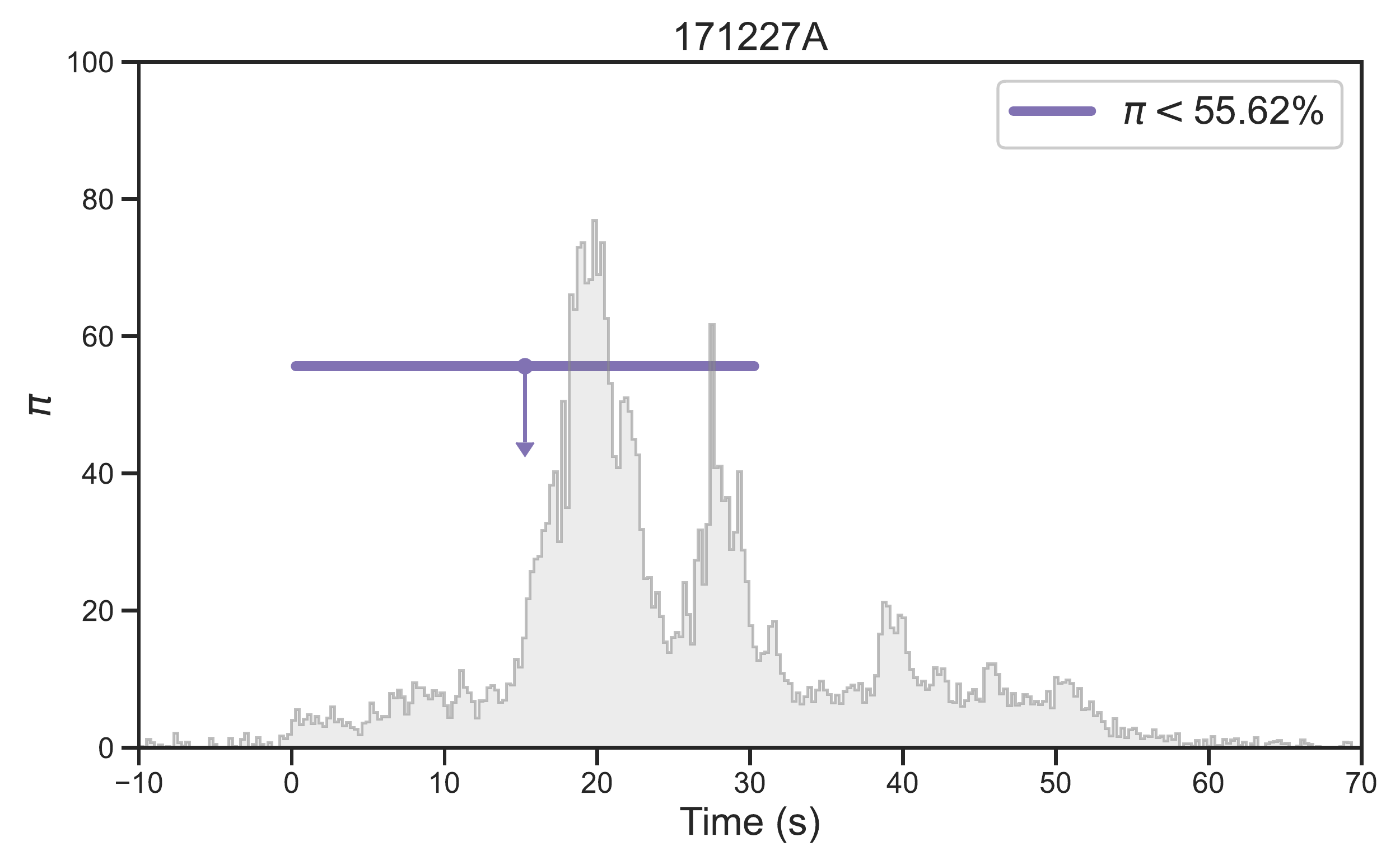}
\includegraphics[width=0.5\hsize,clip]{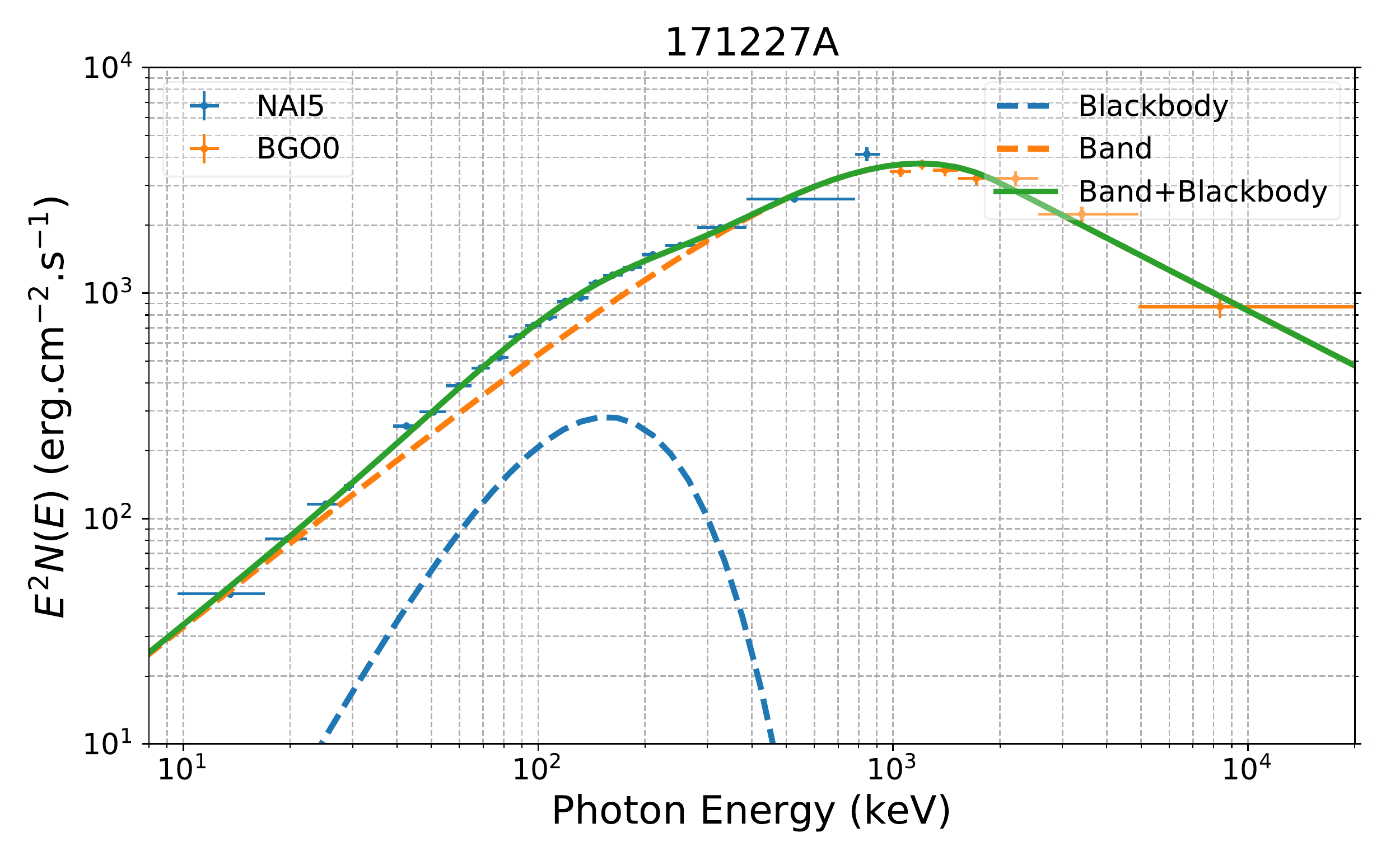}
\includegraphics[width=0.5\hsize,clip]{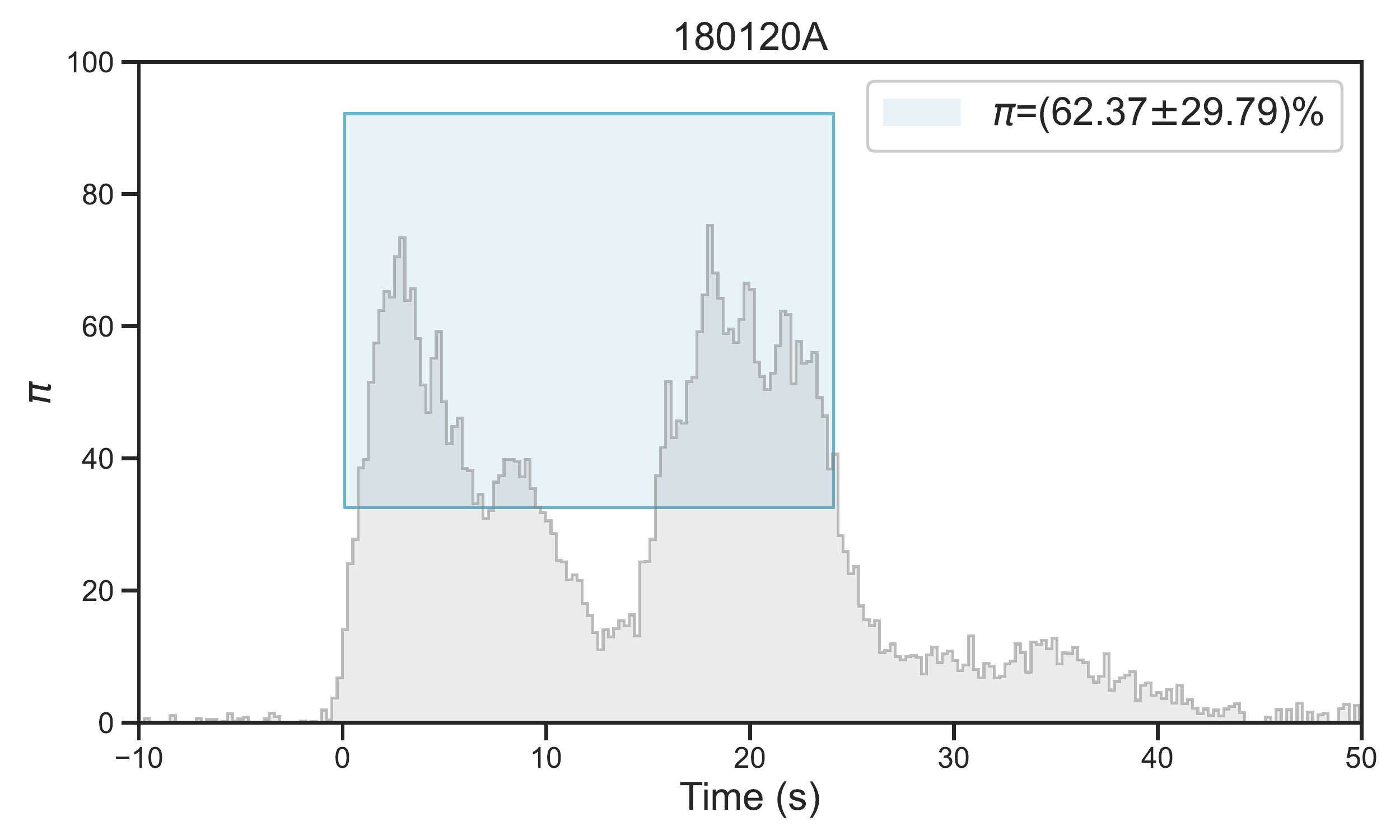}
\includegraphics[width=0.5\hsize,clip]{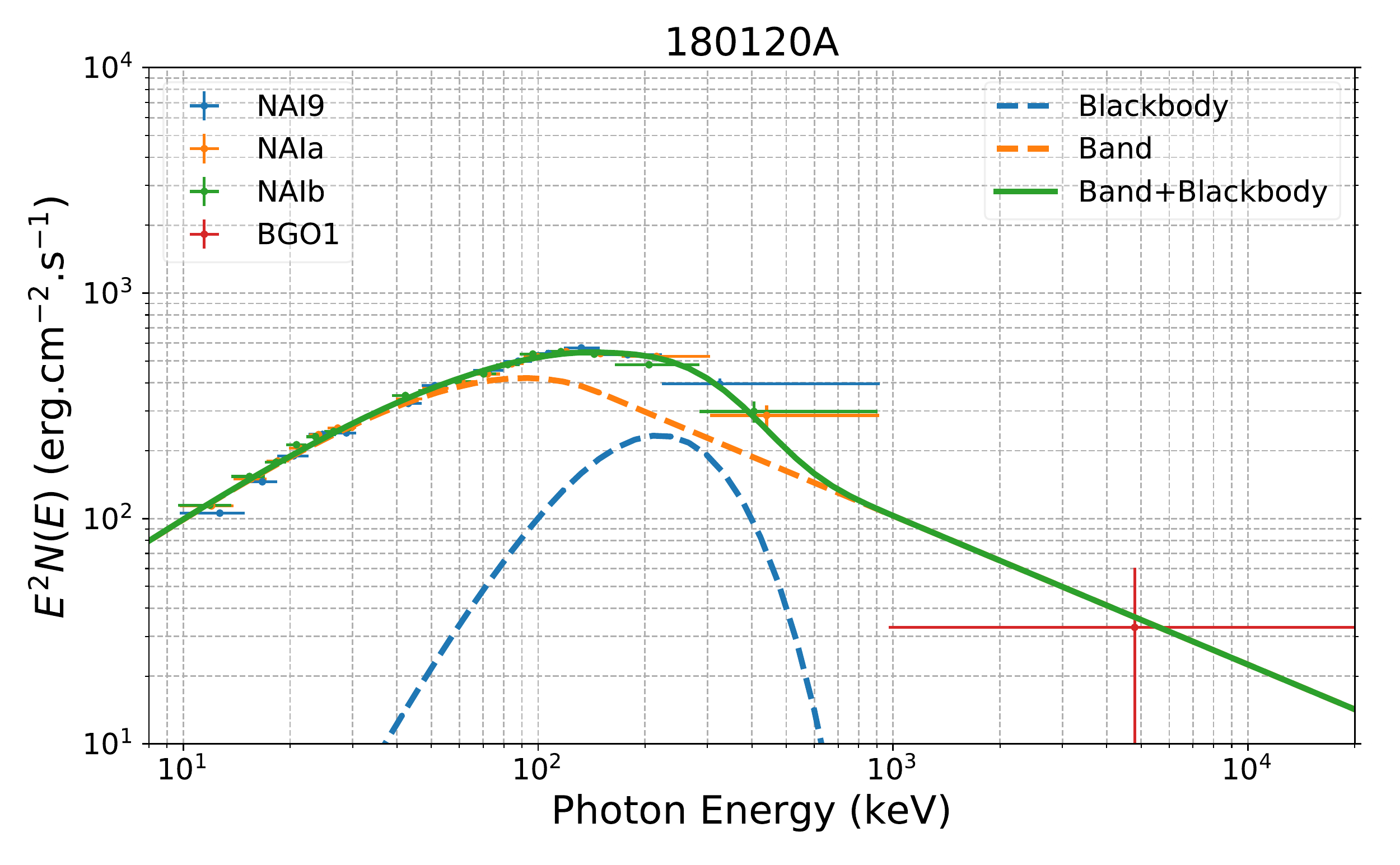}
\includegraphics[width=0.5\hsize,clip]{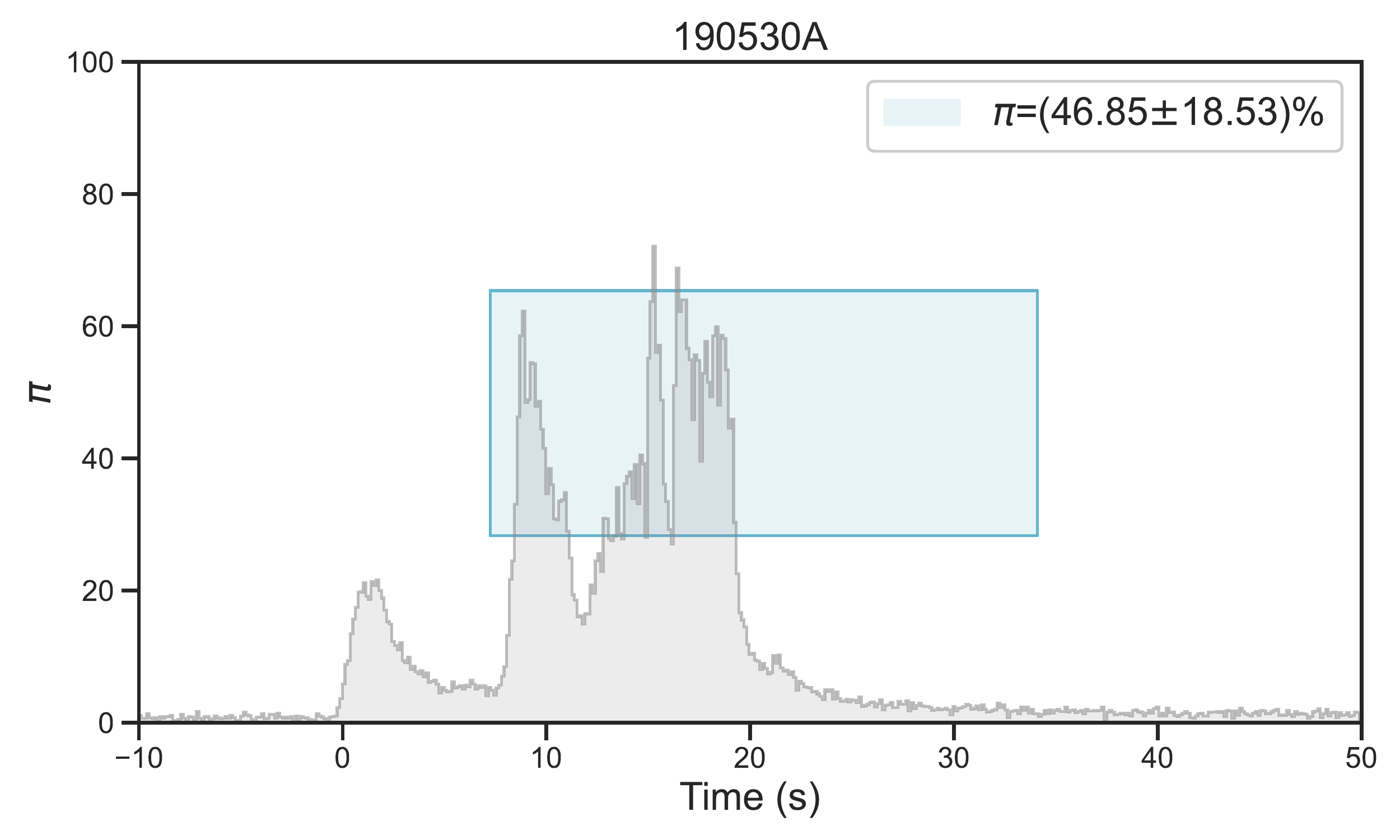}
\includegraphics[width=0.5\hsize,clip]{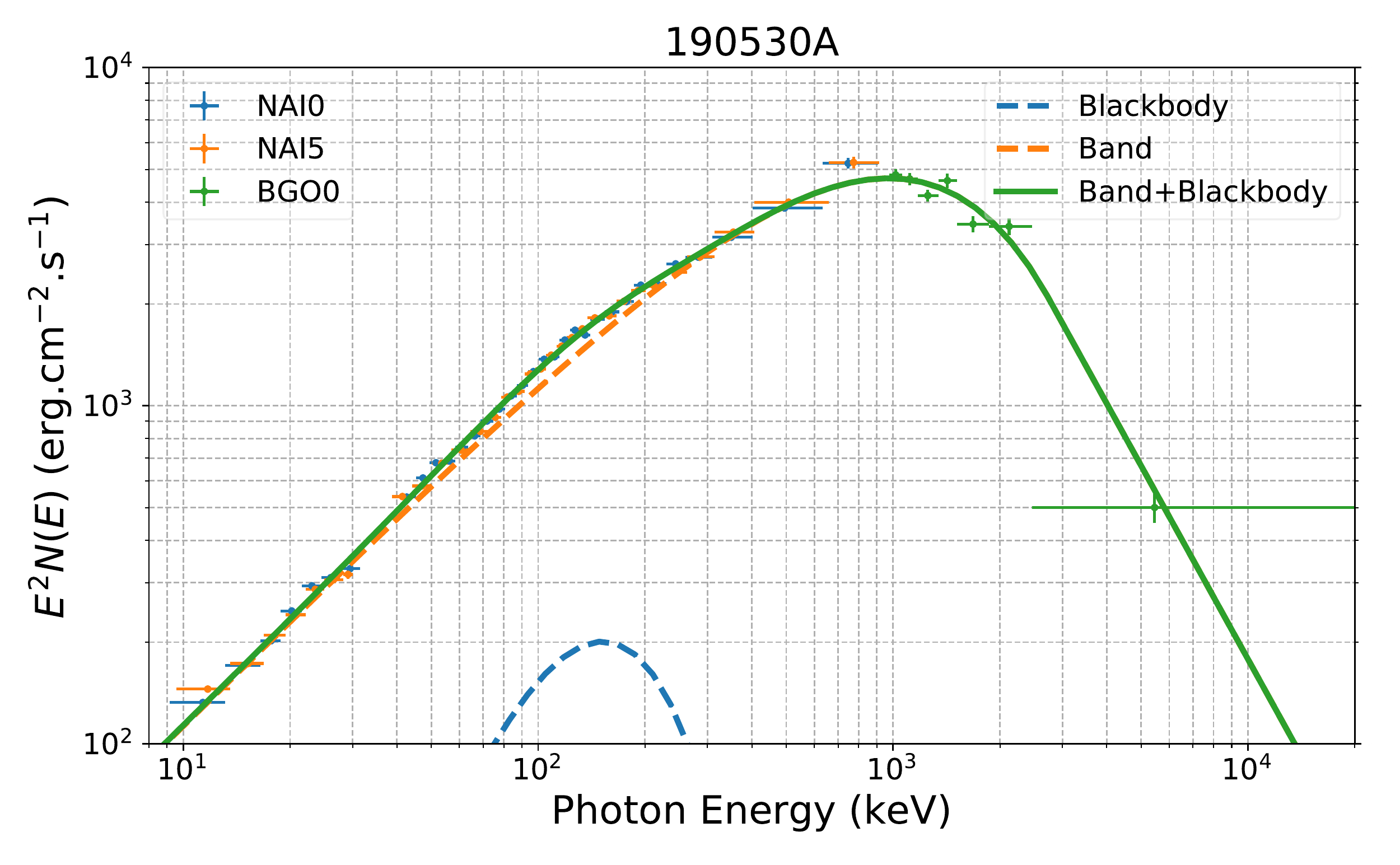}
\center{Fig. \ref{fig:Spectrum_LC_HD}--- Continued}
\end{figure*}

\clearpage
\begin{figure*}
\includegraphics[width=0.5\hsize,clip]{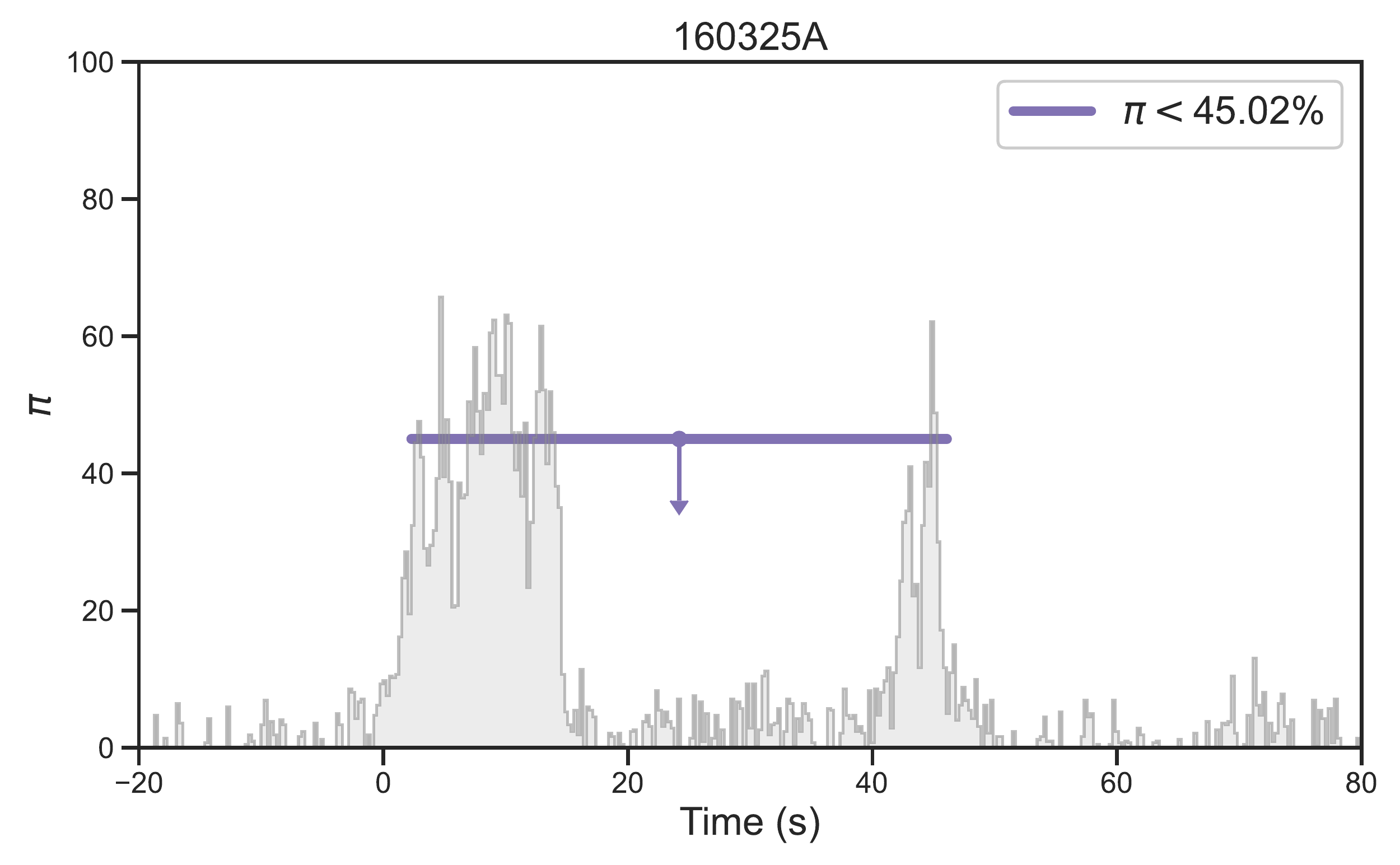}
\includegraphics[width=0.5\hsize,clip]{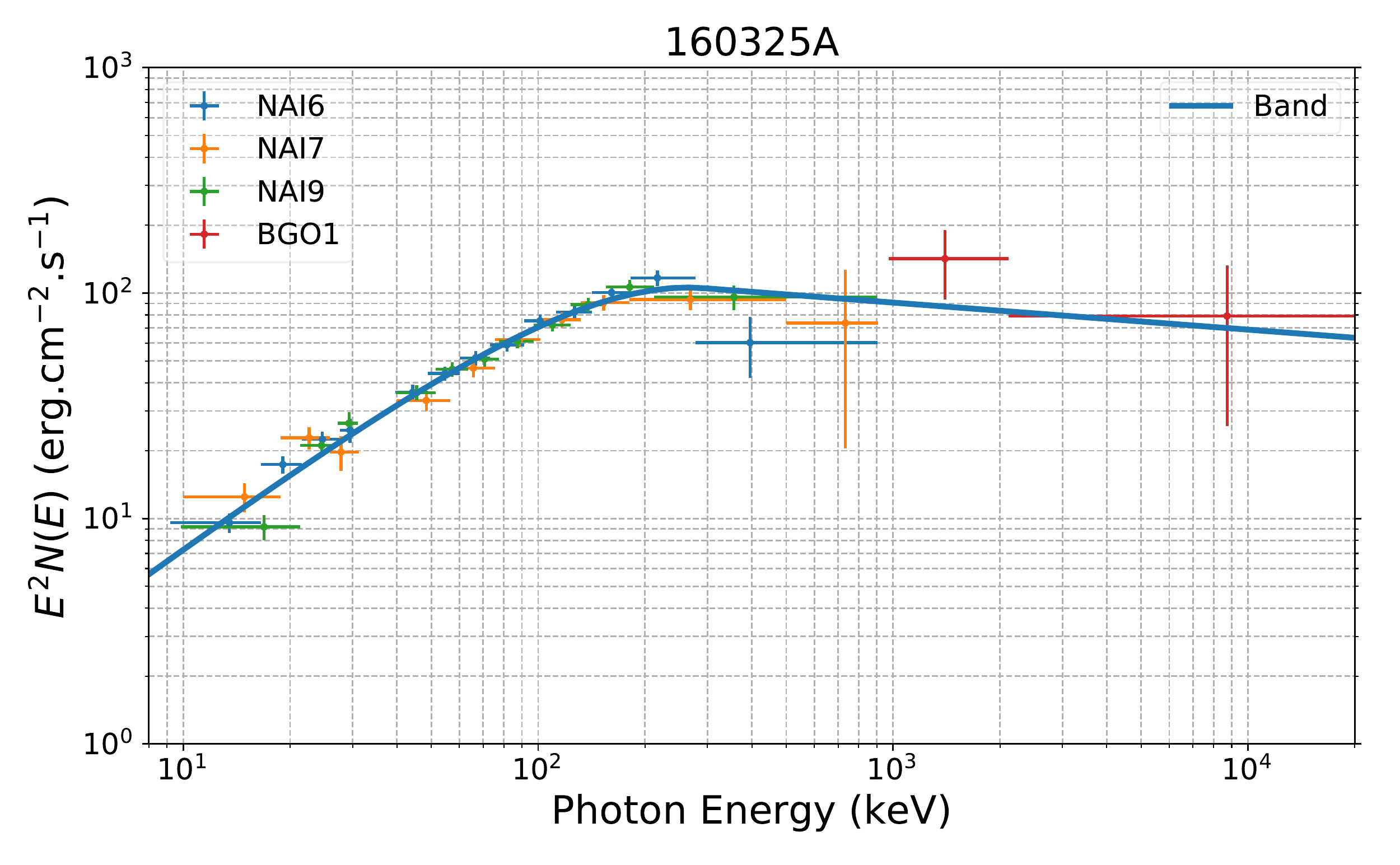}
\includegraphics[width=0.5\hsize,clip]{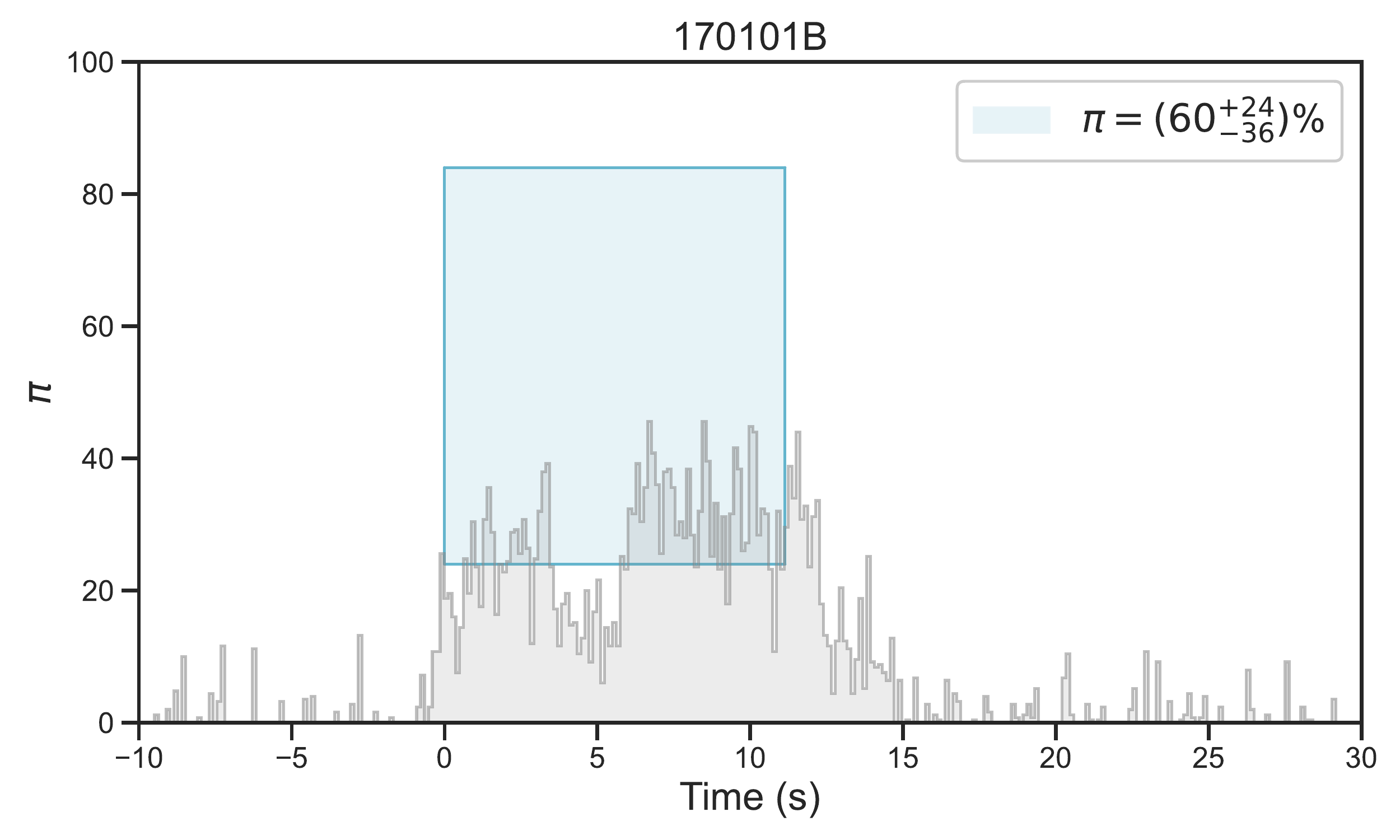}
\includegraphics[width=0.5\hsize,clip]{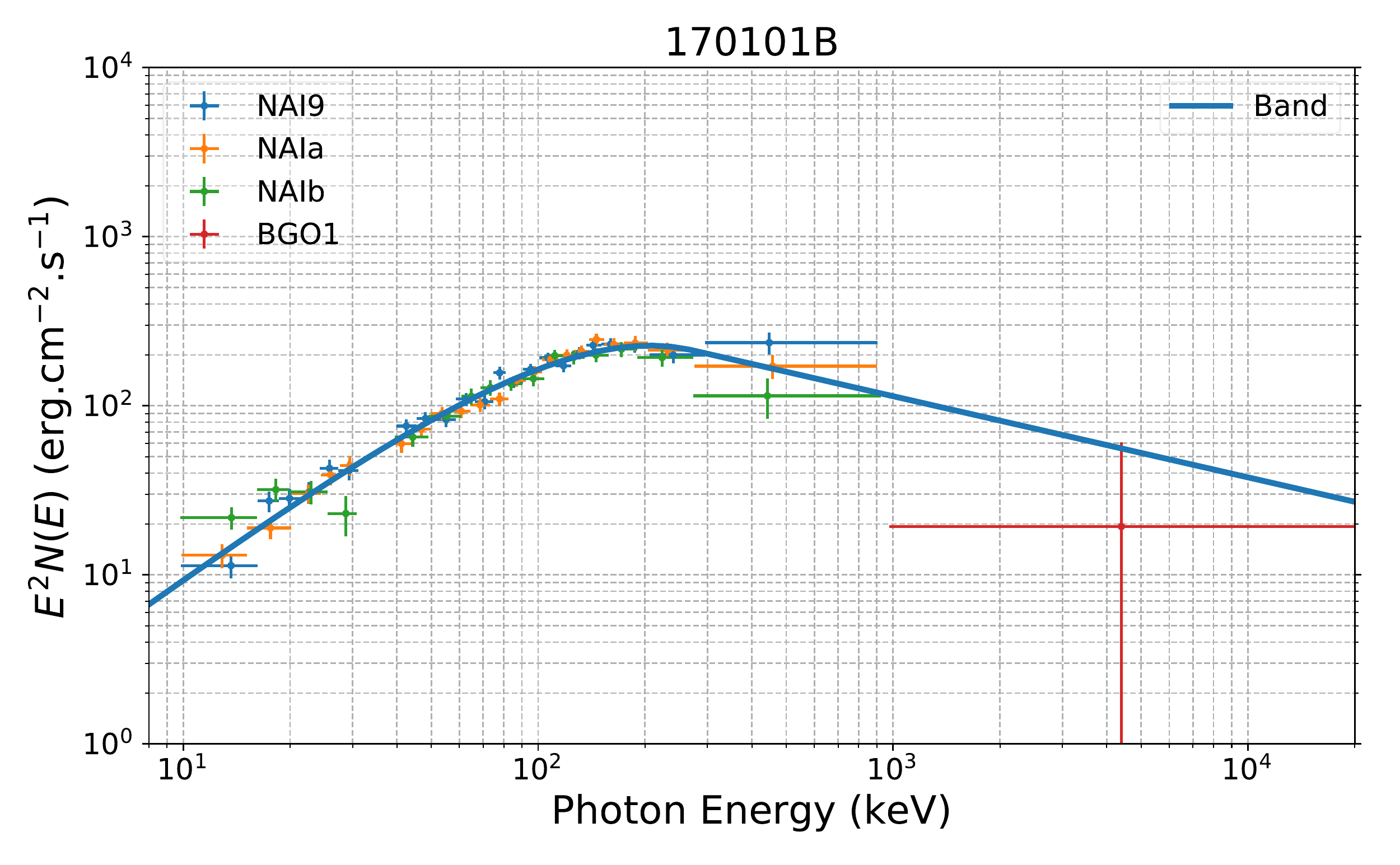}
\includegraphics[width=0.5\hsize,clip]{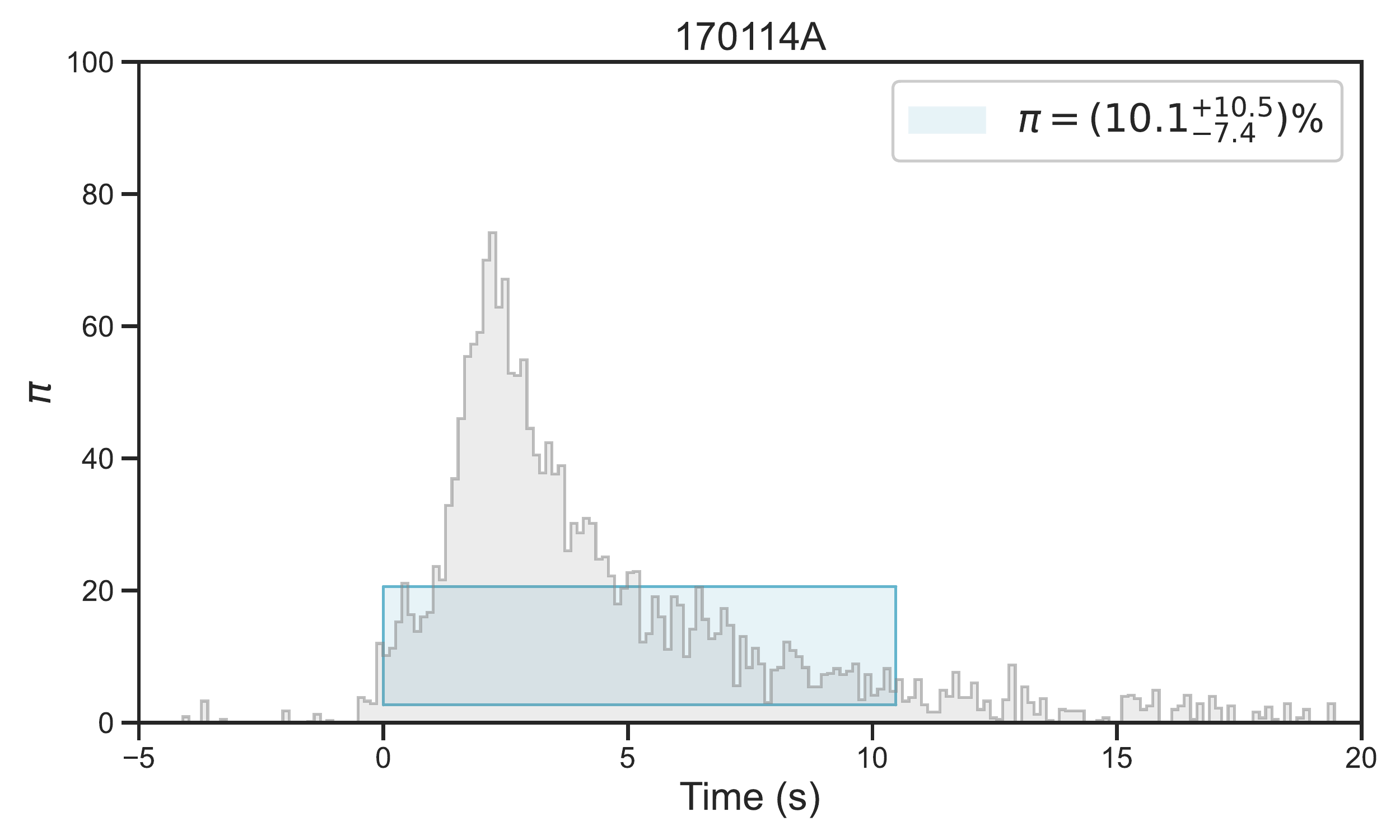}
\includegraphics[width=0.5\hsize,clip]{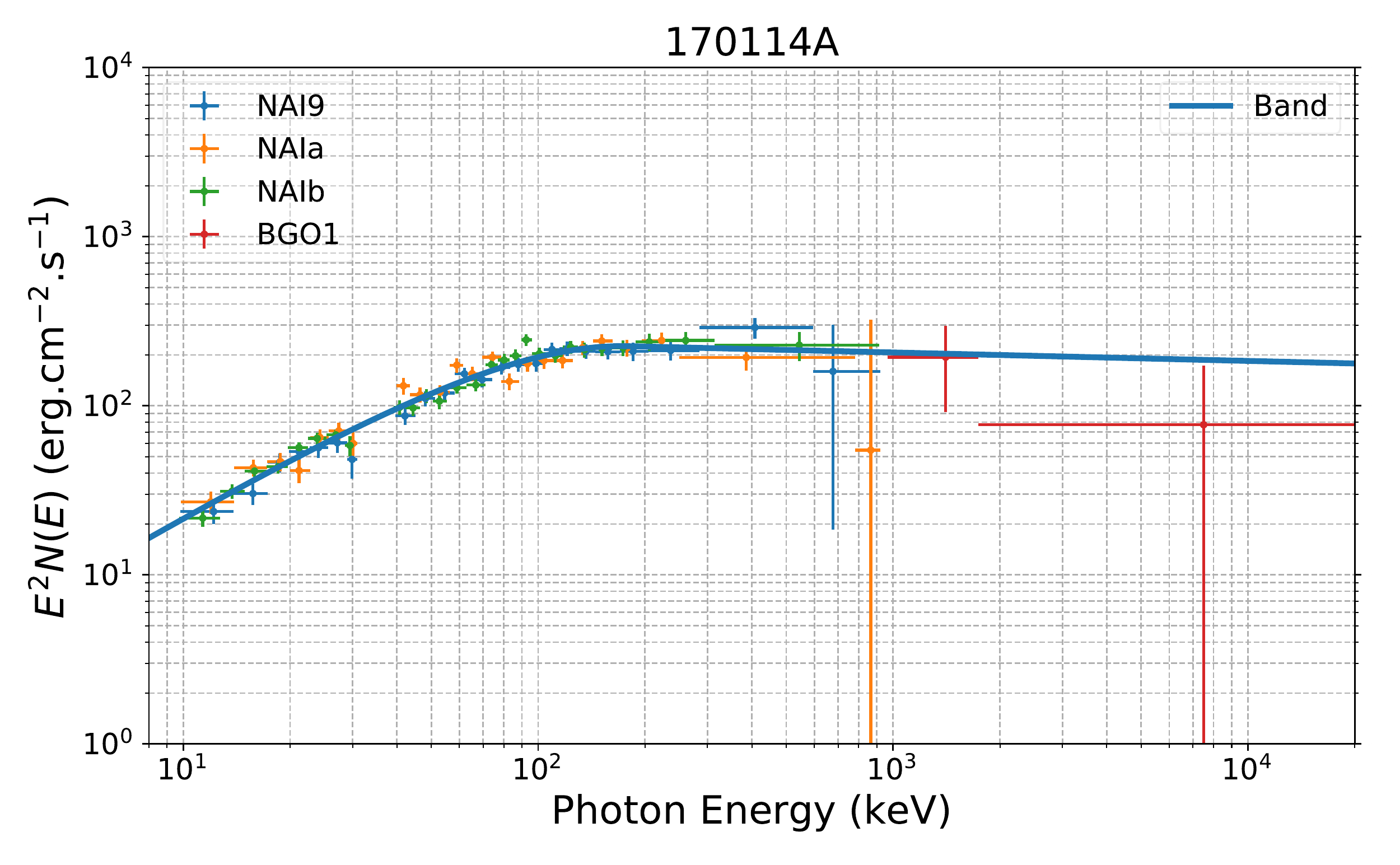}
\caption{Same as Figure \ref{fig:Spectrum_LC_HD}, but for bursts classified as PFD-type with Band-like spectra.}\label{fig:Spectrum_LC_PFD_Band}
\end{figure*}
\begin{figure*}
\includegraphics[width=0.5\hsize,clip]{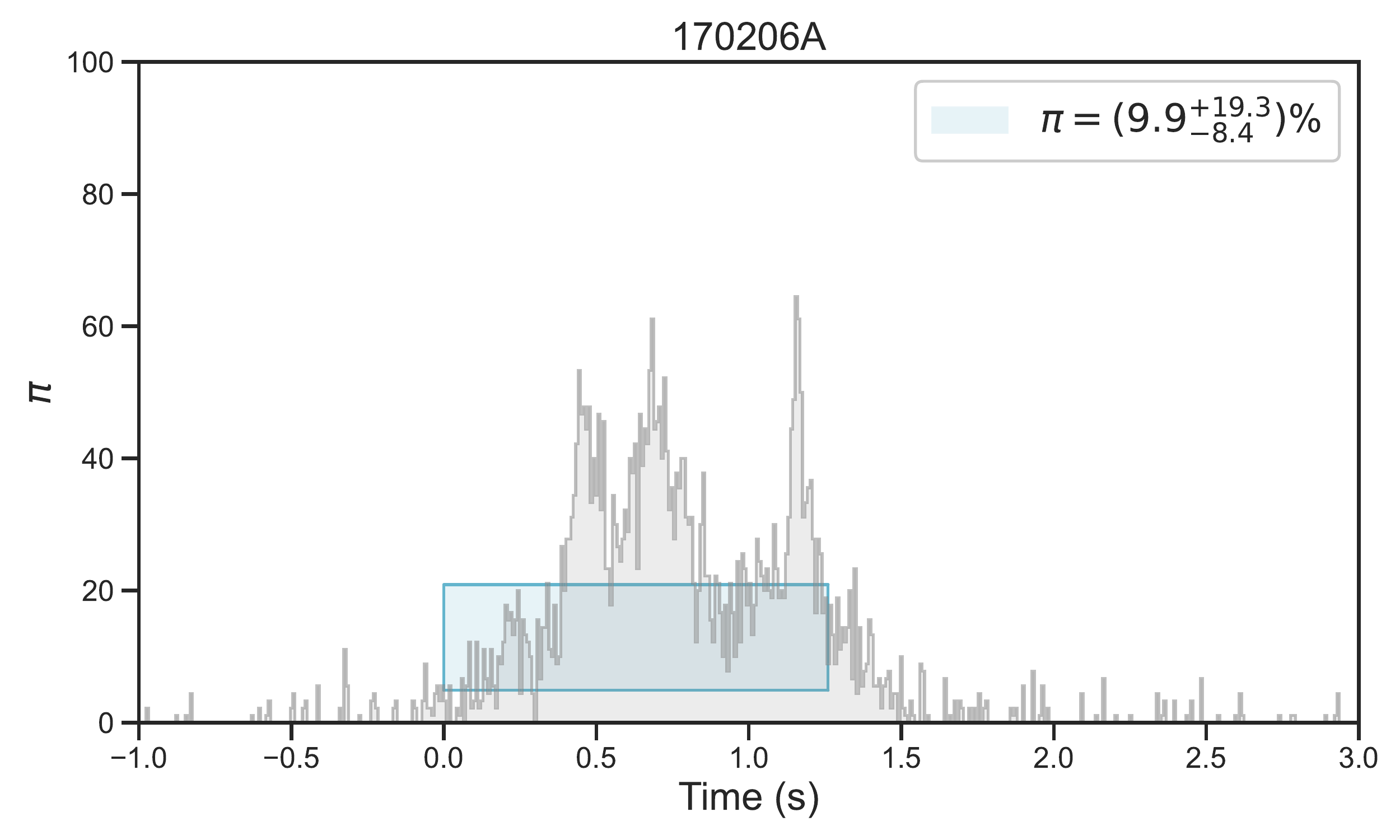}
\includegraphics[width=0.5\hsize,clip]{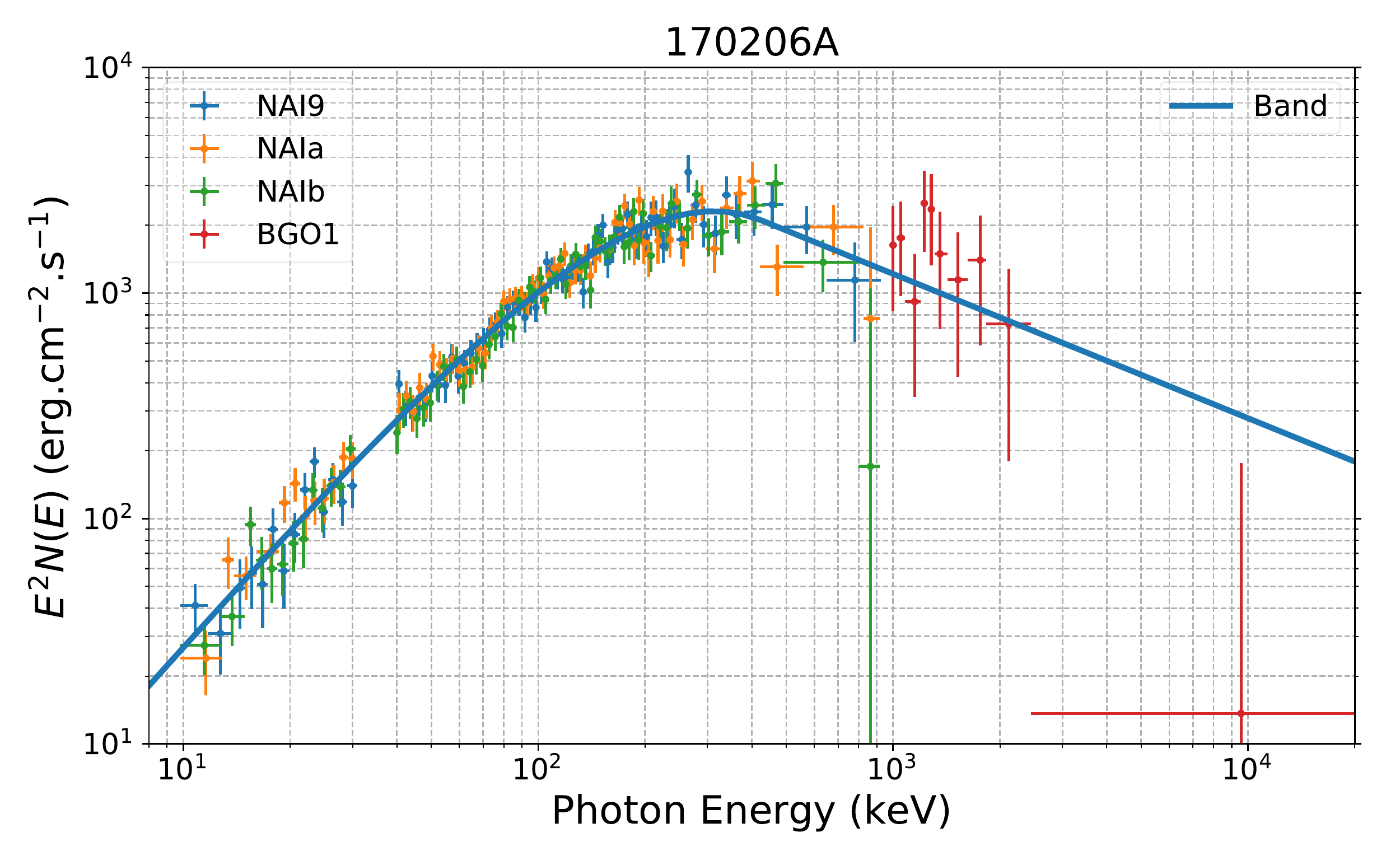}
\includegraphics[width=0.5\hsize,clip]{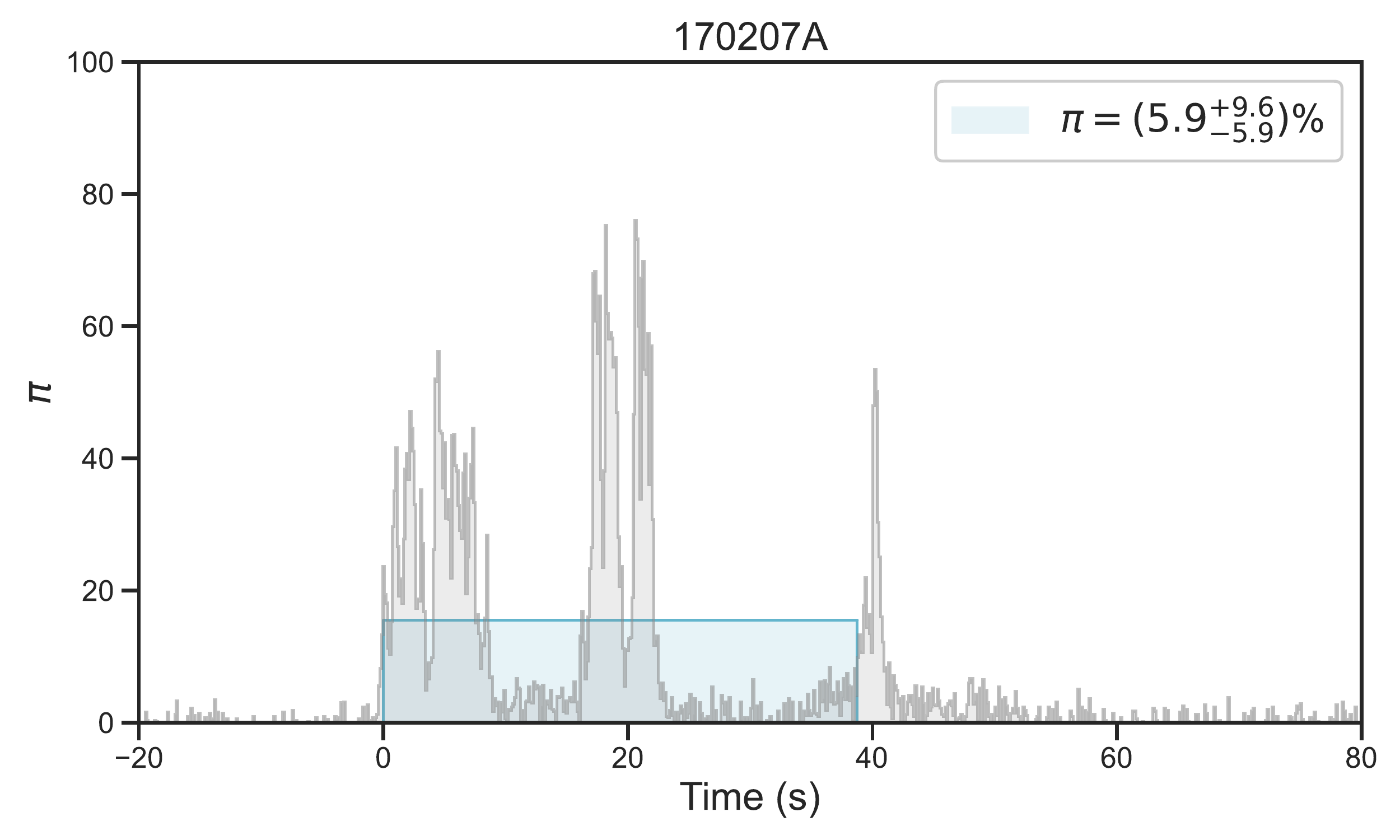}
\includegraphics[width=0.5\hsize,clip]{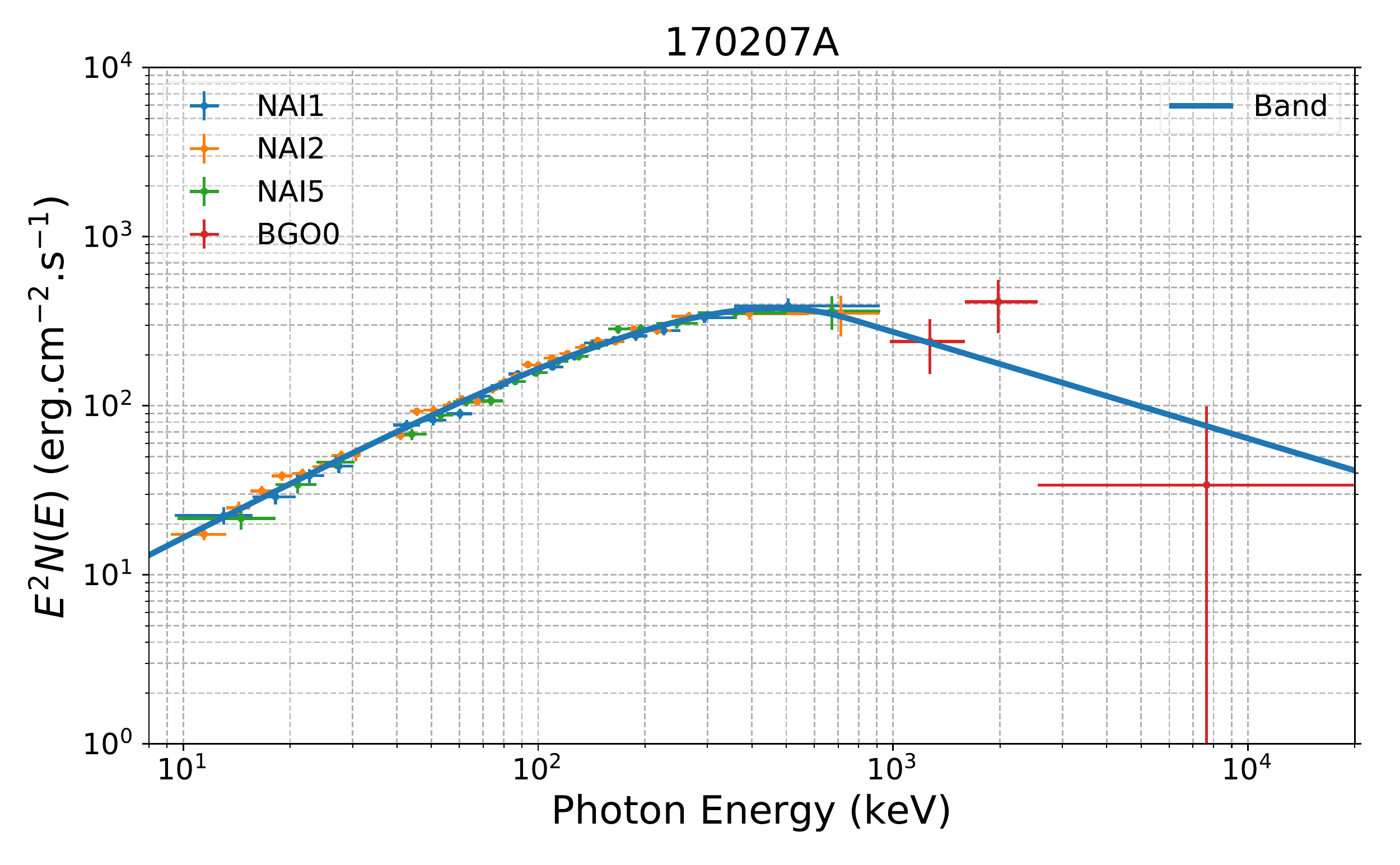}
\includegraphics[width=0.5\hsize,clip]{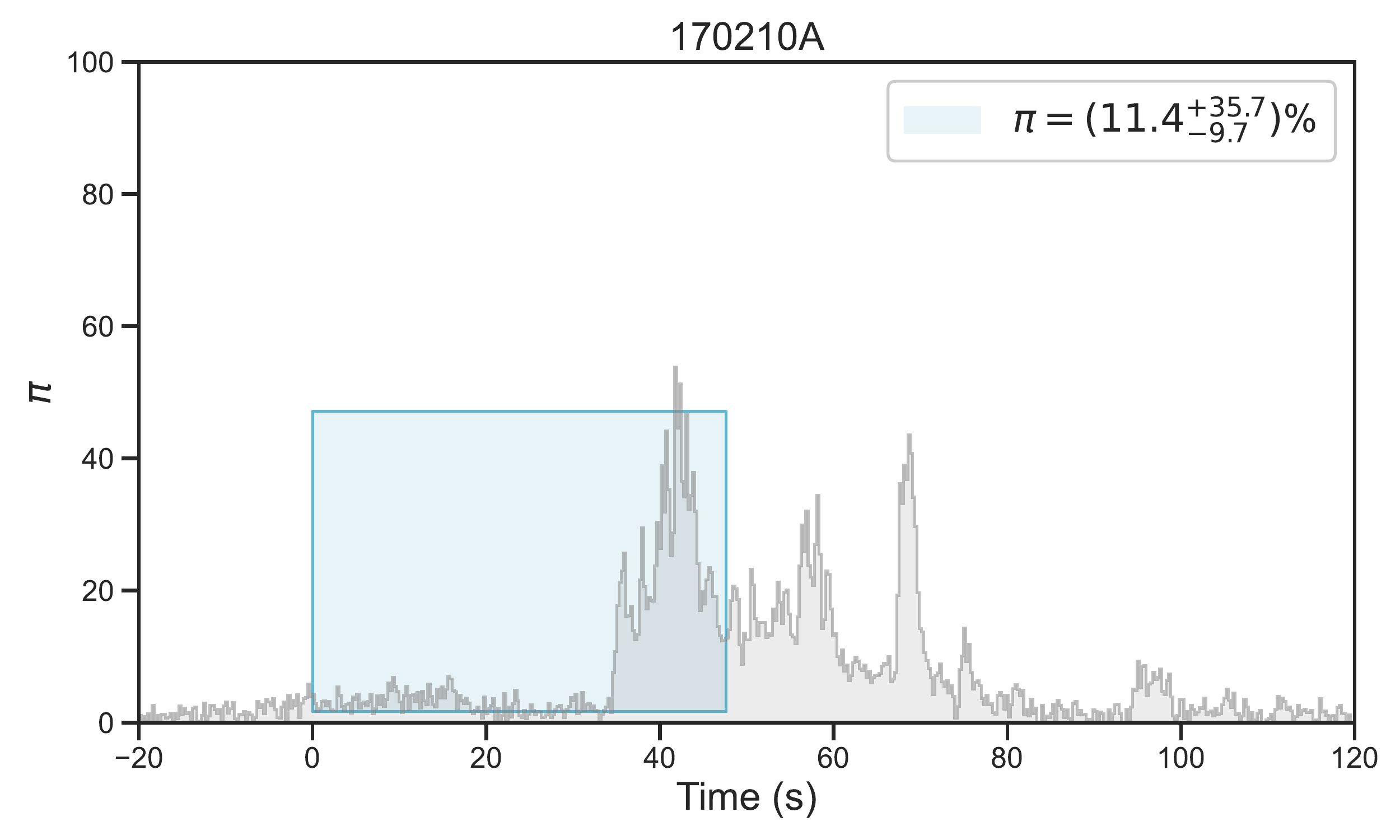}
\includegraphics[width=0.5\hsize,clip]{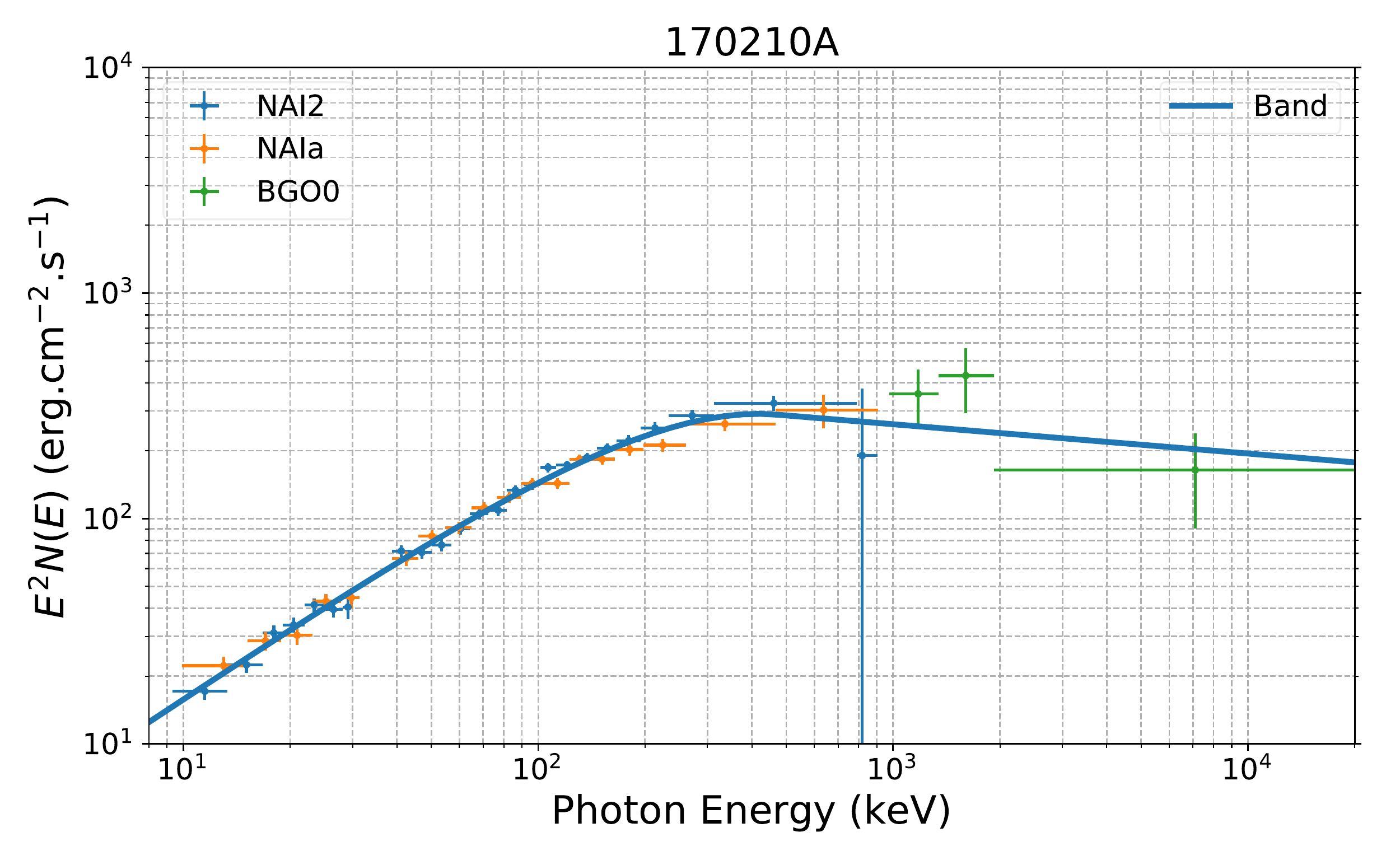}
\includegraphics[width=0.5\hsize,clip]{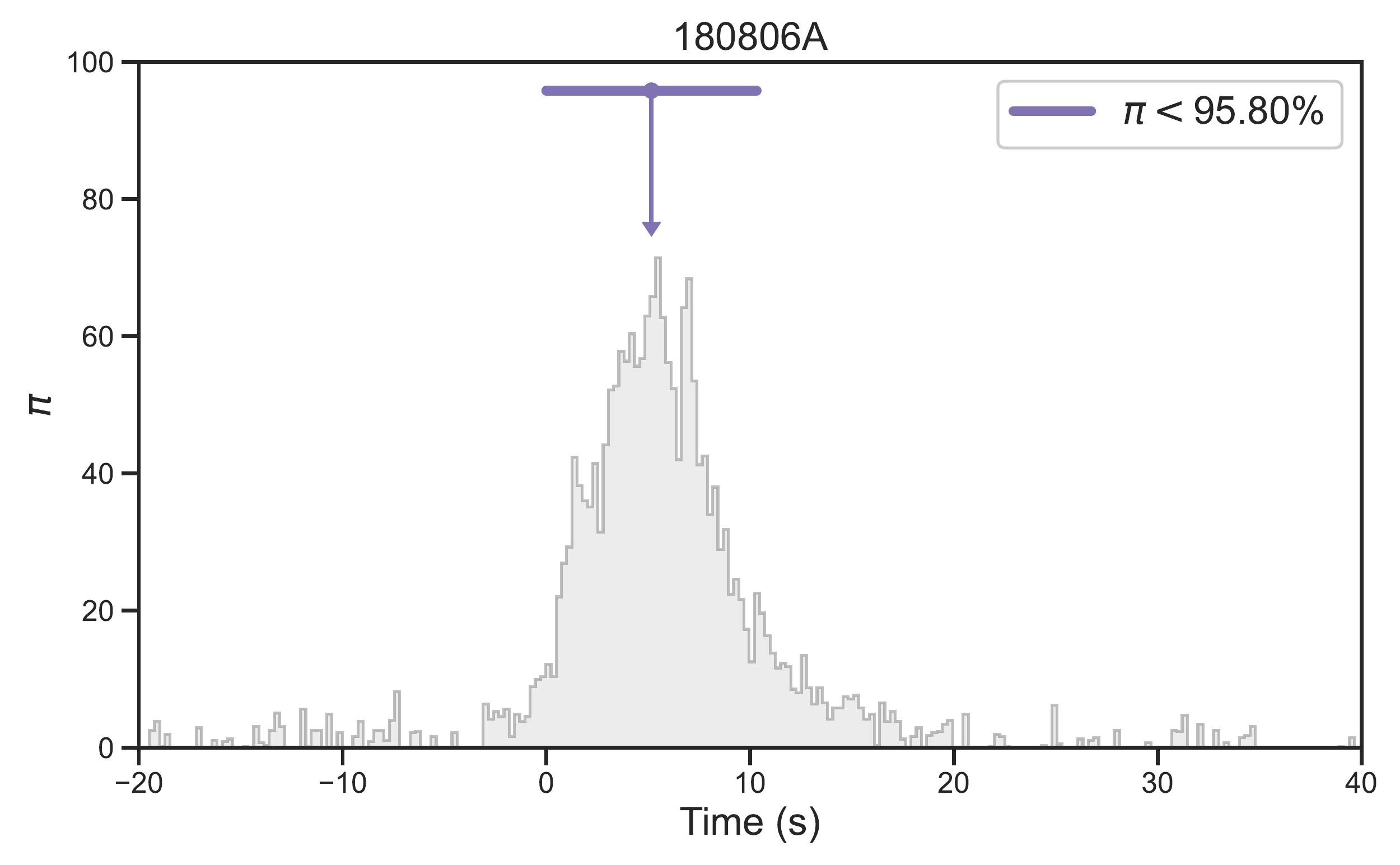}
\includegraphics[width=0.5\hsize,clip]{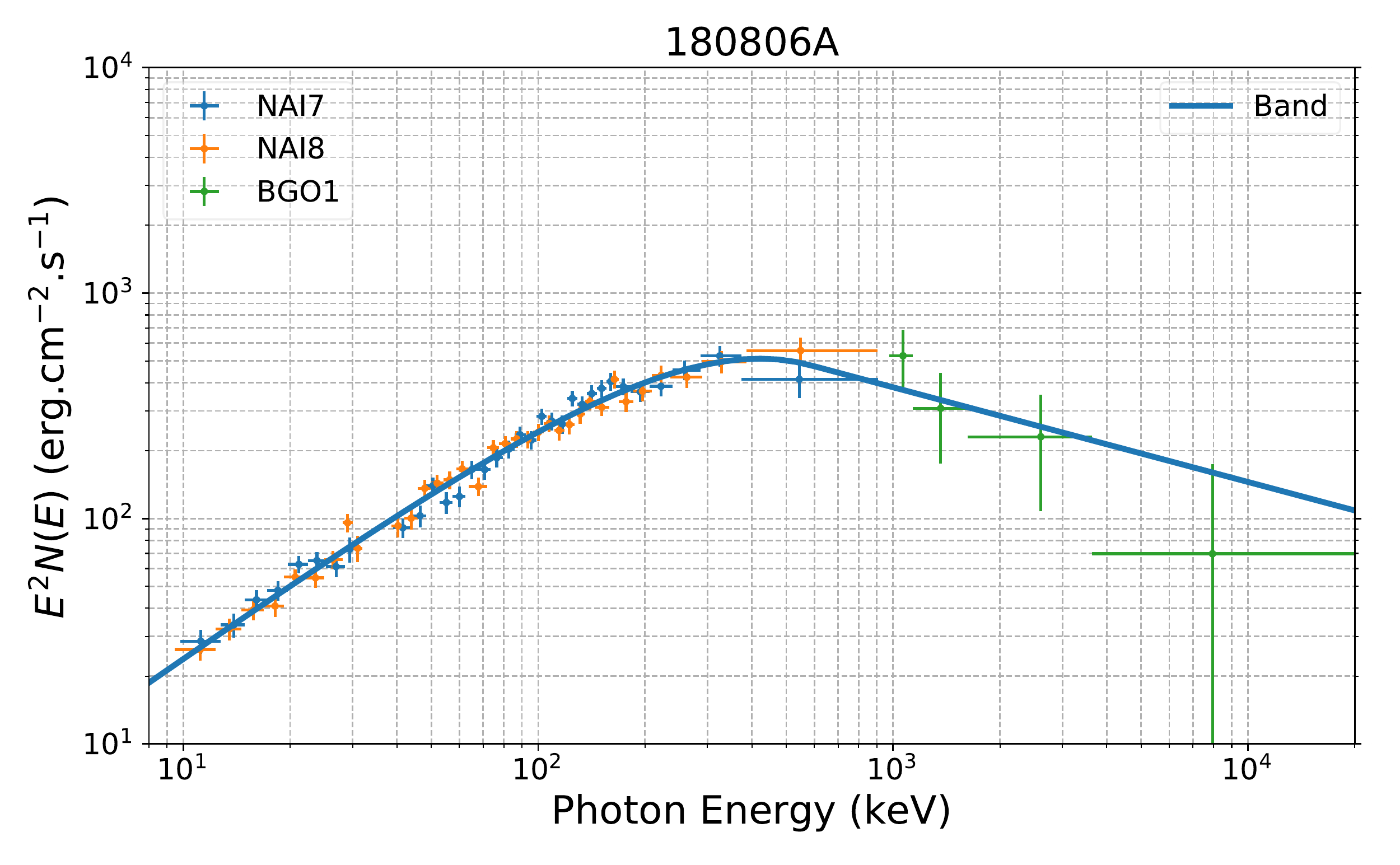}
\center{Fig. \ref{fig:Spectrum_LC_PFD_Band}--- Continued}
\end{figure*}
\begin{figure*}
\includegraphics[width=0.5\hsize,clip]{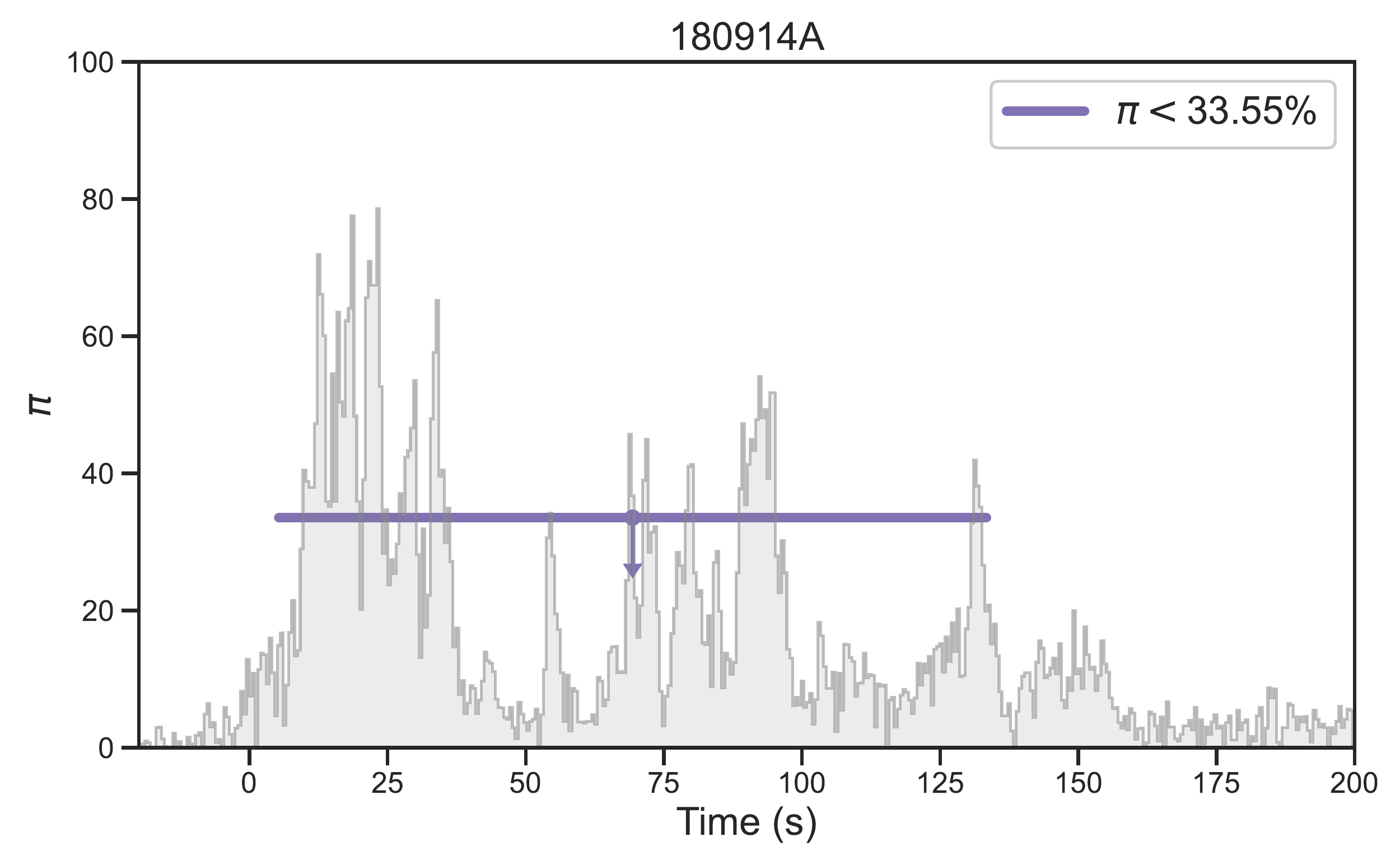}
\includegraphics[width=0.5\hsize,clip]{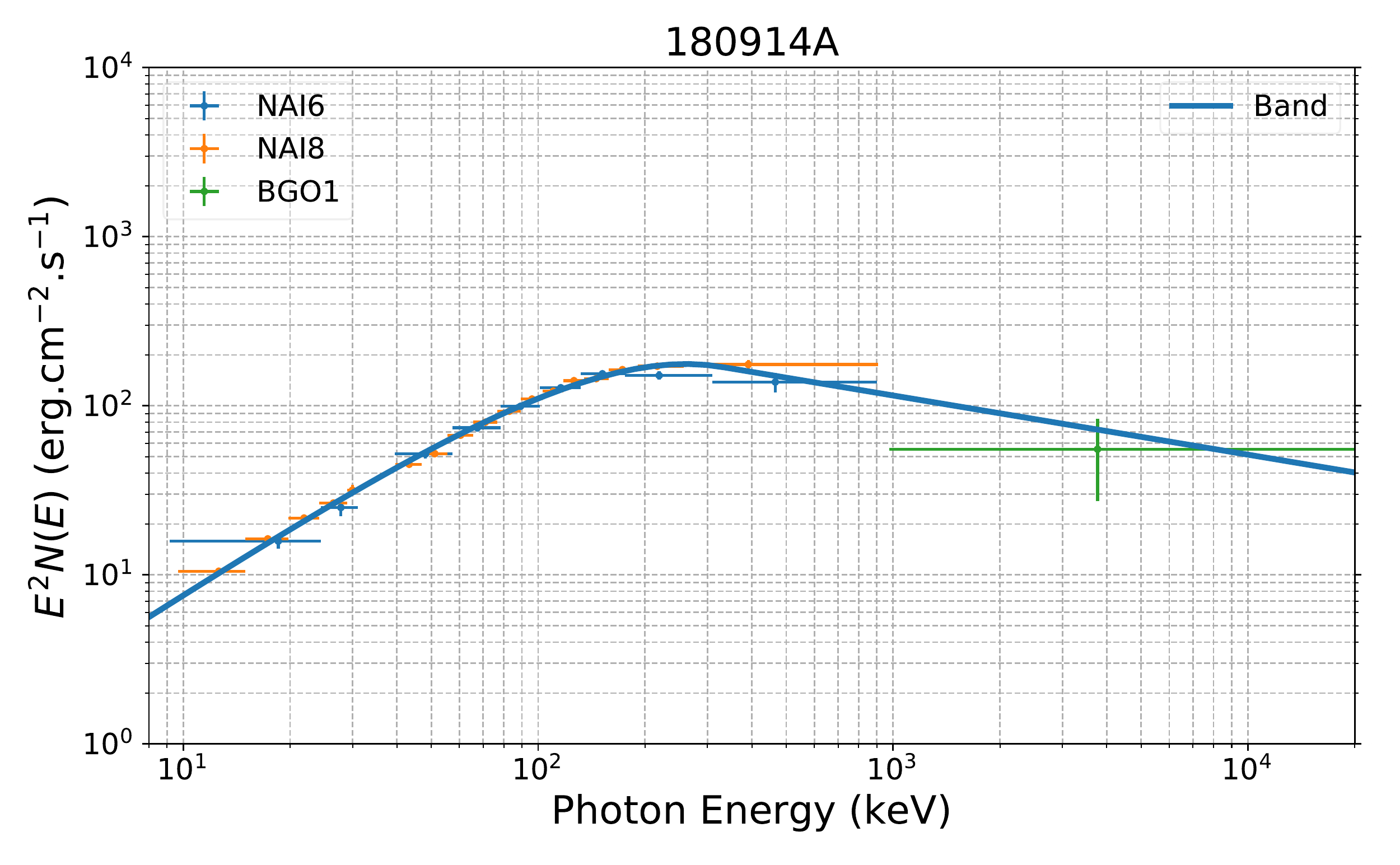}
\includegraphics[width=0.5\hsize,clip]{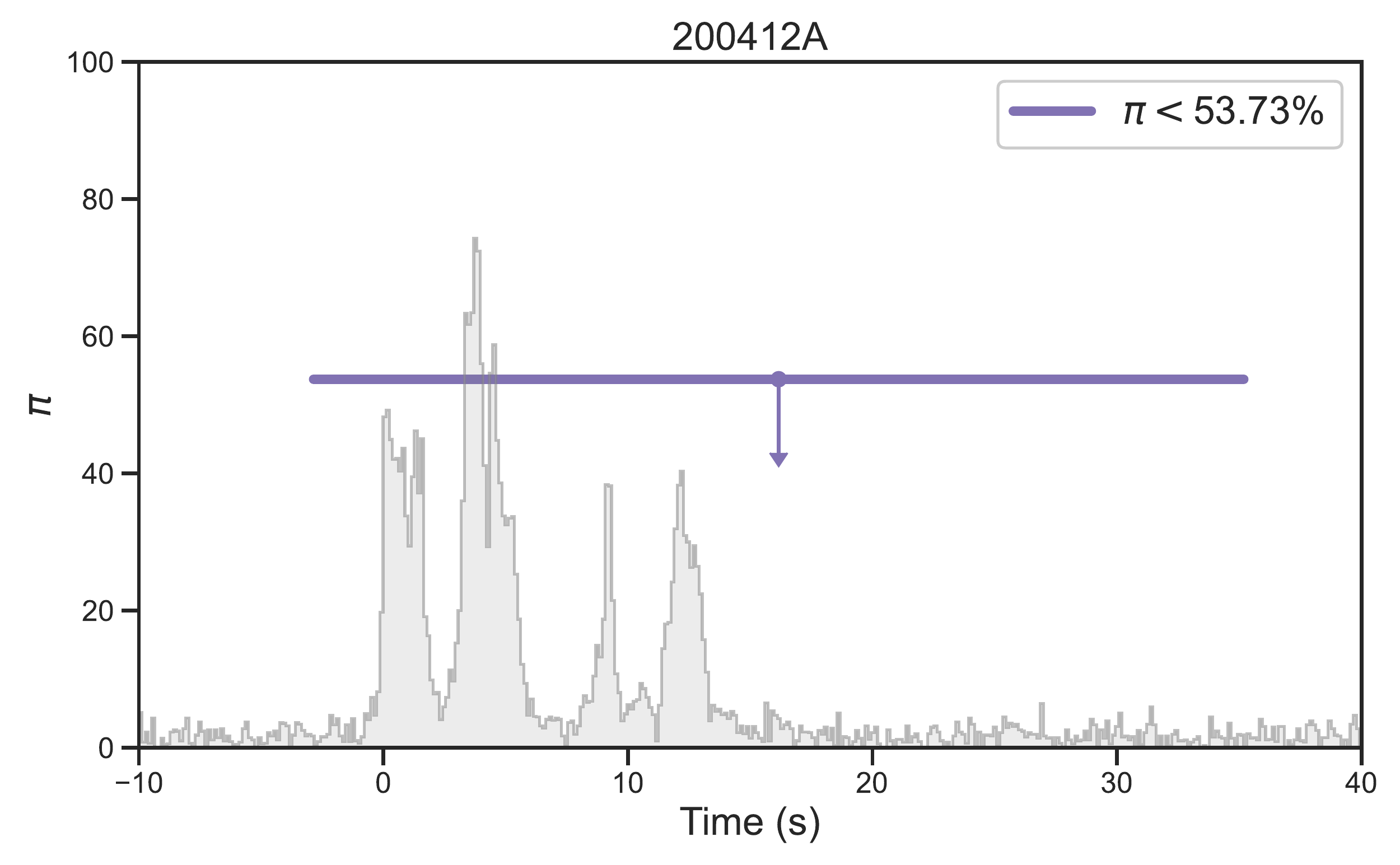}
\includegraphics[width=0.5\hsize,clip]{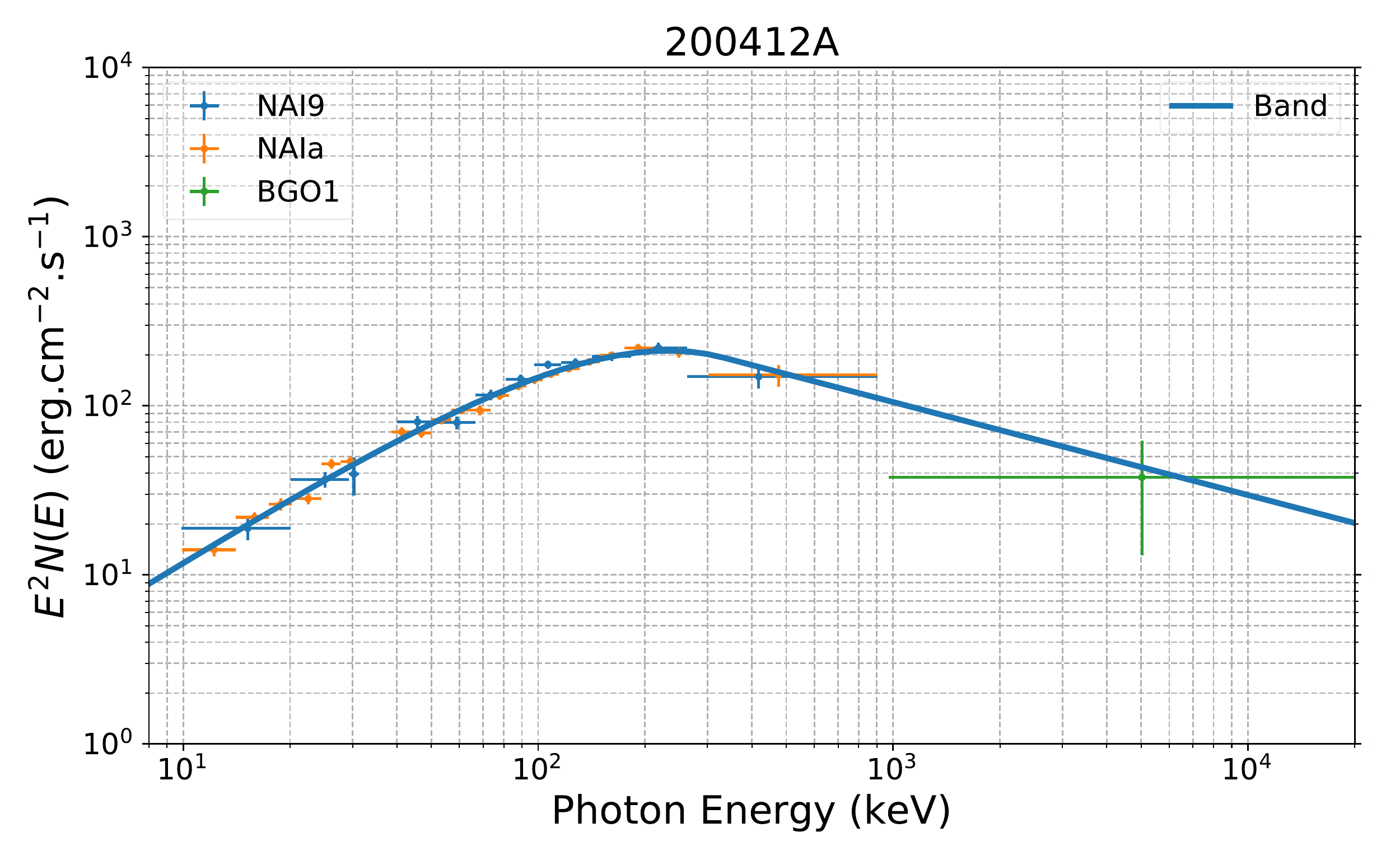}
\center{Fig. \ref{fig:Spectrum_LC_PFD_Band}--- Continued}
\end{figure*}

\clearpage
\begin{figure*}
\includegraphics[width=0.5\hsize,clip]{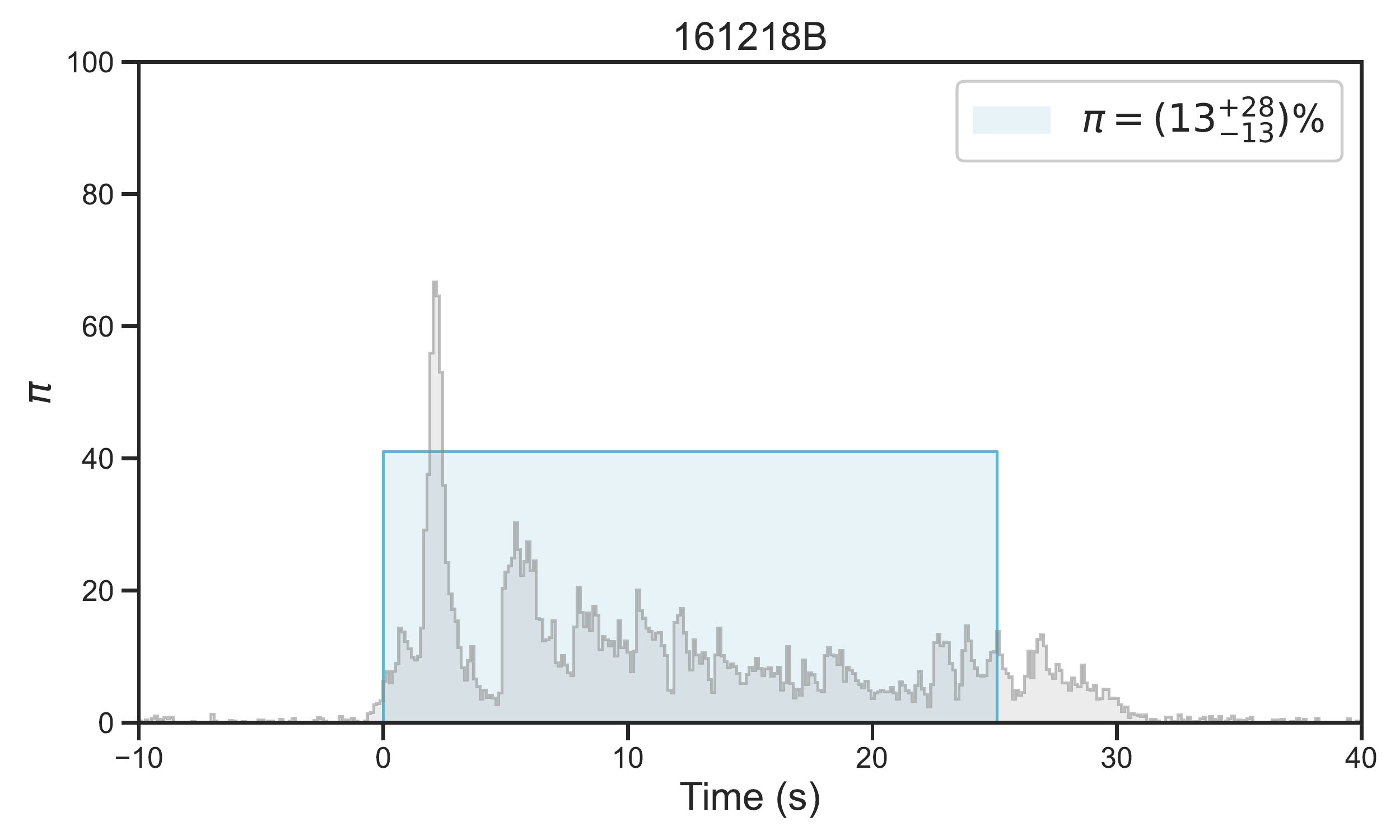}
\includegraphics[width=0.5\hsize,clip]{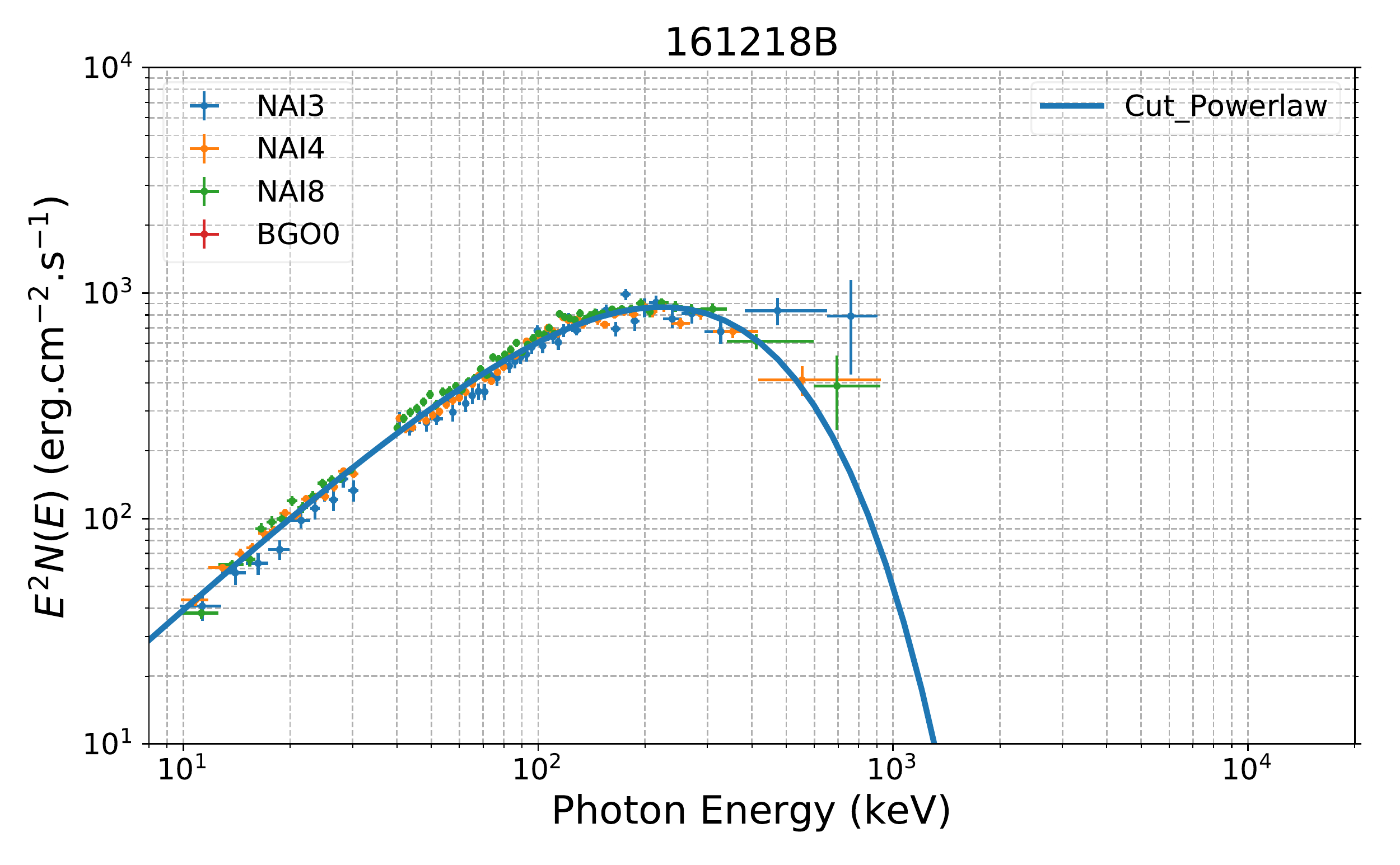}
\includegraphics[width=0.5\hsize,clip]{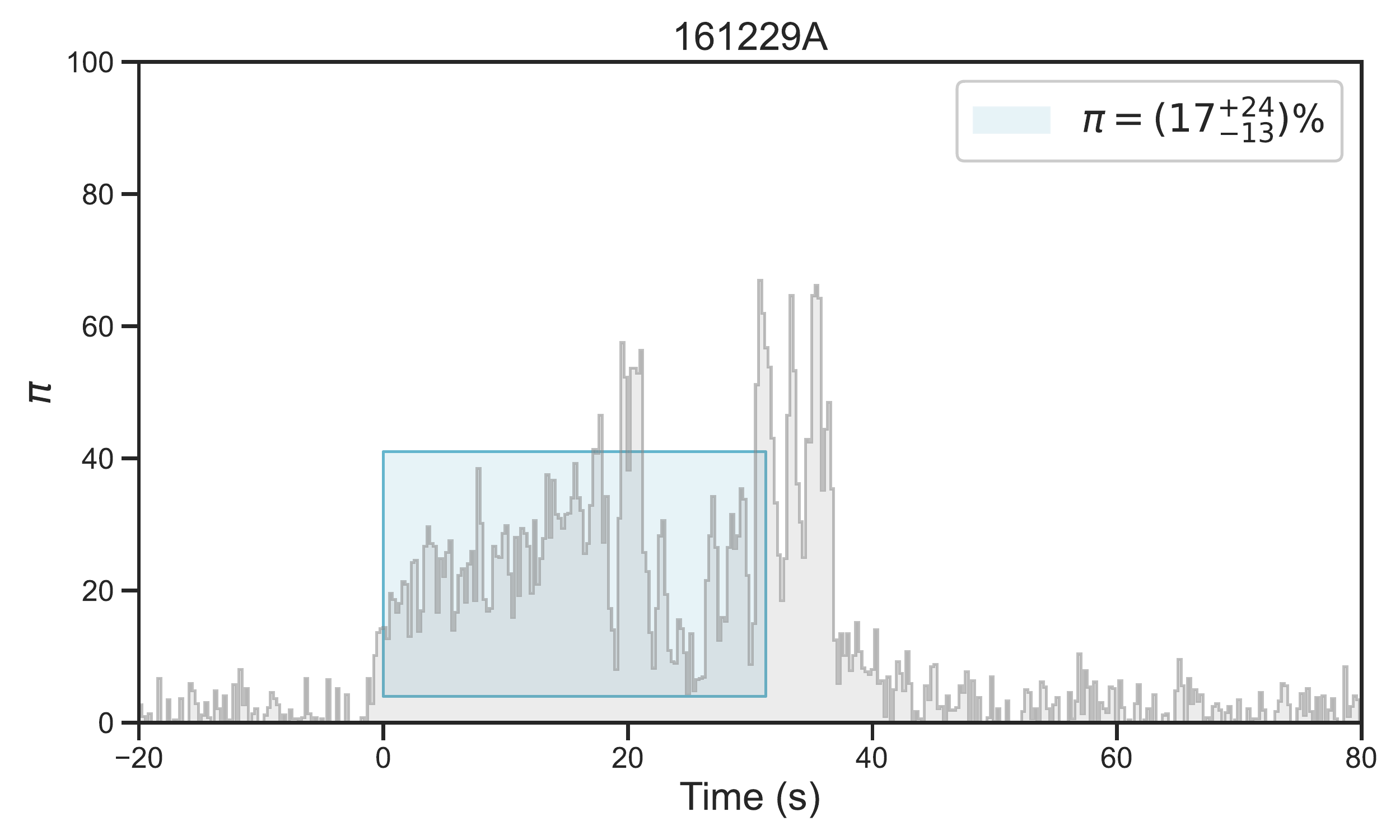}
\includegraphics[width=0.5\hsize,clip]{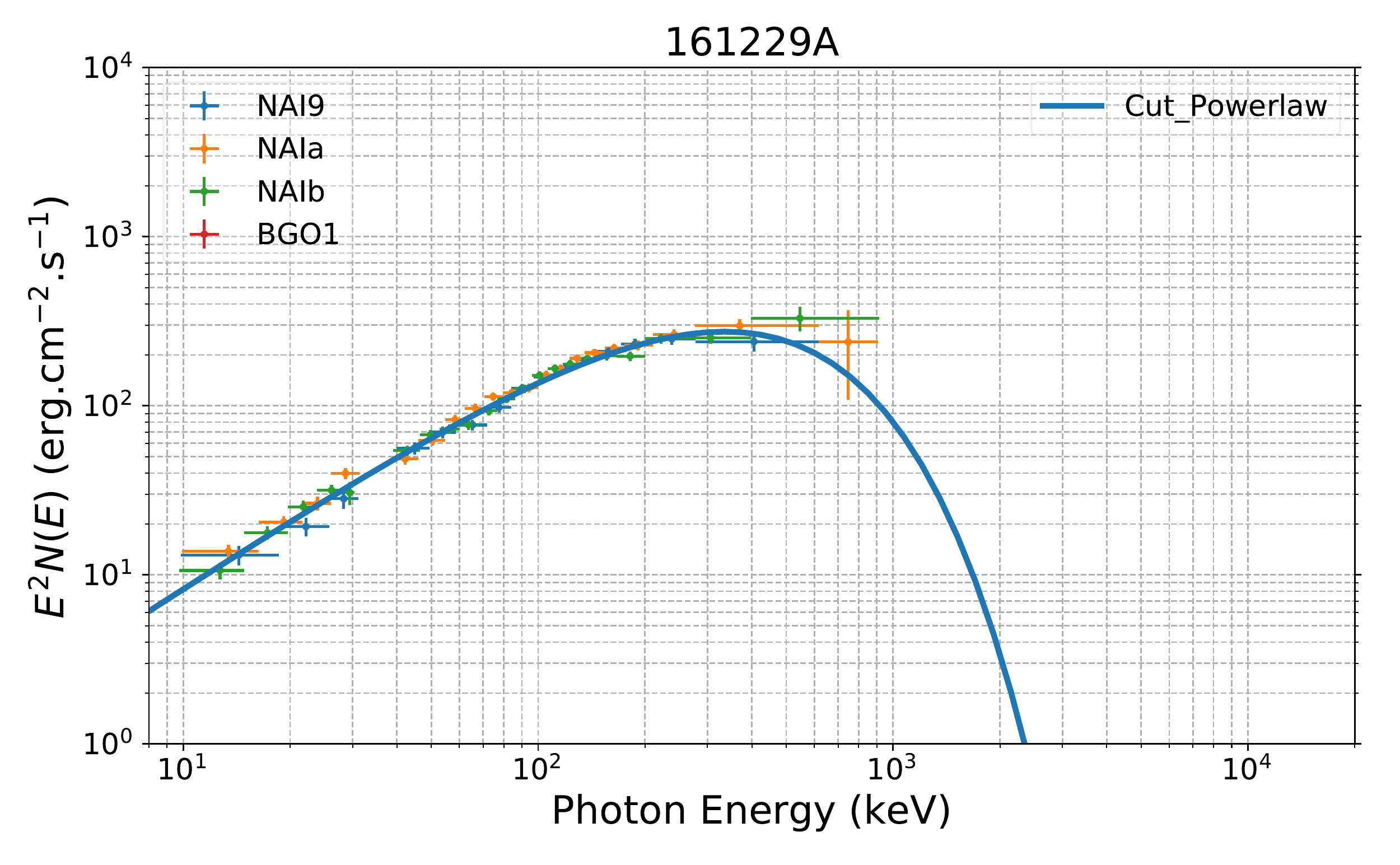}
\includegraphics[width=0.5\hsize,clip]{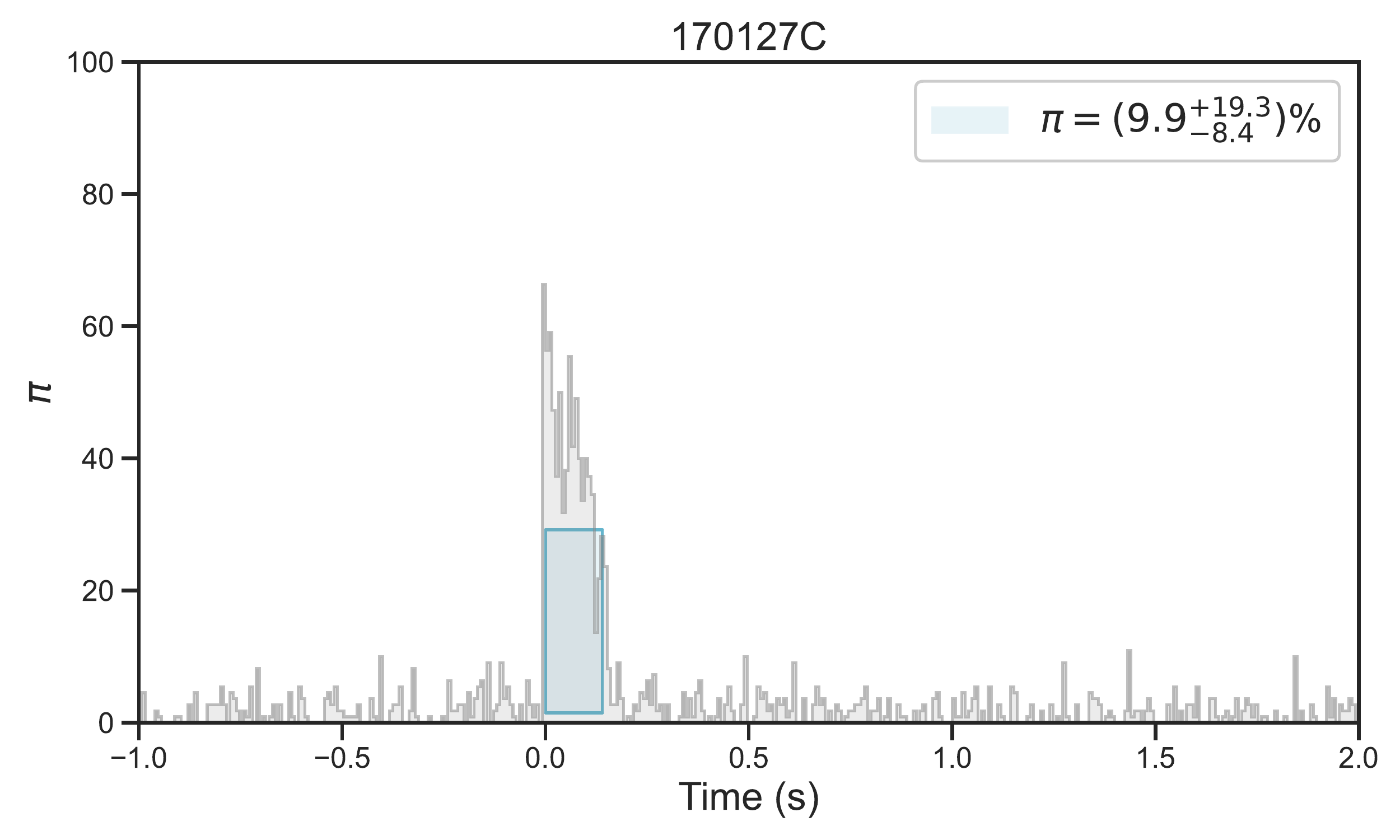}
\includegraphics[width=0.5\hsize,clip]{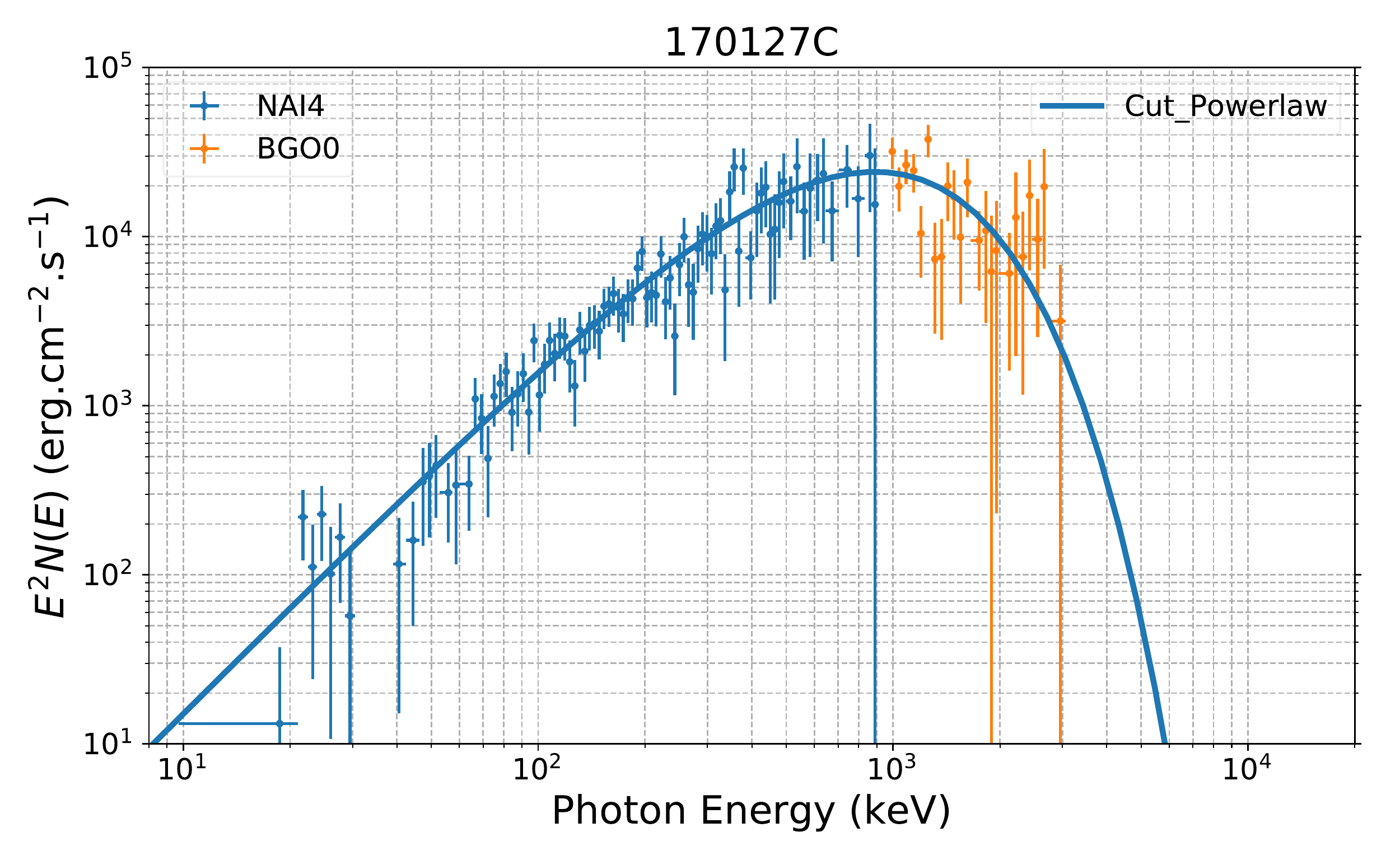}
\caption{Same as Figure \ref{fig:Spectrum_LC_PFD_Band}, but for bursts classified as PFD-type with CPL-like spectra.}
\label{fig:Spectrum_LC_PFD_CPL}
\end{figure*}
\begin{figure*}
\includegraphics[width=0.5\hsize,clip]{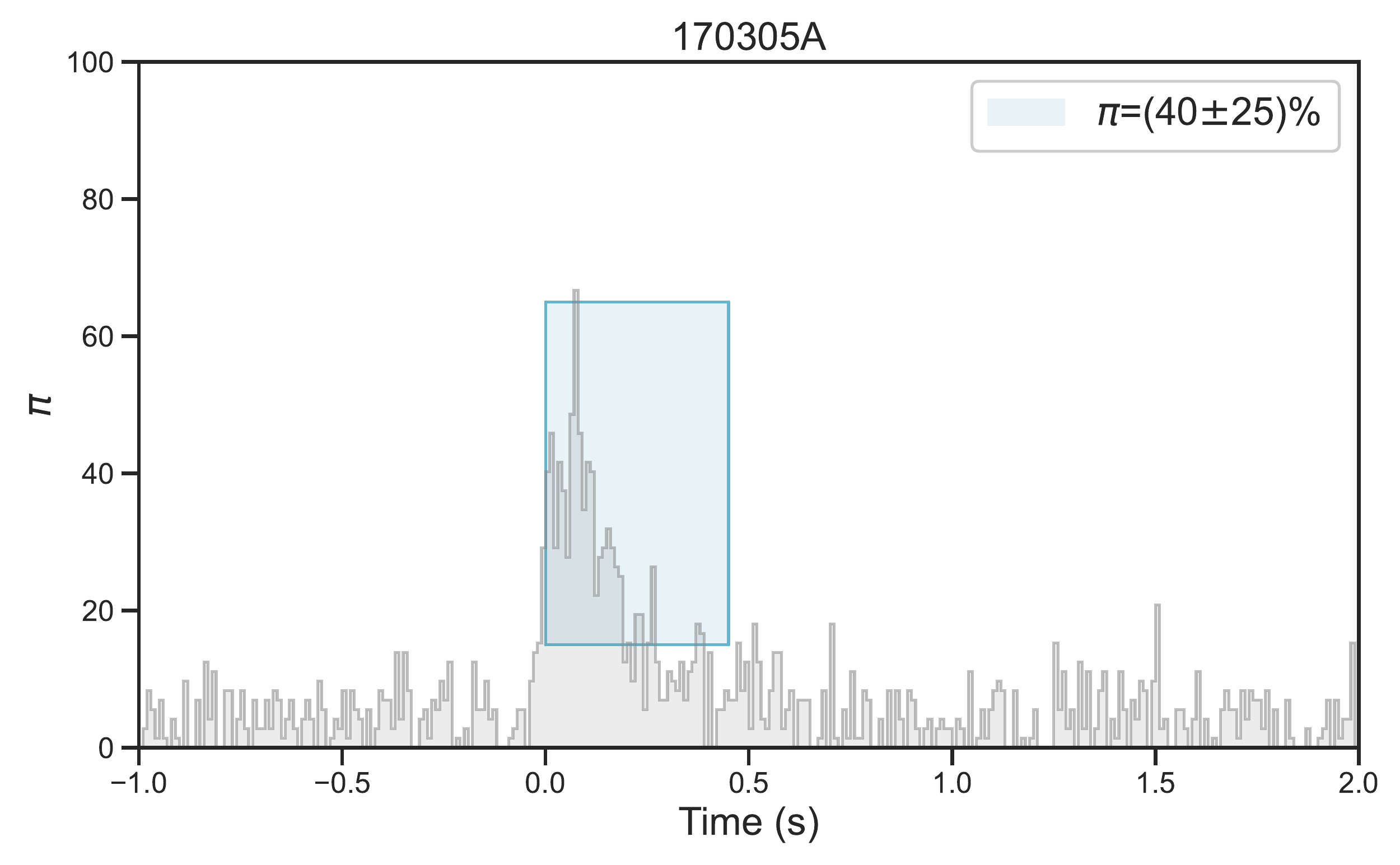}
\includegraphics[width=0.5\hsize,clip]{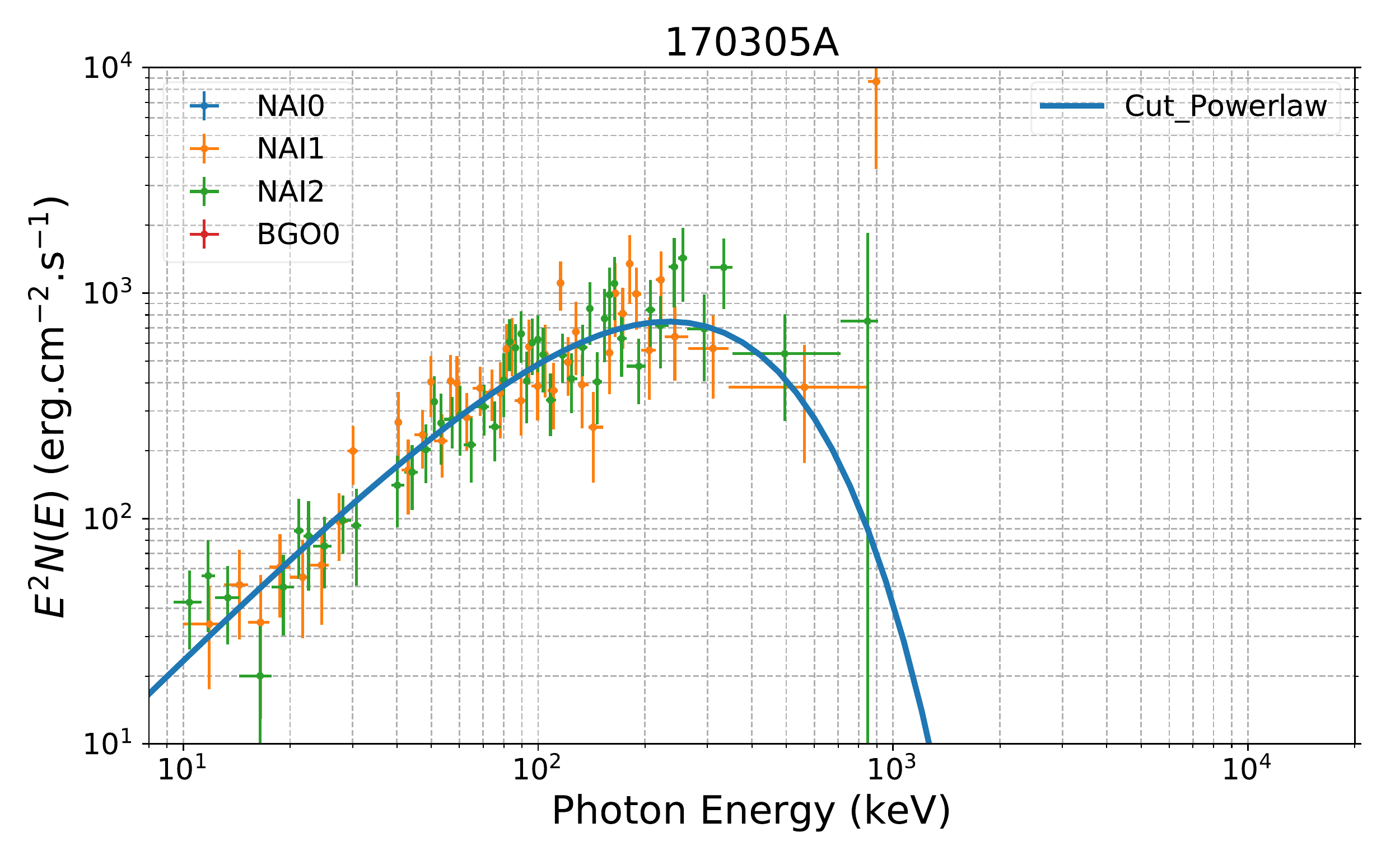}
\includegraphics[width=0.5\hsize,clip]{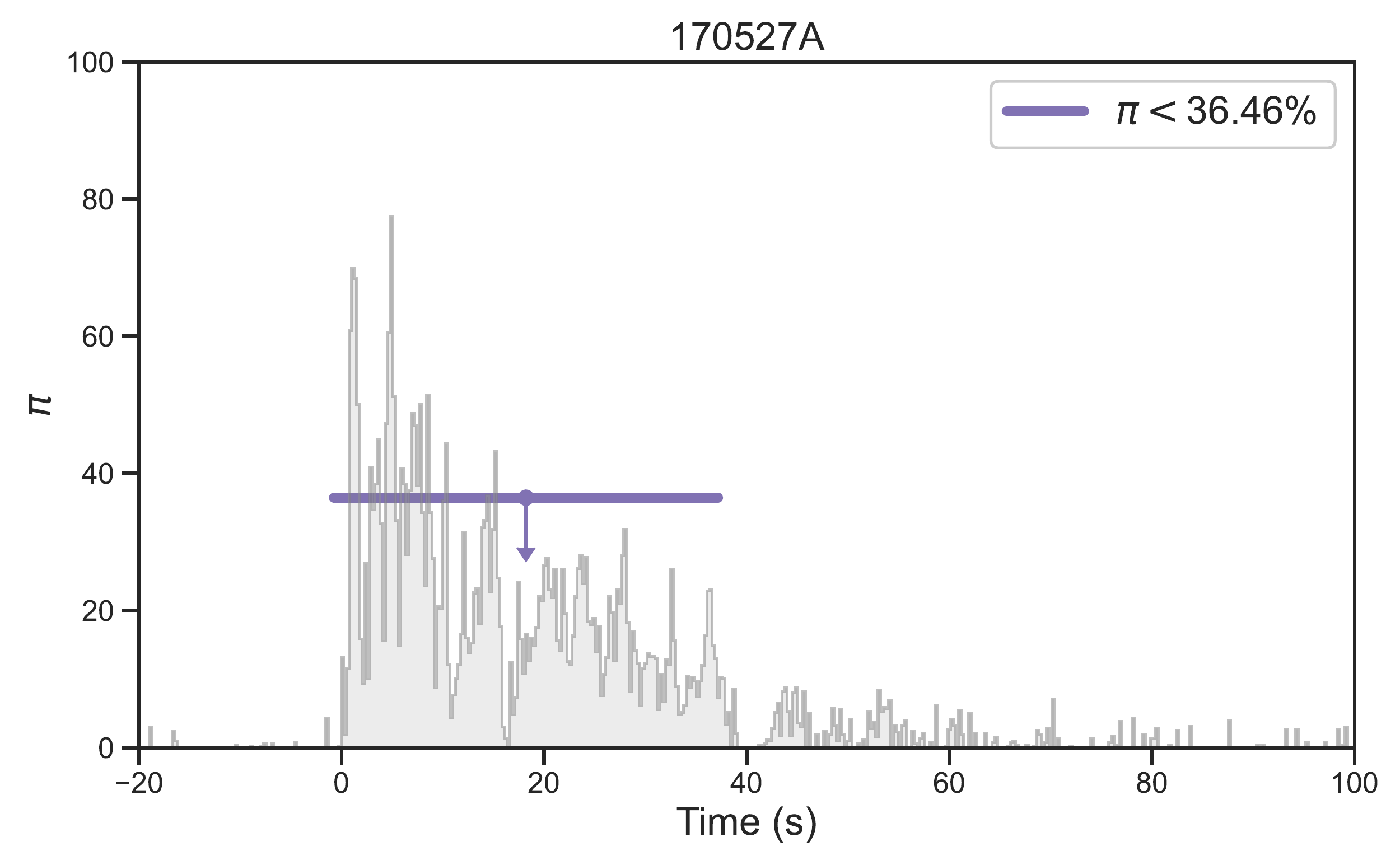}
\includegraphics[width=0.5\hsize,clip]{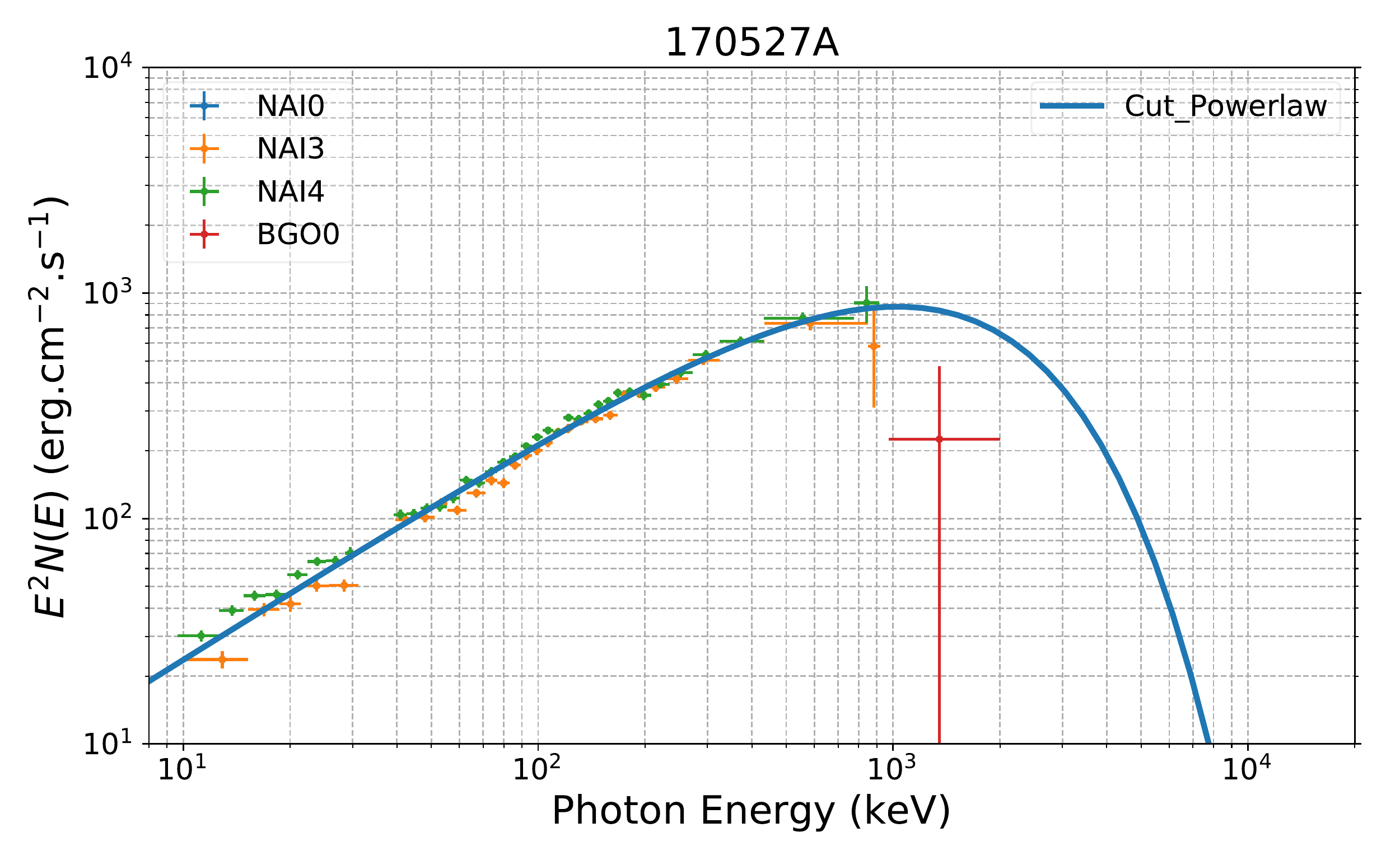}
\includegraphics[width=0.5\hsize,clip]{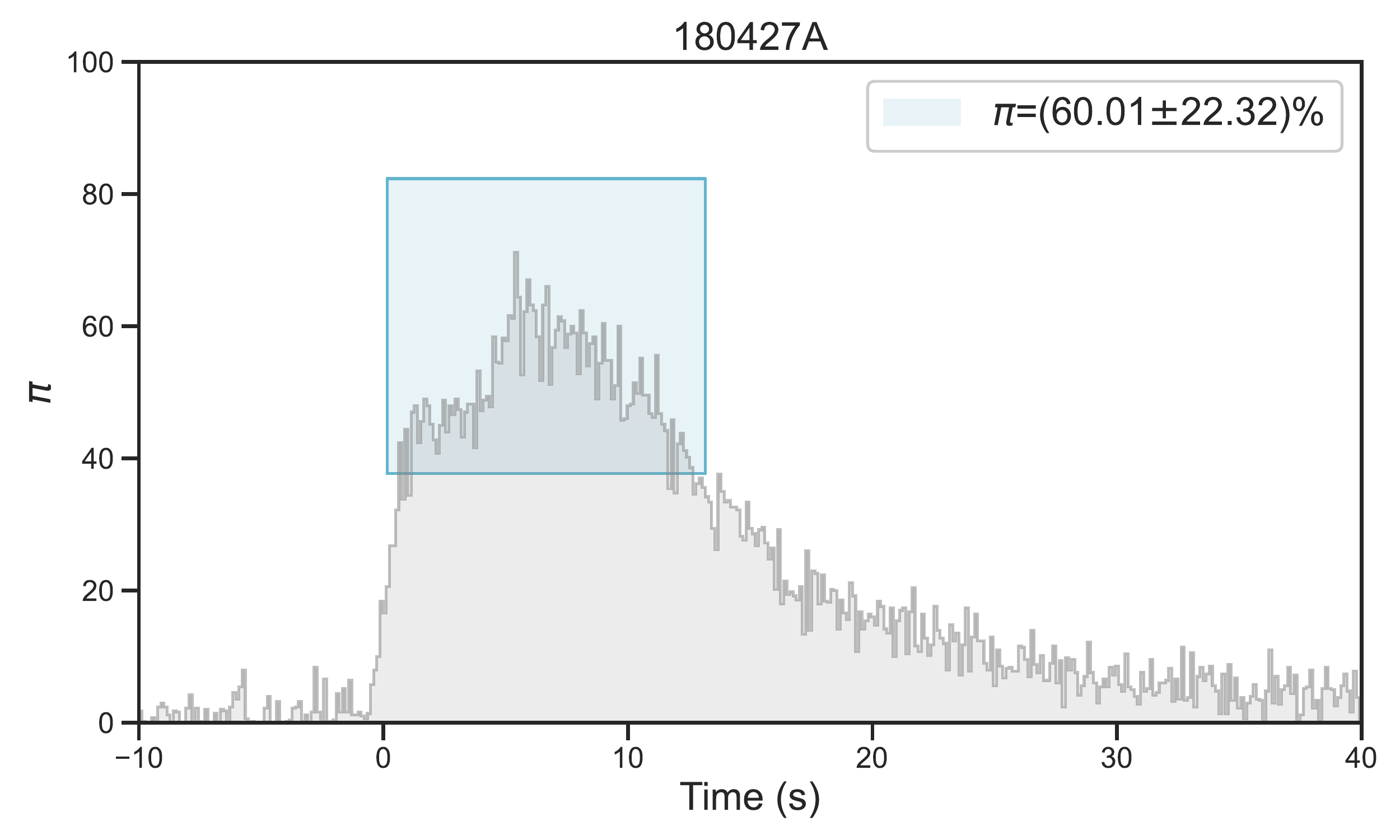}
\includegraphics[width=0.5\hsize,clip]{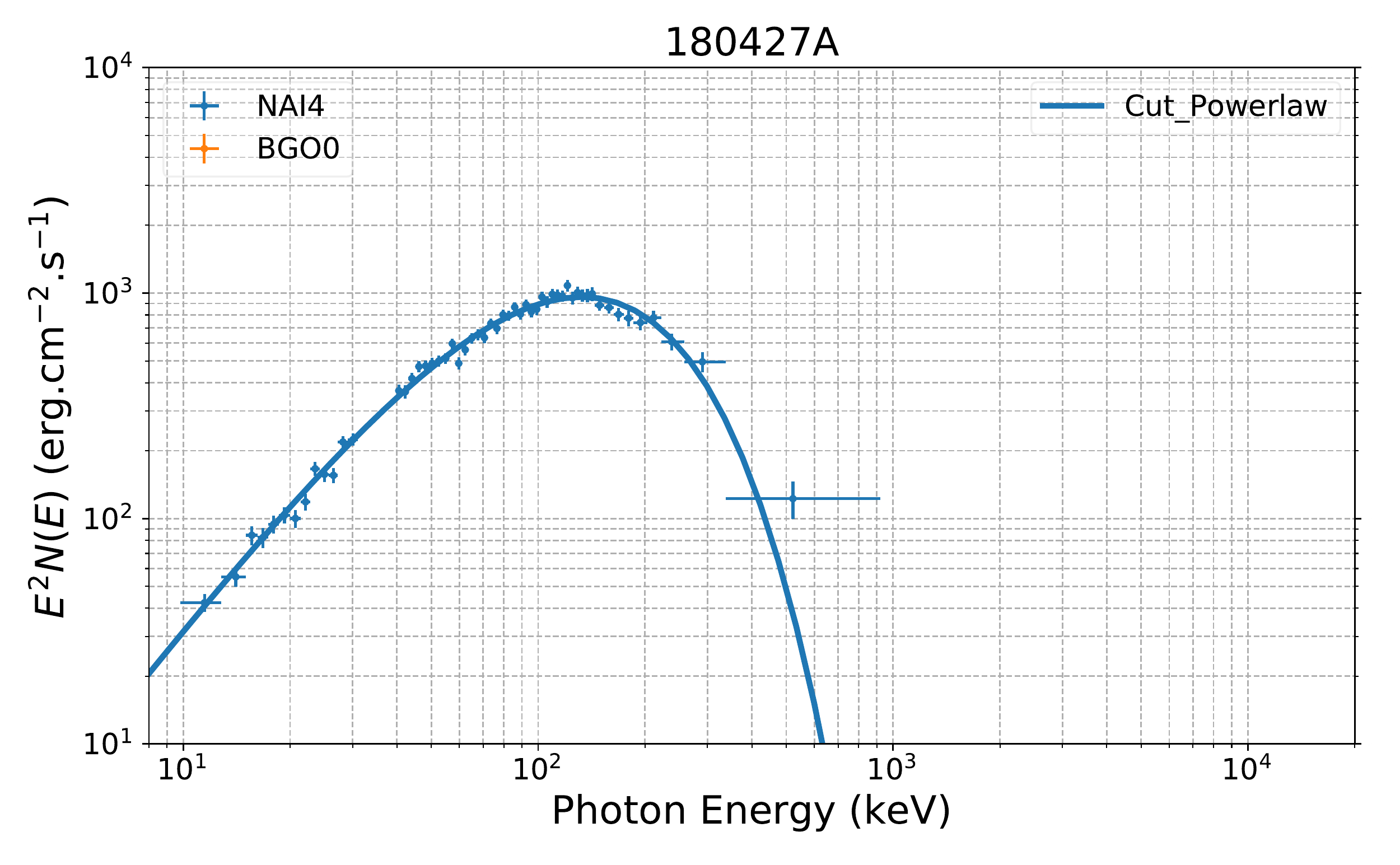}
\includegraphics[width=0.5\hsize,clip]{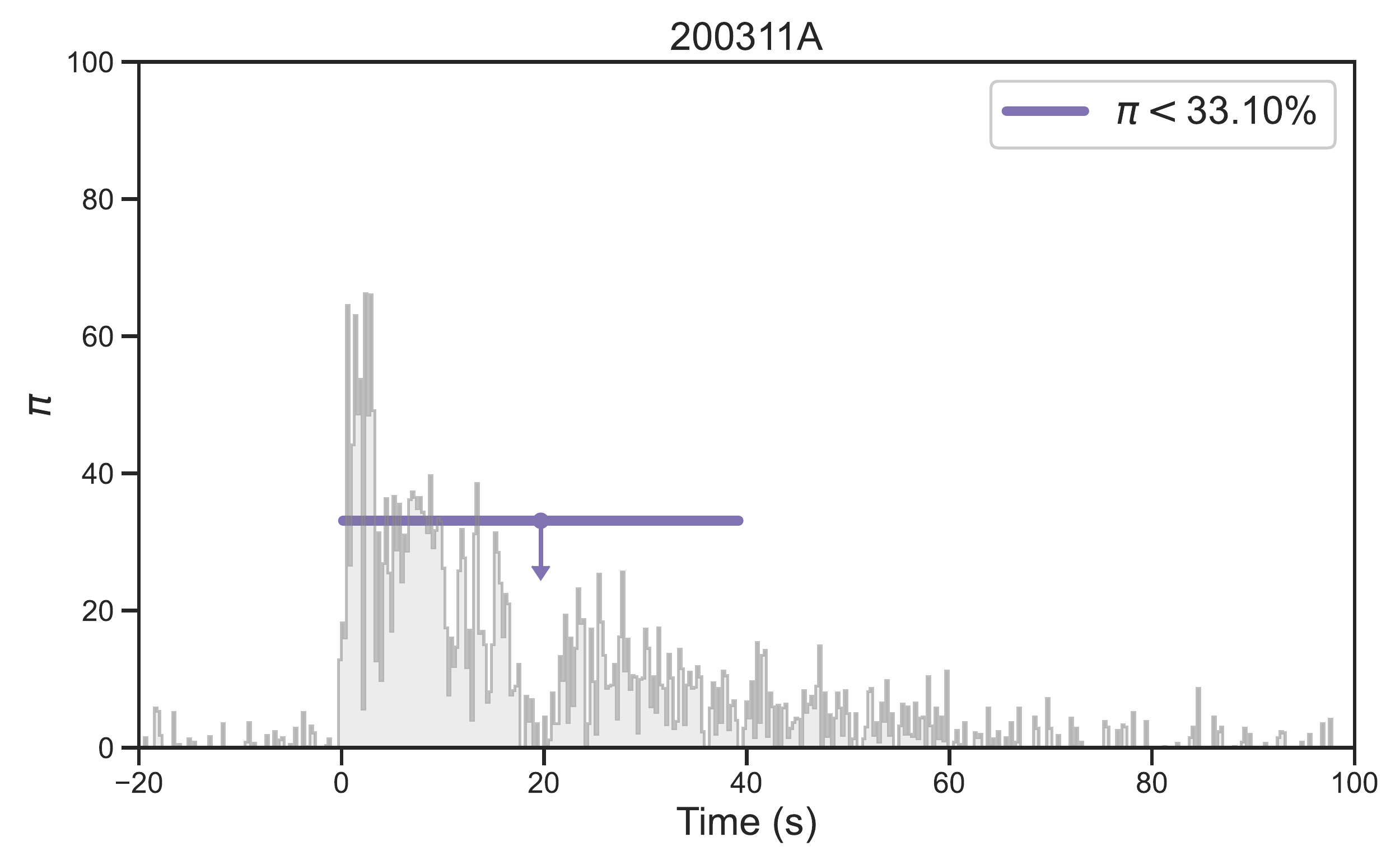}
\includegraphics[width=0.5\hsize,clip]{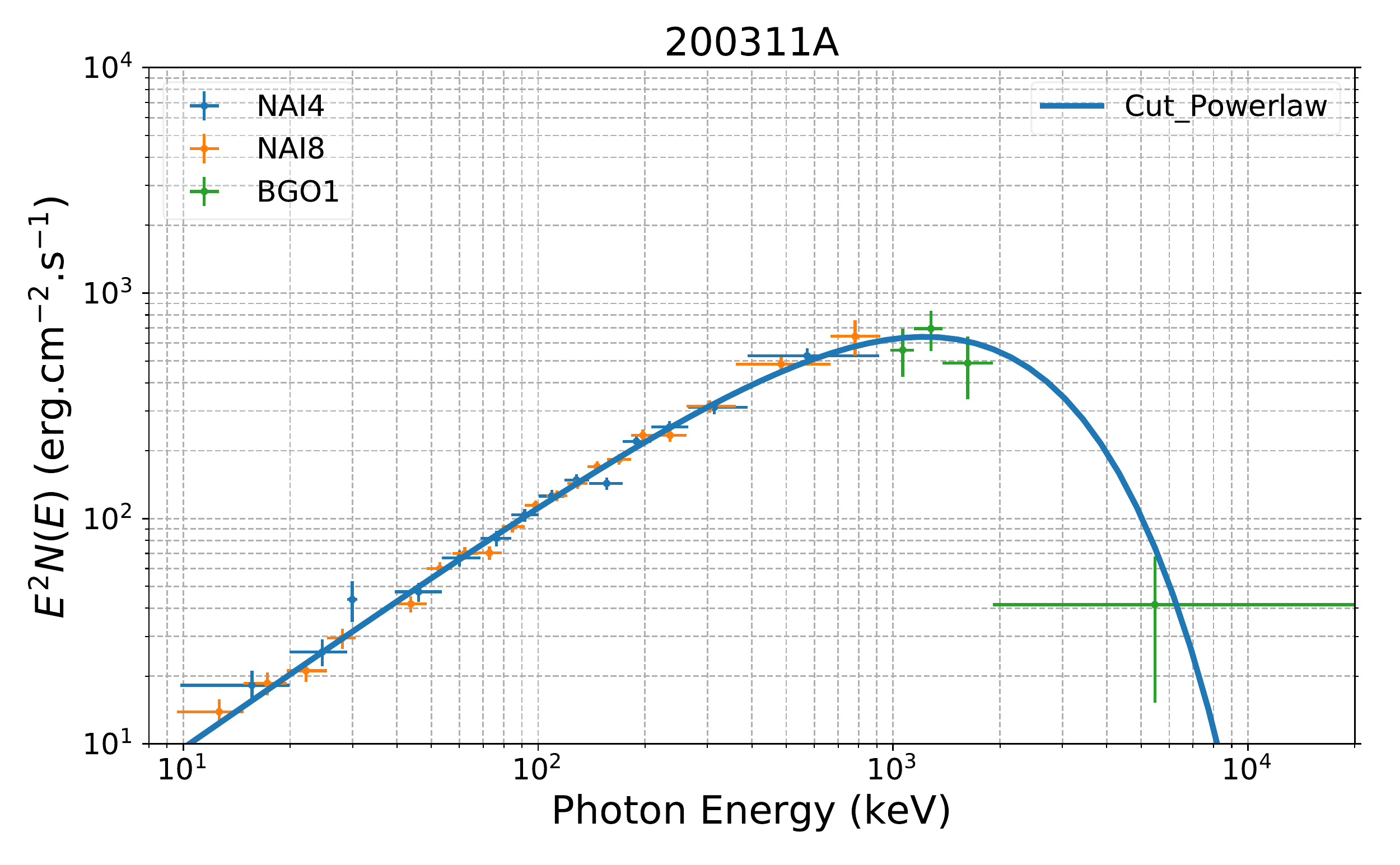}
\center{Fig. \ref{fig:Spectrum_LC_PFD_CPL}--- Continued}
\end{figure*}

\clearpage
\begin{figure*}
\includegraphics[width=1.\hsize,clip]{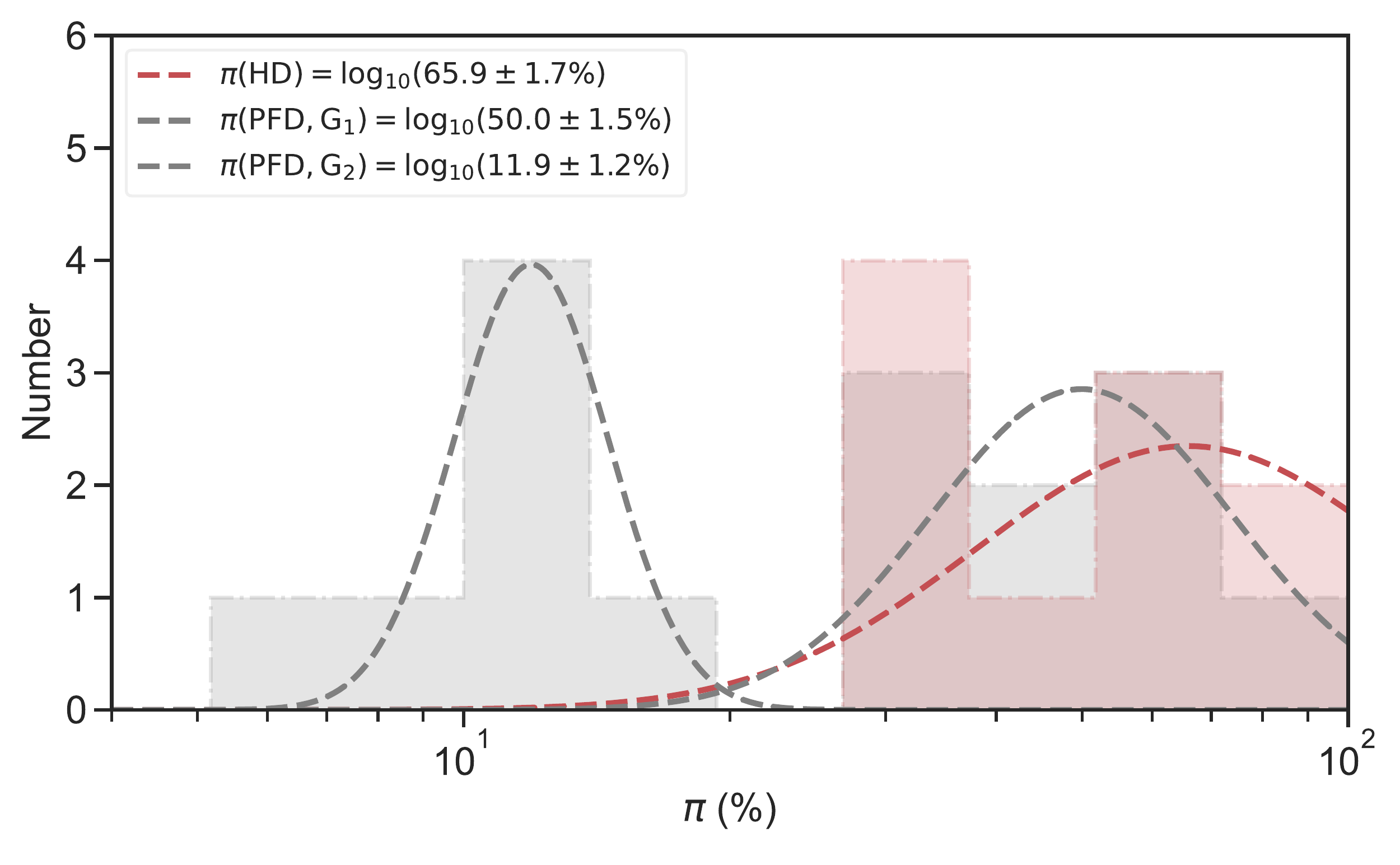}
\caption{Distribution of polarization degree $\pi$, comparing the PFD-type bursts (gray color) with the HD-type bursts (red color). The best Gaussian fits give $\pi$=log10(65.9$\pm$1.7)\% for the HD-type bursts (dashed red line) and $\pi$=log10(50.0$\pm$1.5)\% (the higher-peak $G_1$) and $\pi$=log10(11.9$\pm$1.2)\% (the lower-peak $G_2$) for the PFD-type bursts (dashed gray line).}
\label{fig:Dis}
\end{figure*}

\clearpage
\begin{figure*}
\includegraphics[width=0.5\hsize,clip]{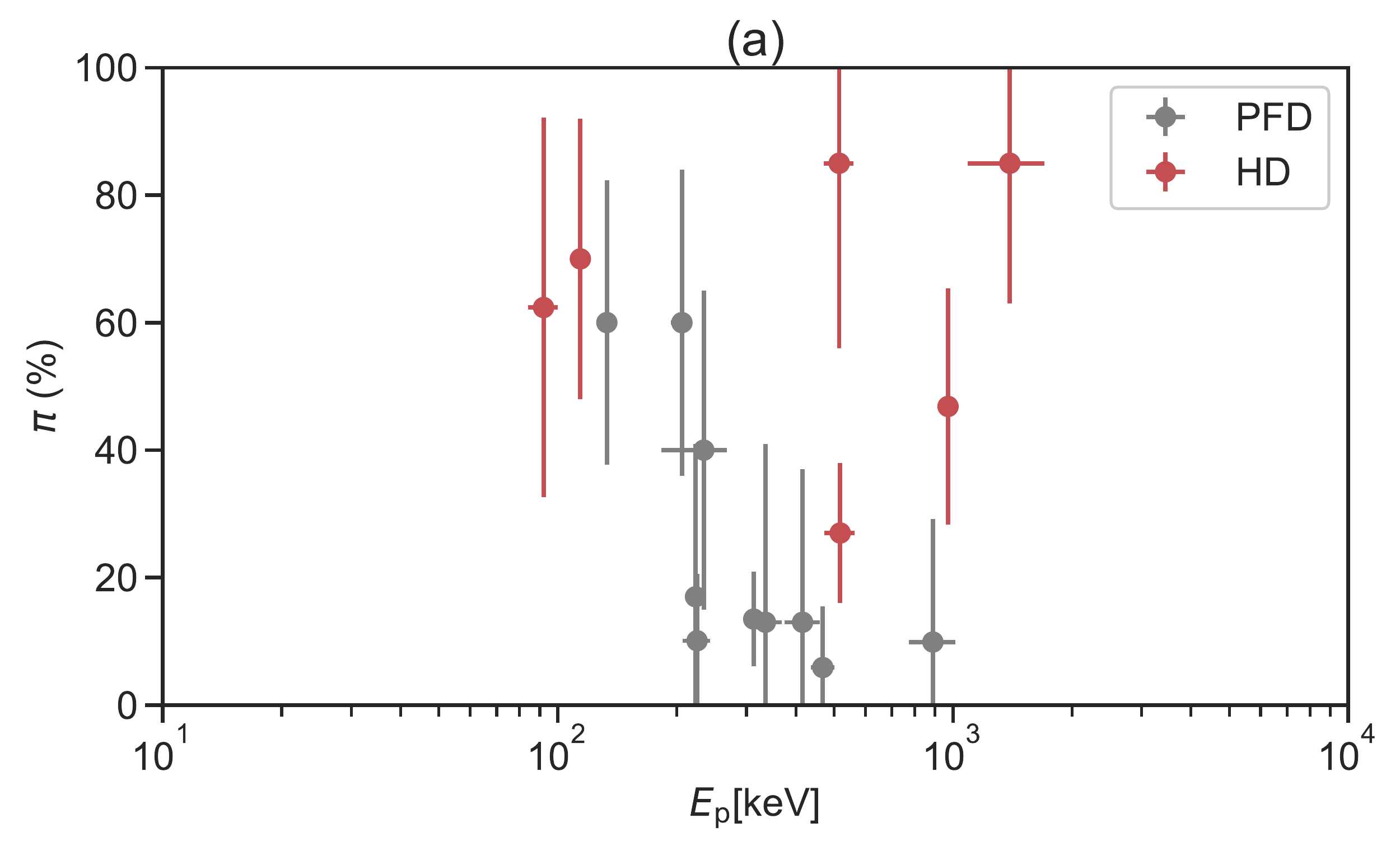}
\includegraphics[width=0.5\hsize,clip]{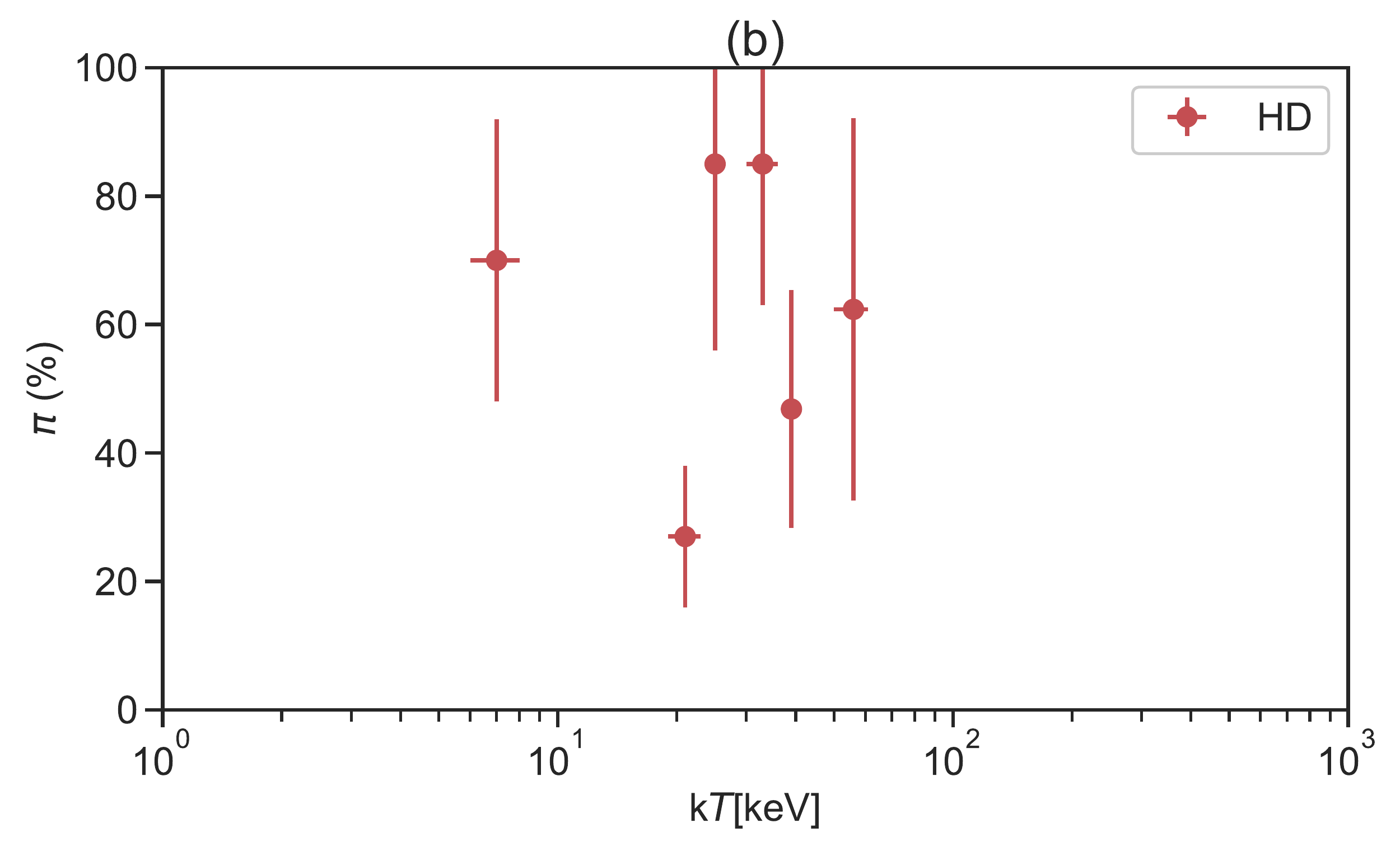}
\includegraphics[width=0.5\hsize,clip]{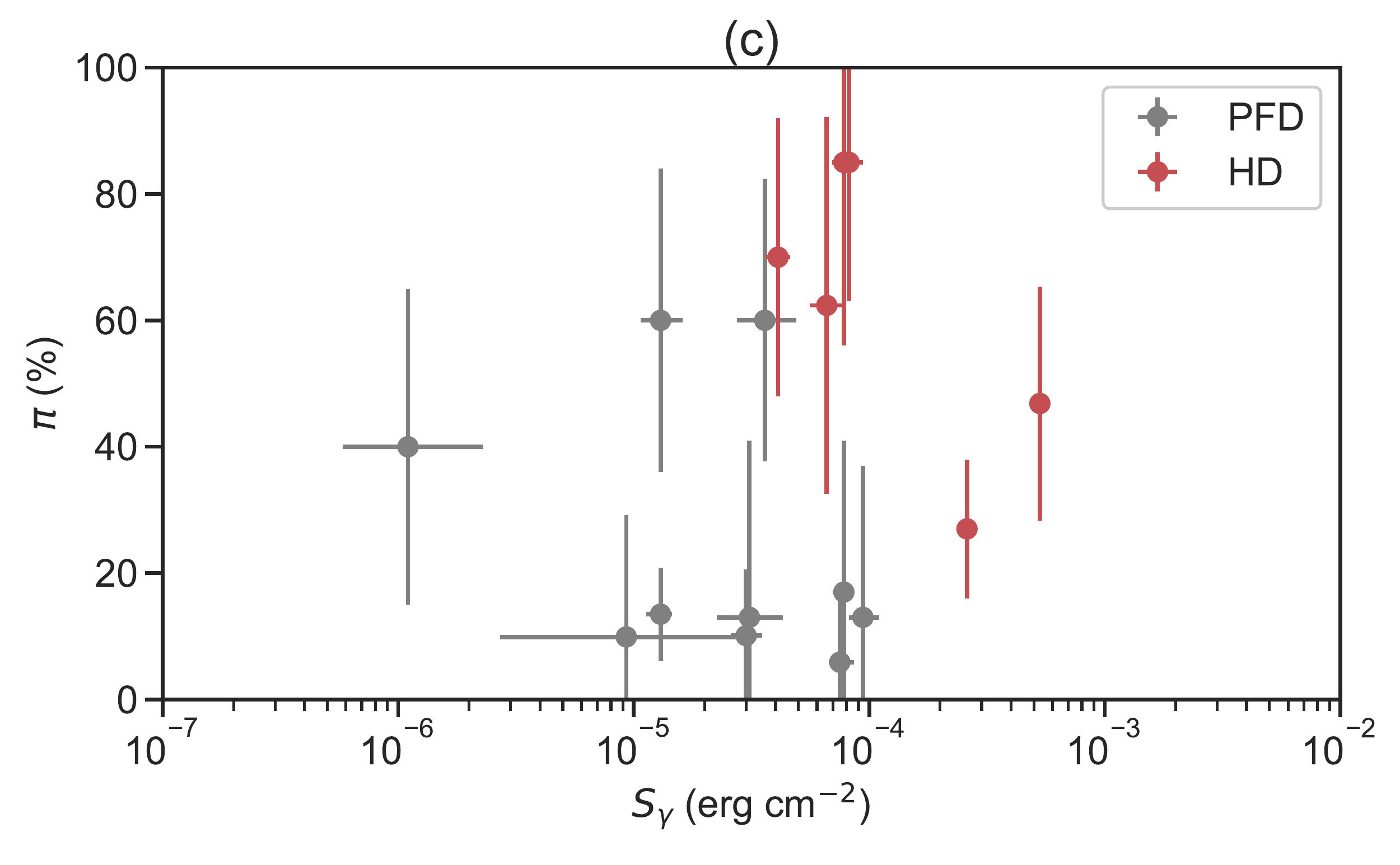}
\includegraphics[width=0.5\hsize,clip]{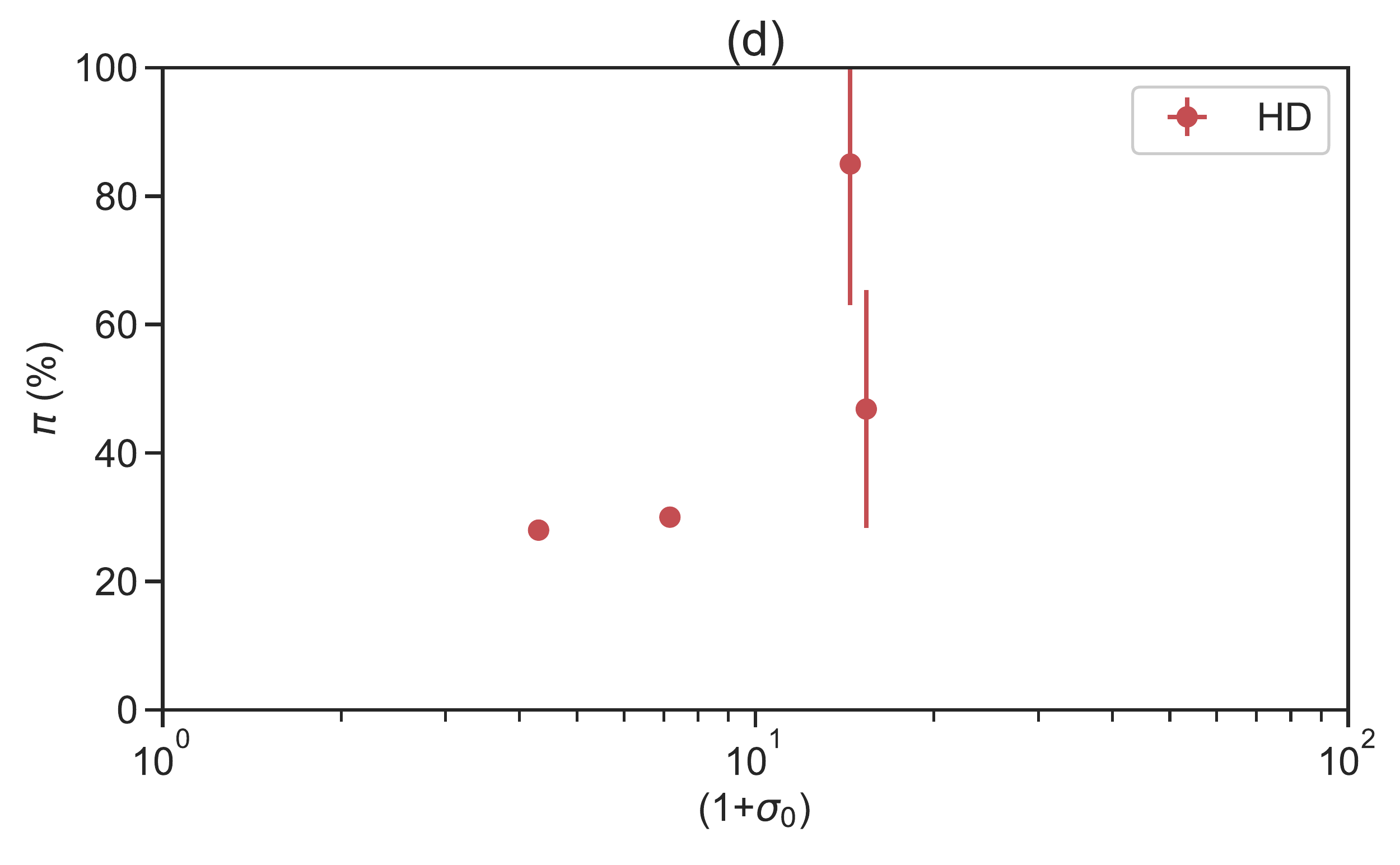}
\includegraphics[width=0.5\hsize,clip]{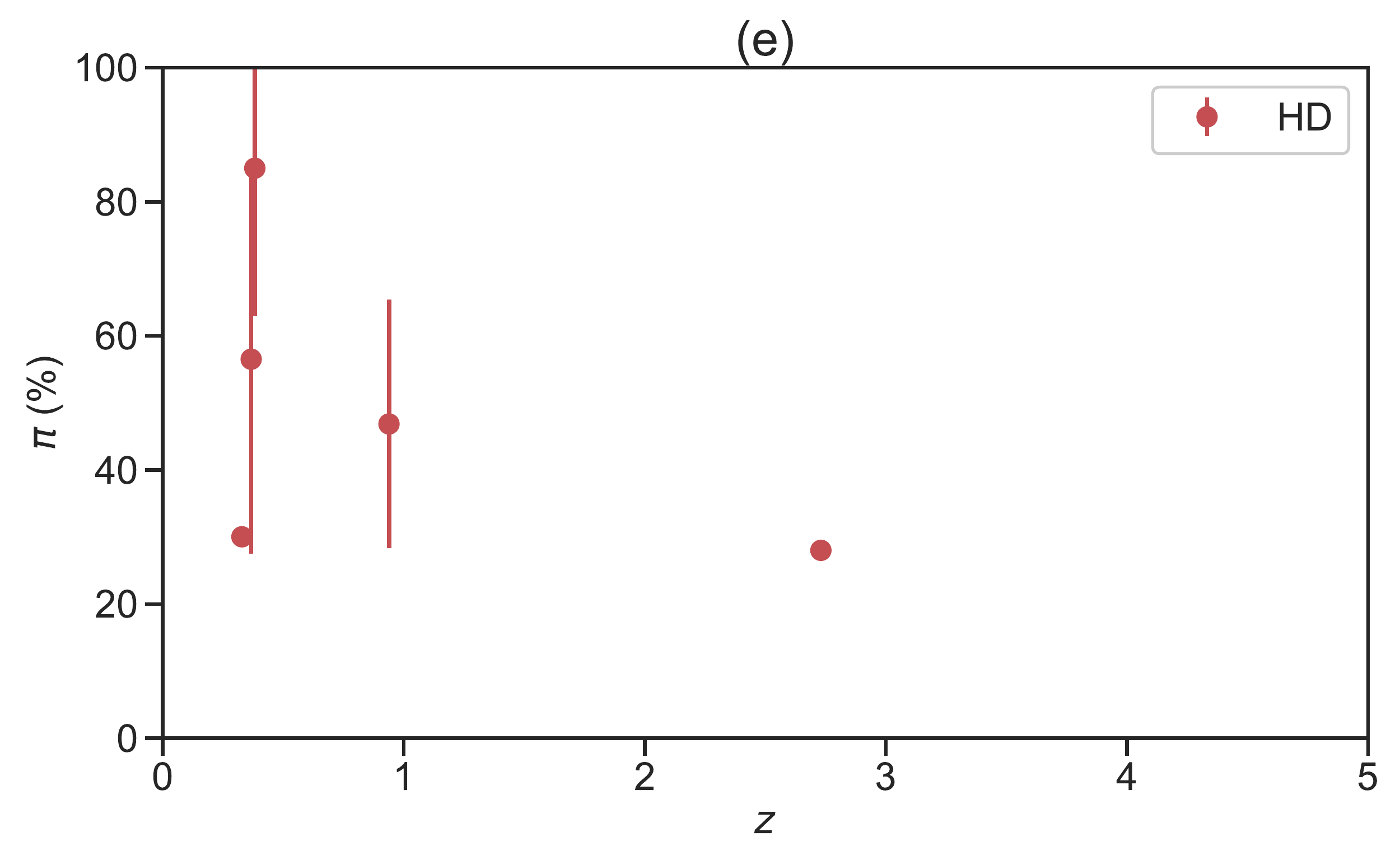}
\caption{Scatter plots of polarization degree $\pi$ versus several other observed quantities: 
(a) the peak energy ($E_{\rm p}$) of the $\nu F_\nu$ prompt emission spectrum, 
(b) the BB temperature $kT$,
(c) the corresponding energy fluence $S_{\gamma}$,
(d) the magnetization parameter $\sigma_{0}$, and
(e) and the redshift $z$. Data points with gray and red colors indicate the PFD and HD bursts, respectively.}.
\label{fig:pipair}
\end{figure*}

\clearpage
\begin{figure*}
\includegraphics[width=0.52\hsize,clip]{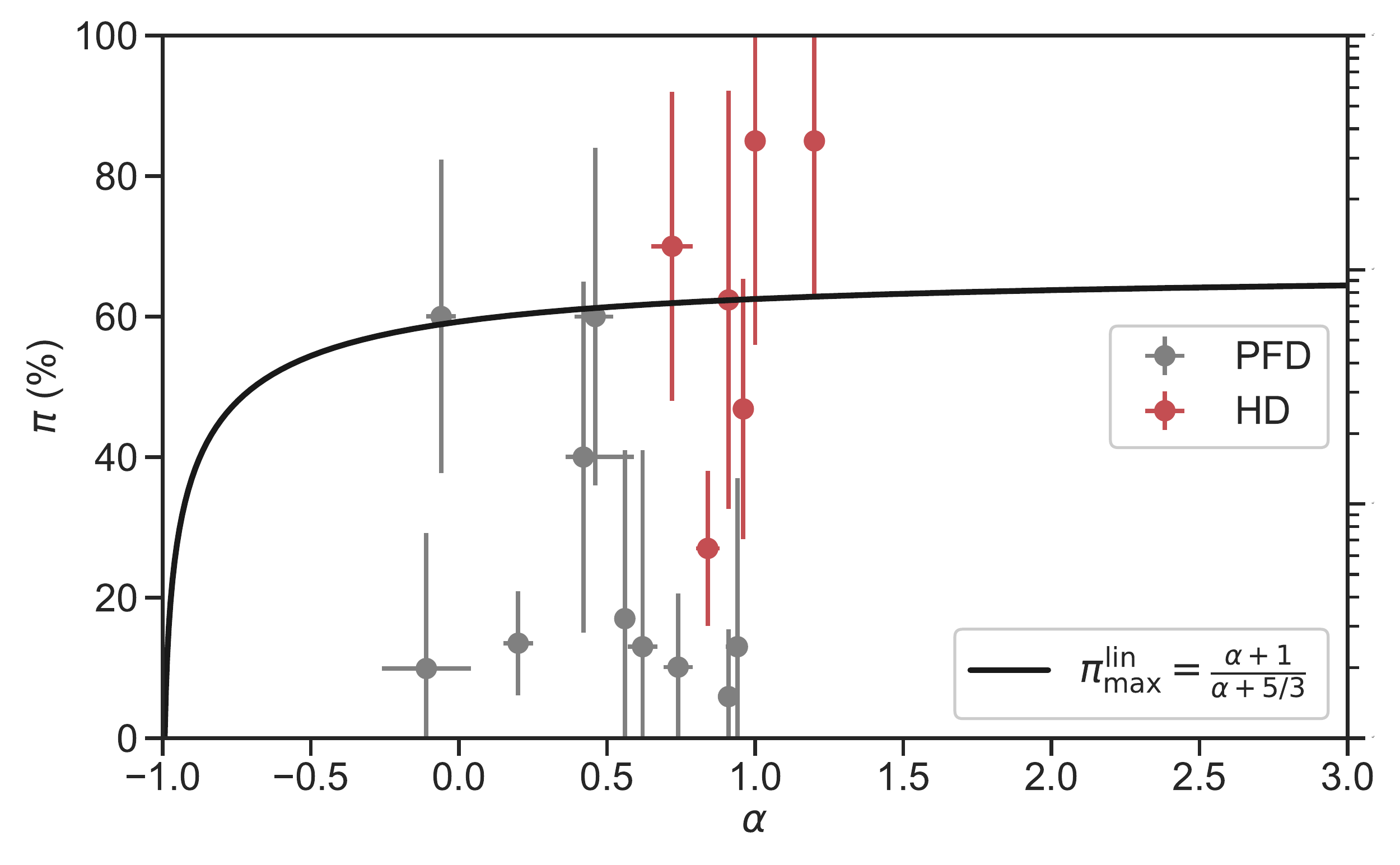}
\includegraphics[width=0.48\hsize,clip]{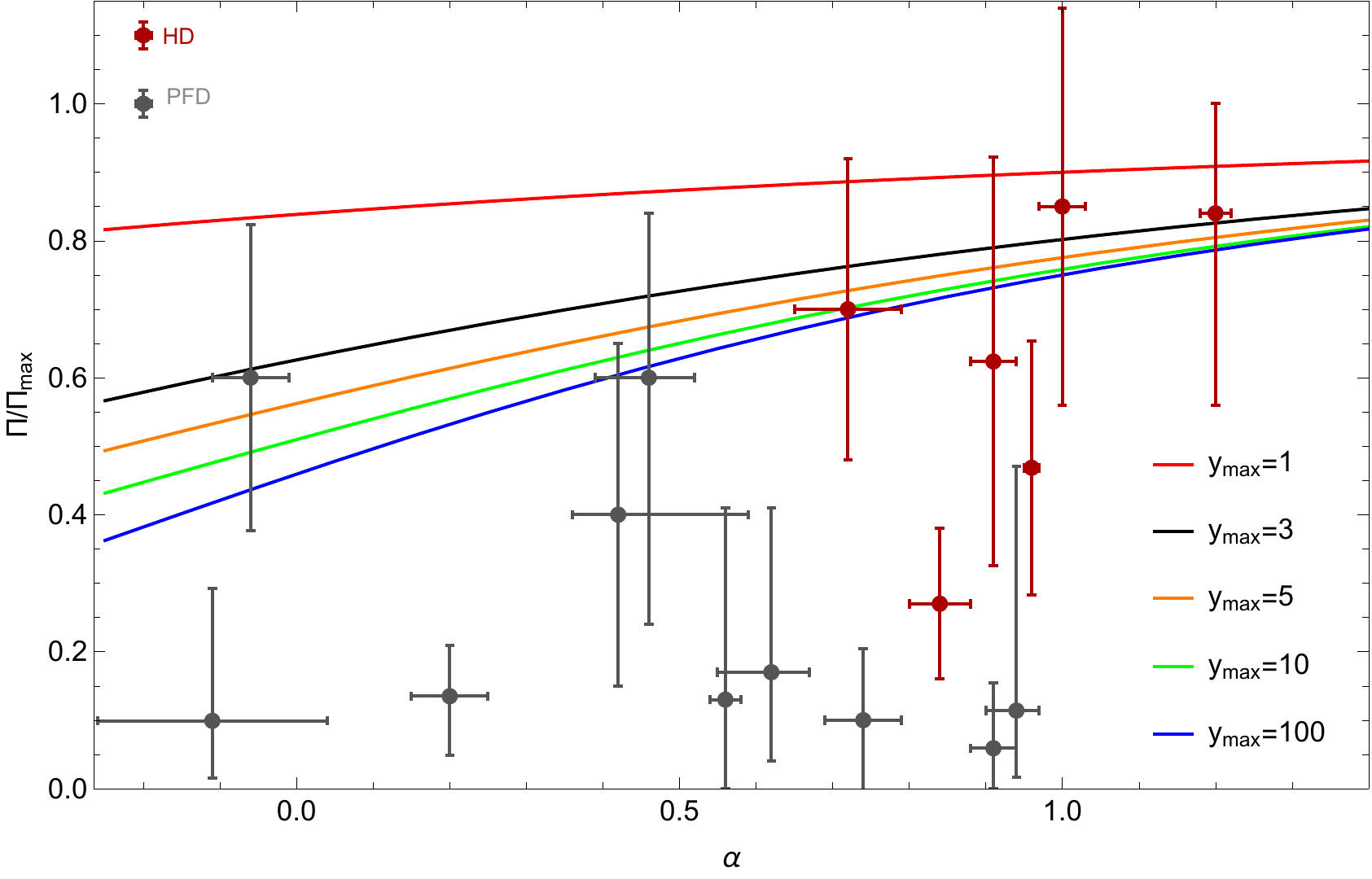}
\caption{Left: the maximum degree of the linear polarization applying synchrotron emission model ($\Pi^{\rm lin}_{\rm max}$  Eq. (\ref{ord})) with observed data using $\alpha$ indices based on a time-integrated spectral analysis. Here, we have plotted the polarization of 16 GRBs out of 27 sources, and we ignored 11 GRBs with only limiting reported values of polarization. Right: time-integrated polarization degree in the presence of an ordered magnetic field $B_{\rm ord}$ within the plane of ejecta (Eq. (\ref{ord})) measured by an on-axis observer ($\theta_{\rm obs}=0$), the evolution of the polarization is plotted in terms of $\alpha$ for different values of $y_{\rm max}=(\Gamma\theta_{\rm max})^2$.}
\label{fig:orderd}
\end{figure*}

\begin{figure*}
\includegraphics[width=0.5\hsize,clip]{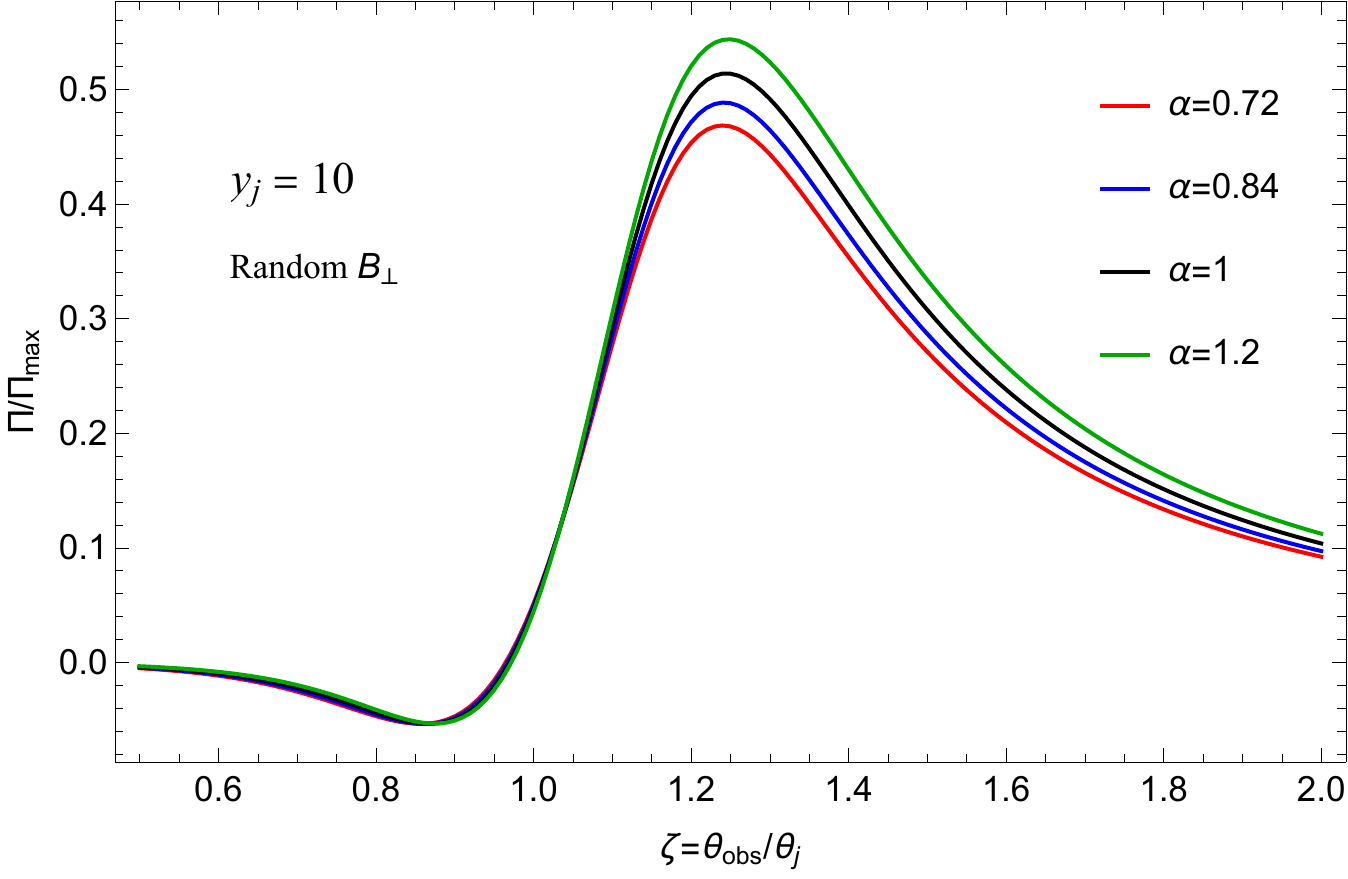}
\includegraphics[width=0.5\hsize,clip]{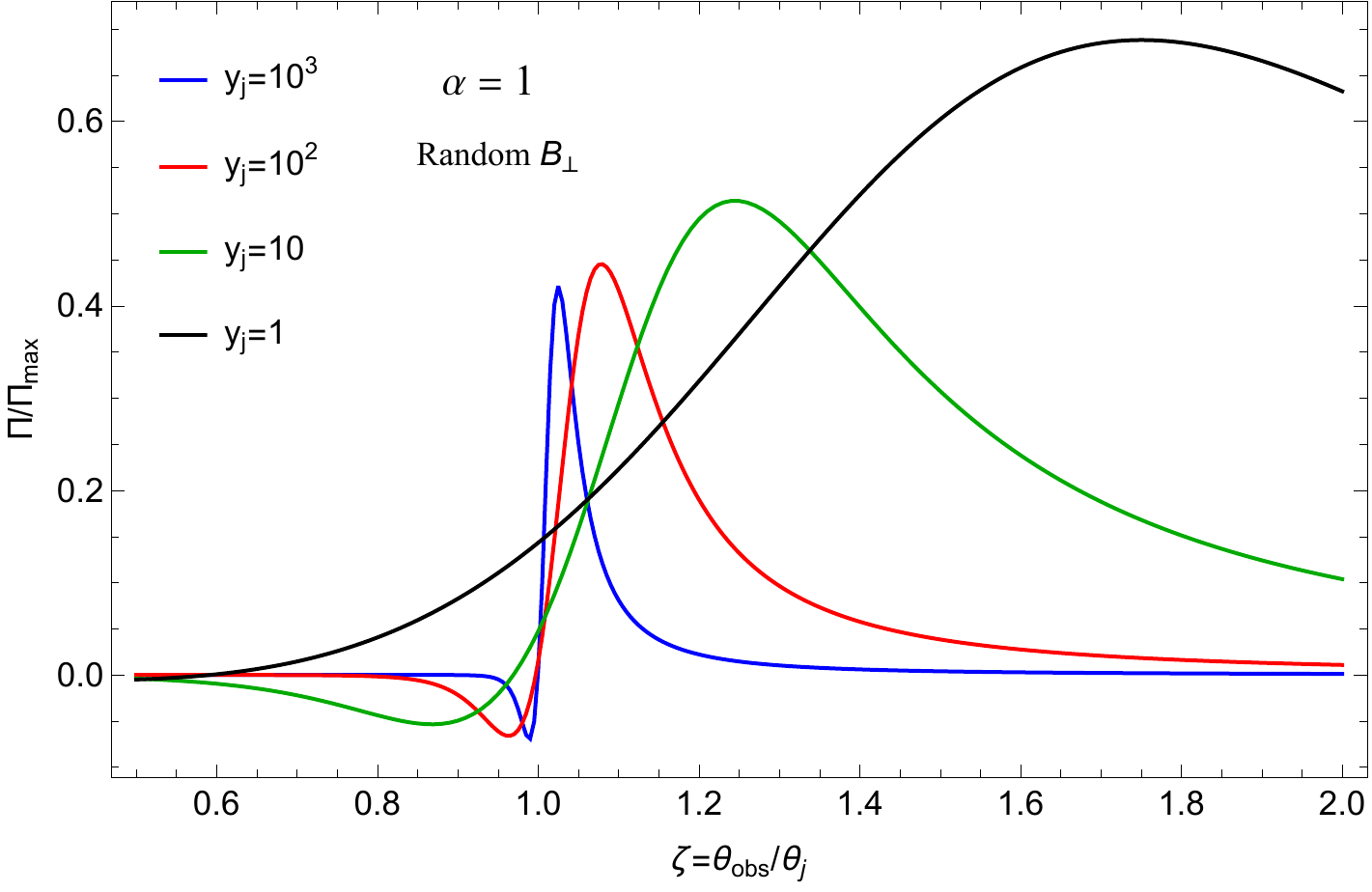}
\caption{The time-integrated polarization for a random magnetic field $B_{\perp}$ that lies entirely in the plane of the shock (Eq. (\ref{orto})) as a function of the off-axis parameter $\zeta=\theta_{obs}/\theta_{j}$ for different values of spectral index $\alpha$ (left) and $y_{j}=(\Gamma \theta_{j})^{2}$ (right) as labeled.}
\label{fig9}
\end{figure*}

\end{document}